\documentclass{article}

\usepackage[a4paper, margin=1.25in]{geometry}

\usepackage{graphicx} 
\usepackage{amssymb, amsmath}
\usepackage{algorithm}
\usepackage{algpseudocode}
\usepackage{subcaption}
\usepackage{caption}
\usepackage{xcolor}
\usepackage{pifont}
\usepackage{alphalph}
\usepackage{tikz}
\usetikzlibrary{positioning}
\usetikzlibrary{fit}
\usepackage{multirow} 
\usepackage{relsize} 
\usepackage{booktabs}
\newcounter{customsubfig}

\usepackage{makecell}
\usepackage{url}

\algnewcommand{\Initialize}[1]{%
  \State \textbf{Initialization:}
  \Statex \hspace*{\algorithmicindent}\parbox[t]{.8\linewidth}{\raggedright #1}
}

\usepackage{rotating}

\newcommand{\verticalsa}[1]{%
  \begin{tabular}{@{}c@{}}
  \rotatebox[origin=bl]{90}{#1}
  \end{tabular}%
}

\usepackage{lipsum}
\usepackage{amsfonts}
\usepackage{epstopdf}
\ifpdf
  \DeclareGraphicsExtensions{.eps,.pdf,.png,.jpg}
\else
  \DeclareGraphicsExtensions{.eps}
\fi

\title{Efficient Bayesian Computation Using Plug-and-Play Priors for Poisson Inverse Problems}

\author{Teresa Klatzer$^{1,4}$ \and Savvas Melidonis$^{2}$ \and Marcelo Pereyra$^{3,4}$ \and Konstantinos C. Zygalakis$^{1,4}$}

\usepackage{amsopn}
\DeclareMathOperator{\diag}{diag}

\date{}

\begin{document}

\maketitle

\begin{abstract}
This paper studies plug-and-play (PnP) Langevin sampling strategies for Bayesian inference in low-photon Poisson imaging problems, a challenging class of problems with significant applications in astronomy, medicine, and biology. PnP Langevin sampling offers a powerful framework for Bayesian image restoration, enabling accurate point estimation as well as advanced inference tasks, including uncertainty quantification and visualization analyses, and empirical Bayesian inference for automatic model parameter tuning. {Herein, we leverage and adapt recent developments in this framework to tackle challenging imaging problems involving weakly informative Poisson data.}
Existing PnP Langevin algorithms are not well-suited for low-photon Poisson imaging due to high solution uncertainty and poor regularity properties, such as exploding gradients and non-negativity constraints. {To address these challenges, we explore two strategies for extending Langevin PnP sampling to Poisson imaging models}: (i) an accelerated PnP Langevin method that incorporates boundary reflections and a Poisson likelihood approximation and (ii) a mirror sampling algorithm that leverages a Riemannian geometry to handle the constraints and the poor regularity of the likelihood without approximations. {The effectiveness of these approaches is evaluated and contrasted through extensive numerical experiments and comparisons with state-of-the-art methods. The source code accompanying this paper is available at \url{https://github.com/freyyia/pnp-langevin-poisson}}.
\end{abstract}

\footnotetext[1]{School of Mathematics, University of Edinburgh, Edinburgh, EH9 3FD, UK}
\footnotetext[2]{Forschungszentrum J\"ulich GmbH, 52425 J\"ulich, Germany}
\footnotetext[3]{School of Mathematical and Computer Sciences, Heriot-Watt University, Edinburgh, EH14 4AS, UK}
\footnotetext[4]{Maxwell Institute for Mathematical Sciences, Bayes Centre, 47 Potterrow,  EH8 9BT, Edinburgh, UK}

\section{Introduction}
Low-photon Poisson imaging problems are ubiquitous in scientific and engineering applications, particularly in scenarios involving low illumination or short acquisition times (see, e.g., \cite{starck06, Bertero_2009, Hohage_2016} for excellent introductions to the topic). Poisson-distributed measurements arise for instance from the use of single-photon detectors that discriminate individual photons within a given time frame \cite{bin_geo,7932527,7150537}, as well as from standard CMOS cameras that operate under poorly illuminated conditions \cite{4472247}. As a result, Poisson imaging problems play a crucial role in astronomy and remote sensing \cite{lanteri-05, figueiredo10, rasti2018}, where imaging systems often operate under limited illumination conditions, in biomedical microscopy, where photon counts are limited to minimize photo-toxicity \cite{SarderMicroReview06}, and in nuclear medical imaging \cite{zhou2020bayesian, melba:2024:001:singh}, where emission-based modalities such as PET and SPECT produce Poisson-distributed measurement data with statistics that are directly related to the amount of radiation used.
  
\begin{figure}[t!]
 \centering
    \begin{minipage}[c]{0.9\textwidth}
    \centering
    \begin{tikzpicture}
    
    \foreach \x/\image in {0/noisy2poisson60, 2/noisy2poisson40, 4/noisy2poisson20, 6/noisy2poisson10, 8/noisy2poisson5} {
        \node[anchor=south] at (\x, 0) {\includegraphics[width=0.16\textwidth]{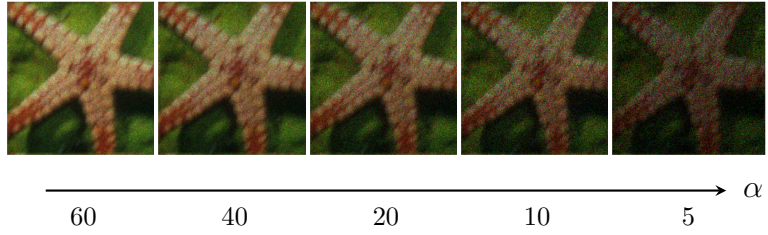}};
    }
    \draw[thick, -{stealth[scale=3.0]}] (-0.5, -0.3) -- (8.5, -0.3) node[midway, below] {};
    \vspace{0.1cm}
    \foreach \x/\val in {0/60, 2/40, 4/20, 6/10, 8/5} {
        \node[anchor=north] at (\x, -0.4) {\val};
    }
    
    \node[anchor=west] at (8.6, -0.3) {\large $\alpha$};

    \end{tikzpicture}
    \caption{Illustration of different photon levels $\alpha$ on a blurred test image (kernel Fig.\ref{fig:levin-1})}
    \label{fig:photon-levels}

    \end{minipage}
\end{figure}

We herein consider Poisson imaging problems where one seeks to recover an unknown image of interest $x \in \mathbb{R}^n$ from a linear measurement $y \in \mathbb{R}^m$ corrupted by Poisson noise, with likelihood function given by
\begin{equation}\label{eq:PoissonLikelihoodNoBeta}
   p(y|x) \propto \prod_{i=1}^m\exp\left\{y_i\log(\alpha(Ax)_i) -\alpha(Ax)_i
- \iota_{\mathbb{R}^+_0} (x_i)\right\}.
\end{equation}
where $A \in \mathbb{R}_+^{m\times n}$  
 is a positive linear measurement operator representing deterministic aspects of the data acquisition process, $\alpha > 0$ is a scalar related to the level of shot noise contaminating the measurements (the smaller the value of $\alpha$, the harder the problem, see Fig. \ref{fig:photon-levels} above), $i$ denotes the $i$-th element of a vector, and $\iota_{\mathbb{R}^+_0}$ is the indicator function enforcing a non-negativity constraint on the elements of $x$. We assume that $A$ is known (i.e., the problem is non-blind) and that $A^TA$ is rank deficient or exhibits a poor conditioning number, making the estimation problem challenging.

A main difficulty in Poisson imaging problems is the signal-dependent nature of Poisson noise, which leads to a signal-to-noise ratio that varies significantly across the image. In addition, in dark regions or low-illumination conditions, the noise distribution is strongly non-Gaussian. As a result, conventional image restoration techniques that rely on $\ell_2$-norm data-fidelity terms struggle with Poisson noise, especially in low-photon settings (for signals in the range $[0,1]$, we see that conventional techniques break when $\alpha \leq 20$ \cite{sanghvi2021photonlimited}). Of course, the above difficulties are amplified in problems that are ill-posed or ill-conditioned, as considered in this paper. Such cases require introducing a significant amount of regularization to deliver accurate solutions.

Modern image reconstruction methods for Poisson imaging problems rely strongly on data-driven regularization techniques derived from machine learning (ML) (see, e.g., \cite{sanghvi2021photonlimited, Chen2022, tachella2025unsure, melidonis2025scorebased}). We focus on so-called Plug-and-Play (PnP) strategies that encode the regularization function through an image denoising operator, which is embedded within an iterative computation algorithm in lieu of the regularizer's gradient or proximal operator, and used in combination with an explicit data-fidelity observation model that is specified during inference time (see \cite{Mukherjee23} for a recent survey on PnP methods with theoretical guarantees). PnP approaches are predominantly derived from optimization schemes that deliver point estimators of $x$ from $y$ \cite{Mukherjee23}, or alternatively from stochastic sampling schemes that deliver a Monte Carlo approximation of a posterior distribution $p(x|y)$ defined implicitly by the image denoiser used \cite{laumont22pnp,Coeurdoux2024,Sun2024,melidonis2025scorebased}. 

In this paper, we study novel PnP Markov chain Monte Carlo (MCMC) techniques suitable for Poisson imaging problems. Such techniques \cite{laumont22pnp,Coeurdoux2024,Sun2024} are valuable because they allow forms of inference that are not accessible with PnP techniques based on optimization \cite{hurault2023} or denoising diffusion models \cite{melidonis2025scorebased}. For example, PnP MCMC techniques can be embedded within empirical Bayesian machinery to tackle semi-blind imaging problems \cite{kemajou24} or to automatically optimize regularization parameters \cite{vidal-2020,tan2024unsupervised}. PnP MCMC techniques are also valuable for performing uncertainty quantification analyses \cite{laumont22pnp, Liaudat2024}. Unfortunately, to the best of our knowledge, the PnP MCMC techniques currently available are not suitable for Poisson imaging problems due to fundamental issues related to the poor regularity properties of \eqref{eq:PoissonLikelihoodNoBeta}. 

{The primary goal of this paper is to make Plug-and-Play (PnP) MCMC a viable and powerful tool for Bayesian inference in challenging Poisson imaging problems, addressing an important gap where existing PnP Langevin methods are ill-suited due to the poor regularity of the likelihood and constraints. Our contributions are threefold. First, we systematically adapt and extend state-of-the-art Euclidean and non-Euclidean samplers to the specific difficulties of the PnP Poisson setting, including the first implementation and investigation of a PnP Mirror Langevin algorithm (PnP-MLA) for an imaging problem. Second, we present the first comprehensive empirical study that directly compares these different geometric and accelerated MCMC methodologies (reflected, projected, mirror) within a unified PnP framework, yielding practical insights beyond the scope of theoretical analysis.  Finally, this comparative analysis provides a set of recommendations for practitioners on selecting the appropriate algorithm and denoiser for various low-photon regimes.}

The remainder of this article is structured as follows:  Section \ref{sec:bayes-comp} reviews key concepts in Bayesian computation relevant to Bayesian imaging with PnP priors in the context of Gaussian observation models, and explains the main challenges in applying this approach for Poisson imaging problems. Section \ref{sec:methodology} presents two methodologies that address these difficulties and extend PnP Bayesian computation to Poisson problems: (i) the reflected PnP-SKROCK, an accelerated sampling scheme that leverages a variant of the Langevin diffusion tailored for constrained domains; and (ii) the PnP Mirror Langevin algorithm, which generalizes PnP-ULA to a non-Euclidean geometry where the Poisson likelihood function is Lipschitz differentiable. In Section \ref{sec:experiments}, we present an extensive experimental analysis that thoroughly explores different neural network denoiser architectures and algorithmic choices for the considered Poisson problems, as well as comparisons with alternative strategies from the state-of-the-art. {A thorough discussion,  conclusions and perspectives for future work are finally reported in Sections \ref{sec:discussion} and \ref{sec:conclusion}, respectively.}

\section{Bayesian imaging with Plug-and-Play priors}\label{sec:bayes-comp}
\subsection{The overdamped Langevin diffusion process}
We consider performing Bayesian computation for Poisson imaging models of the form
\begin{equation} \label{eq:bayes}
	p(x|y) = \frac{p(x)p(y|x)}{\int p(y|\tilde{x}) p(\tilde{x}) d\tilde{x}}\,
\end{equation}
where $p(y|x)$ is given by \eqref{eq:PoissonLikelihoodNoBeta} and where $p(x)$ is an image prior encoded by a ML model, which will be discussed in detail in Section \ref{Sec:PnP}.
The canonical approach to computing posterior probabilities, expectations and estimators in Bayesian imaging problems is to employ a Markov Chain Monte Carlo (MCMC) method to draw samples from the posterior distribution of interest $p(x|y)$, followed by Monte Carlo integration \cite{Robert2010-lh}. In particular, Bayesian imaging methods often rely on discretizations of the overdamped Langevin diffusion process, which scales efficiently to high dimensional problems \cite{durmus2017}. This diffusion is governed by the following stochastic differential equation (SDE)
\begin{equation} \label{eq:Langevin_plain}
dX_{t}=\nabla \log p(X_t)dt + \nabla\log p(y|X_t)dt+\sqrt{2}dW_{t}\, ,
\end{equation}
where $W_t$ is a $d$-dimensional Brownian motion \cite{durmus2017}. Several discrete-time approximations of \eqref{eq:Langevin_plain} can be considered, but the most common choice is the Euler Maruyama (EM) approximation, leading to the so-called Unadjusted Langevin Algorithm (ULA)
\begin{equation*}
    X_{k+1}= X_k +\delta\nabla\log p(X_k) +\delta\nabla\log p(y|X_k) + \sqrt{2\delta}\xi_{k+1},
\end{equation*}
with a time step $\delta$ and $\xi_k$ being i.i.d. standard Gaussian random variables. There exists a vast literature analyzing the properties of ULA under the assumption that $\nabla \log p{(x|y)}$ is Lipschitz, with detailed convergence guarantees in the log-concave \cite{Dalalyan17, roberts1996, durmus2017, LM12} and non log-concave settings \cite{majka20notconcave, erdogdu2021convergence, cheng18noncvxlangevin}. Similar results have been established for variants of ULA suitable for non-smooth models \cite{langevinmeetsmoreau18, chatterji2020langevin, lehec23nonsmooth, GHADERI2024-nonsmooth}, as well as for models and ULAs using machine-learning priors under mild regularity conditions such as Lipschitz continuity and boundedness (see, e.g., \cite{laumont22pnp, cai2023nfula}). In the following section, we introduce the PnP Langevin framework \cite{laumont22pnp} which we adopt in this paper.

\subsection{Plug-and-Play priors}\label{Sec:PnP}
Bayesian imaging methods have traditionally relied on model-driven priors that enforce specific expected properties in the solution, such as smoothness, sparsity, or piecewise regularity. However, these are often not informative enough to deliver accurate reconstructions in challenging Poisson imaging problems. Therefore, we consider data-driven priors encoded by ML models, which can leverage large data sets of clean images to derive highly informative statistical image priors. In particular, herein we focus on priors encoded by ML-based image denoising operators, which can be embedded within iterative Monte Carlo sampling or optimization schemes in a PnP manner (see, e.g., \cite{ryu19a, Pesquet2020LearningMM, hurault22a, hauptmann_2024,  prov-qnewton-pnp-24} for examples of PnP optimization schemes and \cite{laumont22pnp, cai2023nfula, renaud2024plugandplay, mbakam2024empiricalbayesianimagerestoration} for examples of PnP sampling schemes). Here, we adopt the Bayesian PnP framework introduced in \cite{laumont22pnp}.

The main idea behind the Bayesian PnP approach \cite{laumont22pnp} is to construct a Bayesian model $p(x|y)$ where the prior knowledge is represented in the form of an image denoising operator $D_{\epsilon}$, rather than through an explicit prior distribution $p(x)$. This operator estimates the posterior expectation of $x$ given a noisy observation $x^\prime \sim \mathcal{N}(x,\epsilon \text{Id})$, with noise variance $\epsilon$. In practice, $D_{\epsilon}$ is implemented by a deep neural network trained on a data set of clean and noisy image pairs  $\{x_i, x'_{i}\}_{i=1}^{N}$. The connection between $D_{\epsilon}$ and the prior $p(x)$ stems from Tweedie's identity, which states that if $D_\epsilon$ is close to the posterior expectation of $x$ given $x^\prime \sim \mathcal{N}(x,\epsilon \text{Id})$, then 
\begin{equation}\label{eq:tweedie-score1}
     \nabla \log p_\epsilon(x) \approx \frac{1}{\epsilon}(D_\epsilon(x) - x).
\end{equation}
where $p_\epsilon(x)$ is a smooth approximation of $p(x)$ defined as $p_\epsilon(x)= \int k_\epsilon(x, \Tilde{x}) p(\Tilde{x})d\Tilde{x}$, obtained via convolution with a Gaussian smoothing kernel $k_\epsilon$ of bandwidth $\sqrt{\epsilon}$. As $\epsilon$ decreases, the approximation $p_\epsilon$ converges to $p$, at the expense of reduced smoothness. 

In the context of PnP Langevin sampling methods, the gradient $\nabla \log p(x)$ in the Langevin SDE is first replaced by $\nabla \log{p_{\epsilon}(x)}$ and subsequently approximated by \eqref{eq:tweedie-score1} by using a denoising operator $D_\epsilon$ that has been trained so that \eqref{eq:tweedie-score1} holds. When combined with an ULA scheme, this leads to the PnP-ULA \cite{laumont22pnp} 
\begin{equation} \label{eq:PnP_ULA}
    X_{k+1}= X_k +\delta\nabla\log p(y|X_k) + \frac{\delta}{\epsilon}(D_\epsilon(X_k) - X_k)+ \sqrt{2\delta}\xi_{k+1}.
\end{equation}

An alternative approach, known as projected PnP-ULA, modifies PnP-ULA by enforcing a hard projection onto a constraint set $C$. The projection ensures better geometric ergodicity properties for the generated Markov chain and also guarantees that the chain stays within the region of the solution space in which $D_\epsilon$ has been trained, avoiding out-of-distribution evaluations where $D_\epsilon$ can potentially behave erratically. In particular, we have the following recursion  
\begin{equation} \label{eq:PPnP-ULA}
X_{k+1} =\Pi_{C}(X_k +\delta\nabla\log p(y|X_k) + \frac{\delta}{\epsilon}(D_\epsilon(X_k) - X_k)+ \sqrt{2\delta}\xi_{k+1}). 
\end{equation}

When dealing with Poisson noise, two main challenges hinder the direct application of PnP-ULA or PPnP-ULA.  First, the continuous-time Langevin process is not well-defined because of the non-negativity constraint on $x$. Second, the gradient $x \mapsto \nabla \log p(y|x)$ for \eqref{eq:PoissonLikelihoodNoBeta} is not globally Lipschitz, which is important for guaranteeing the convergence of the Langevin process to $p(x|y)$.  In the next section, we discuss two approaches to deal with these challenges, each leading to new PnP algorithms.

\section{Proposed methodology} \label{sec:methodology}
We present two strategies to extend the conventional PnP Langevin approach to Poisson Bayesian imaging problems. Instead of directly modifying PnP algorithms, we introduce key adjustments at the level of the continuous-time SDE. These ensure that the algorithms resulting from time discretizations of the SDE are robust to the non-smoothness and constraints of \eqref{eq:PoissonLikelihoodNoBeta}. Note that, for presentation clarity, in a slight abuse of notation, we use $\nabla \log p_\epsilon(x)$ as a shorthand for the prior score function, although the algorithms are in practice implemented by using denoisers that do not verify Tweedie's identity exactly. This allows us to consider a wide range of denoiser architectures, including non-Euclidean denoisers, without having to redefine the algorithms.

\subsection{Langevin diffusion for constrained domains}
\subsubsection{The reflected overdamped Langevin diffusion process}
Building on \cite{savvas23}, we first consider the following reflected Langevin SDE (RSDE) to sample approximately from $p(x|y)$
\begin{equation}\label{eq:sde-ct}
    dX_t = \nabla \log p^\beta(y|X_t)dt + \nabla\log p_\epsilon(X_t)dt  + \sqrt{2}dW_t + d\kappa_t
\end{equation}
where $x\mapsto p^\beta(y|x)$ with $\beta >0$ is a regularized approximation of the original likelihood \eqref{eq:PoissonLikelihoodNoBeta}, given by
\begin{equation}\label{eq:PoissonLikelihood}
p^\beta(y|x) \propto \prod_{i=1}^{m}\exp\left\{y_i\log(\alpha(Ax)_i + \beta) - \alpha(Ax)_i - \beta - \iota_{\mathcal{R}_0^+}(x_i)\right\}\,,
\end{equation}
where $\kappa_t$ is a local time that increases only on the boundary $\partial\mathbb{R}^n_+$. The local time enforces the non-negativity constraint, as required by \eqref{eq:PoissonLikelihoodNoBeta} \cite{pilipenko2014introduction, savvas23}, while taking $\beta > 0$ ensures that $\nabla \log{p}^{\beta}(y|x)$ is Lipschitz-continuous on the positive orthant\footnote{The constant $\beta > 0$ can be interpreted as a constant background noise level.}. Note that setting $\beta = 0$ recovers the original likelihood \eqref{eq:PoissonLikelihoodNoBeta}, and that the bias stemming from using $\beta > 0$ can be made arbitrarily small by reducing the value of $\beta$, at the expense of convergence speed (see \cite[Section 4.4]{savvas23} for recommendations regarding setting $\beta$).

The RSDE \eqref{eq:sde-ct} can be approximated by a standard EM scheme, leading to a constrained ULA. Two standard ways to discretize this RSDE ensuring that the samples are always in $\mathbb{R}^n_{++}$ are using a reflected \cite{savvas23} or a projected Euler scheme \cite{laumont22pnp}.  The reflected EM scheme is defined by
\begin{equation}\label{eq:reflected-euler}
 X_{k+1}=\left|X_k -  \delta \nabla \log p^\beta(y|X_k) - \delta\nabla \log p_\epsilon(X_k) + \sqrt{2\delta} \xi_{k+1}\right|
\end{equation}
where $\delta$ and $\xi_k$ remain as previously, and $|.|$ denotes the component-wise absolute value. The projected EM scheme is defined by 
\begin{equation}\label{eq:projected-euler}
    X_{k+1}=\left(X_k -  \delta \nabla \log p^\beta(y|X_k) - \delta\nabla \log p_\epsilon(X_k) + \sqrt{2\delta} \xi_{k+1}\right)^+
\end{equation}
where $(.)^+$ is the projection to the positive orthant. Under mild regularity assumptions on $p_\epsilon$, both approaches are by construction well-posed and converge exponentially fast to a neighborhood of the regularized target $p_{\beta,\epsilon}(x|y)\propto p^\beta(y|x)p_\epsilon(x)$, with the reflected EM schemes \eqref{eq:reflected-euler} typically exhibiting a smaller bias \cite{savvas23}.

\subsubsection{Reflected and Projected PnP Langevin algorithms} \label{sec:RPnP-PPnP}
PnP priors encoded by neural network denoisers can become unreliable when evaluated on out-of-distribution data \cite{laumont22pnp}. Therefore, to enable the reliable use of recursions \eqref{eq:reflected-euler} and \eqref{eq:projected-euler} for sampling, we impose an additional constraint to ensure that the iterates remain within the domain of $D_\epsilon$, which we henceforth denote by $C \subset \mathbb{R}^d$ {and assume to be convex and compact. This constraint can be enforced using either reflection on the boundary $\partial C$ or projection onto $C$, leading to two variants of the PnP-ULA. In both cases, the Langevin update step targets a posterior with log-density of the form
\begin{equation*}
    \log p_{\beta, \epsilon}(x|y) \propto \log p^\beta(y|x) +  \rho  \log p_\epsilon(x)\,,\quad\forall x \in C 
\end{equation*} 
where $\rho >0 $ is a regularization parameter. The prior score, $\nabla \log p_\epsilon(x)$, is related to the denoiser $D_\epsilon$ using Tweedie's identity \eqref{eq:tweedie-score1} as $\nabla \log p_\epsilon(x) = \frac{1}{\epsilon}(D_\epsilon (x) - x)$. For specific network architectures, such as gradient step denoisers discussed in Section \ref{subsec:data-driven-priors}, this expression can be further simplified. 

The resulting algorithms are PnP-ULA with reflection (RPnP-ULA, see Algorithm \ref{alg:pnpula-reflect}), and PnP-ULA with projection (PPnP-ULA, see Algorithm \ref{alg:ppnpula}). These algorithms require specifying the number of iterations $N$, the observation $y$, the photon level $\alpha$ (which appears within $\nabla \log p^\beta$), a noise level $\epsilon$ for the used denoiser, and a step size $\delta$.
For RPnP-ULA, a reflection operator $\mathcal{R}_C$ is applied after each update step to keep the iterates within the set $C$. The choice of the step size $\delta$ for RPnP-ULA is governed by stability bounds given in \cite{laumont22pnp}, which requires $\delta < \frac{1}{3}\operatorname{Lip}(\nabla \log p_{\beta, \epsilon}(x|y))^{-1}$ where $\operatorname{Lip}(.)$ denotes the Lipschitz continuity constant. In contrast, PPnP-ULA uses a projection operator $\Pi_C$ to enforce the constraint. PPnP-ULA admits a larger step size than its reflected counterpart, at the expense of introducing additional estimation bias \cite{laumont22pnp}. For a constraint set $C \in [a,b]^d$, the projection operator is applied component-wise. The projection operator is given as $\Pi_C(x)_i = \min(\max(a,x_i),b)$, and the reflection operator can be defined in terms of the projection as $\mathcal{R}_C(x)_i = 2\Pi_C(x)_i-x_i$.
}

\begin{algorithm}[t]
\caption{PnP-ULA with Reflection (RPnP-ULA)}\label{alg:pnpula-reflect}
\begin{algorithmic}
\Require $N \in \mathbb{N}, y \in \mathbb{R}^m, \epsilon, \delta > 0, \rho > 0, C \subset \mathbb{R}^n$ convex and compact
\Initialize {Set $X_0 \in \mathbb{R}^n_{++}$ and $k=0.$}
\For{$k = 0 : N $}
\begin{align*}
Z_{k+1} &\sim \mathcal{N}(0,\text{Id}) \\
X_{k+1}&= \mathcal{R}_C \left(X_k+\delta\nabla\log p^\beta(y|X_k) + \delta \rho \nabla \log p_\epsilon(X_k) + \sqrt{2\delta}Z_{k+1}\right)
\end{align*}
\EndFor
\end{algorithmic}
\end{algorithm}
\begin{algorithm}[t]
\caption{PnP-ULA with Projection (PPnP-ULA)}\label{alg:ppnpula}
\begin{algorithmic}
\Require $N \in \mathbb{N}, y \in \mathbb{R}^m, \epsilon, \delta > 0, \rho > 0, C \subset \mathbb{R}^n$ convex and compact
\Initialize {Set $X_0 \in \mathbb{R}^n_{++}$ and $k=0.$}
\For{$k = 0 : N $}
\begin{align*}
Z_{k+1} &\sim \mathcal{N}(0,\text{Id}) \\
X_{k+1}&=  \Pi _C \left(X_k+\delta\nabla\log p^\beta(y|X_k) + \delta \rho \nabla \log p_\epsilon(X_k) + \sqrt{2\delta}Z_{k+1} \right)  
\end{align*}
\EndFor
\end{algorithmic}
\end{algorithm}

One potential drawback of EM-based algorithms such as Algorithms  \ref{alg:pnpula-reflect} and \ref{alg:ppnpula} is that they might converge slowly due to a step size restriction \cite{PVZ20}. One algorithm that alleviates this step size restriction is the stochastic orthogonal Runge-Kutta-Chebyshev (SKROCK) method \cite{PVZ20}, which can be shown to behave similarly to ``accelerated'' optimization algorithms in terms of convergence to equilibrium for Gaussian targets.  A follow-up method called reflected SKROCK (RSKROCK) was proposed in \cite{savvas23}  to deal with Poisson noise and analytical model-based priors by adding a reflection to SKROCK, and was shown to behave in an accelerated manner. 

Following on from this, in this paper we propose a so-called ``accelerated'' variant of RPnP-ULA: PnP-SKROCK with reflection (RPnP-SKROCK, see Algorithm \ref{alg:euclidean-skrock}). Unlike ULAs that perform a single gradient evaluation per iteration, SKROCK requires $s \in \mathbb{N}$ gradient evaluations per iteration, allowing SKROCK to take much longer integration steps. While the cost per iterations increases by a factor $s$, in problems that are ill-conditioned or ill-posed, this can lead to an improvement in convergence speed of the order of $s^2$. We refer the reader to \cite{PVZ20} for more details. In all our experiments, we use $s=10$ and Runge-Kutta expansion parameter $\eta = 0.05$.

\begin{algorithm}[t]
\caption{Reflected PnP-SKROCK (RPnP-SKROCK)}\label{alg:euclidean-skrock}
\label{alg:r_SKROCK}
\begin{algorithmic}
\Require $N \in \mathbb{N}, y \in \mathbb{R}^m, \epsilon, \delta > 0, \rho > 0, \eta, C \subset \mathbb{R}^n$ convex and compact
\State{\textbf{Compute} $l_{s}=(s-0.5)^{2}(2-4/3\eta)-1.5\vspace{0.2cm}$}
\State{\textbf{Compute} $\omega_{0} = 1+\dfrac{\eta}{s^{2}},\hspace{0.35cm} \omega_{1} = \dfrac{T_{s}(\omega_{0})}{T^{'}_{s}(\omega_{0})},\hspace{0.35cm} \mu_{1}  = \dfrac{\omega_{1}}{\omega_{0}},\hspace{0.35cm} \nu_{1}=s\omega_{1}/2,\hspace{0.35cm} k_{1} = s\omega_{1}/\omega_{0} \vspace{0.2cm}$}
\State{\textbf{Choose} $\delta\in(0,\delta_{s}^{max}]$, where $\delta_{s}^{max}= l_{s}/\operatorname{Lip}(\nabla \log p_{\beta,\epsilon}(x|y))$; $\rho > 0$}
\State{\textbf{Initialization} $X_{0}\in \mathbb{R}^{n}_{++}$ and $k=0$ }
\For{$k = 0:(N-1)$}
\State{$Z_{k+1} \sim \mathcal{N}(0,\mathbb{I}_{n})$}
\vspace{0.1cm}
\State{$K_{0} = X_{k}$}
\State{$W_{1} = \mathcal{R}_C\left(X_{k} + \nu_{1}\sqrt{2\delta}Z_{k+1}\right)$}
\State{$Y_{1} = X_{k} - \mu_{1}\delta\nabla \log p^\beta(y|W_{1}) + \mu_{1}\delta \rho \nabla \log p_\epsilon( W_{1}) 
+ k_{1}\sqrt{2\delta}Z_{k+1}$}
\State{$K_{1} = \mathcal{R}_C\left(Y_{1}\right)$}
\vspace{0.1cm}
\For{$j = 2:s$}
\State{\textbf{Compute}}
$\mu_{j} = \dfrac{2\omega_{1}T_{j-1}(\omega_{0})}{T_{j}(\omega_{0})},\hspace{0.35cm} \nu_{j}=\dfrac{2\omega_{0}T_{j-1}(\omega_{0})}{T_{j}(\omega_{0})},\hspace{0.35cm} k_{j} = -\dfrac{T_{j-2}(\omega_{0})}{T_{j}(\omega_{0})}= 1-\nu_{j}\vspace{0.2cm}$
\State{$K_{j} =\mathcal{R}_C(-\mu_{j}\delta\nabla \log p^\beta(y|K_{j-1}) +  \mu_{j}\delta \rho \nabla \log p_\epsilon( K_{j-1}) $}
\hspace{1cm}\State{$ + \nu_{j}K_{j-1}+k_{j}K_{j-2})$}
\EndFor
\State{$X_{k+1} = K_{s}$}
\EndFor \\
\Return $\{X_{k}: k\in\{1,\ldots,N\}\}$
\end{algorithmic}
\end{algorithm}

\subsection{A Riemannian overdamped Langevin diffusion process}
An interesting alternative to the RSDE approach described previously is to tackle the constraints and non-Lipschitz gradients by modifying the geometry underpinning the Langevin SDE. This can be achieved by using the so-called mirror Langevin scheme \cite{zhang20mirror}, analogue to mirror descent optimization algorithms \cite{nemirovskij1983mirror}. Mirror optimization algorithms have been introduced by \cite{nemirovskij1983mirror} and can be viewed as projected sub-gradient (or Bregman proximal gradient) methods, derived from using a Bregman divergence instead of the usual Euclidean squared distance to alter the geometry of the problem \cite{BECK03mirror}. In mirror schemes, this modification from Euclidean to (Hessian) Riemannian geometry is encoded through the so-called mirror map \cite{BECK03mirror}. Following \cite{zhang20mirror}, we consider the following (Riemannian) mirror Langevin SDE to sample from a generic distribution $\pi$
\begin{equation}\label{eq:MLA}
\begin{split}
        X_t &= \nabla \phi^*(Y_t) \\
    dY_t &= \nabla \log \pi (X_t) dt + \sqrt{2}[\nabla^2\phi(X_t)]^{1/2}W_t 
    \end{split}
\end{equation}
where $\phi$ is the mirror map, which we assume to be $\mathcal{C}^2(\mathcal{X})$ Legendre-type convex where $\mathcal{X}$ denotes the support of $\pi$, and where $\phi^*$ the Legendre-Fenchel conjugate of $\phi$, i.e., 
\begin{equation*}
    \phi^*({y})=\sup_{x\in \mathcal{X}}\langle {x}, {y}\rangle-\phi({x})\,,
\end{equation*}
and where $\pi$ is assumed differentiable on $\mathcal{X}$. This mirror Langevin SDE stems from endowing $\mathcal{X}$ with a Riemannian metric, derived from the Hessian $\nabla^2\phi(x)$ (see \cite{zhang20mirror} for details). 

{It is important to choose a mirror map $\phi(x)$ that is well-suited to the structure of the potential {$U(x) = - \log \pi(x)$}. The convergence analysis in \cite{zhang20mirror} is based on the notions of relative Lipschitz smoothness and relative strong convexity of $U$ with respect to $\phi$. This condition is satisfied if there exists constants $0 < m \leq L < \infty$ such that $\forall x \in \mathcal{X}$
\begin{equation}
    m \nabla^{2}\phi(x) \;\preccurlyeq\; \nabla^{2}U(x) \;\preccurlyeq\; L \nabla^{2}\phi(x),
\end{equation}
where $\preccurlyeq$ is the Loewner order. This means the curvature of the potential $U$ is bounded by the curvature of the mirror map $\phi$.}

For  Poisson inverse problems, {the negative log-likelihood \eqref{eq:PoissonLikelihoodNoBeta} is not globally Lipschitz smooth in the Euclidean sense, however} a natural choice for {the mirror map} $\phi$ is the Burg's entropy (or maximum entropy) \cite{burg_maximum_1981}, given by
\begin{equation}\label{eq:burg}
    \phi(x)=\sum_{i=1}^n \log(x_i).
\end{equation}
{The Hessian of this map, given by $\nabla^2\phi(x)=\diag (1/x_1^2, \dots, 1/x_n^2)$, strongly penalizes values near the boundary of the positive orthant. In addition, \eqref{eq:PoissonLikelihoodNoBeta} is relatively smooth and strongly convex with respect to this choice of $\phi$ (see \cite{bauschke2017descent, hurault2023} for details), and thus the conditions on the likelihood for convergence of the mirror Langevin method are satisfied. This relationship extends to the contractive and Lipschitz-constrained PnP priors that we consider (see Section \ref{Sec:PnP}) due to the effective Lipschitz constant.}

There are many ways to obtain a discrete sampling scheme from a mirror Langevin SDE \cite{li22b-mirror, pmlr-v162-lau22a, zhang20mirror, ahn2021efficient, hsieh2018mld, girolami11riemann}. We choose the EM scheme studied in \cite{zhang20mirror}
\begin{equation}\label{eq:mld}
    {X}_{k+1} := \nabla \phi^*\left(\nabla\phi({X}_k) + \delta\nabla \log \pi ({X}_k)+ \sqrt{2\delta[\nabla^2\phi({X}_k)]}\xi_{k+1}\right)\, .
\end{equation}
Crucially, \eqref{eq:mld} can be applied directly to $\pi$ without the need for smoothing or reflections, as the constraints and non-smoothness of \eqref{eq:PoissonLikelihoodNoBeta} are being dealt with by \eqref{eq:burg}.

Algorithm \ref{alg:b-pnp} below summarizes our proposed PnP mirror Langevin algorithm (PnP-MLA) to sample from $p(x|y)$. Because the positivity constraints and smoothness issues are addressed by using the Burg's entropy \eqref{eq:burg} as mirror map $\phi$, PnP-MLA uses the correct likelihood $\nabla \log p(y|x)$, not the approximation $\nabla \log p^\beta(y|x)$ as previously. 
\begin{algorithm}[t]
\caption{PnP Mirror Langevin Algorithm (PnP-MLA)}\label{alg:b-pnp}
\begin{algorithmic}
\Require $N \in \mathbb{N}, y \in \mathbb{R}^m, \delta > 0, \rho > 0$
\Initialize {Set $X_0 \in \mathbb{R}^n$ and $k=0.$}
\For{$k = 0 : N $}
\begin{align*}
Z_{k+1} &\sim \mathcal{N}(0,Id) \\
X_{k+1}&=  \nabla \phi^*\left(\nabla\phi(X_k)+\delta\nabla\log p(y|X_k) + \delta \rho \nabla  \log p_{\epsilon}(X_k) + \sqrt{2\delta[\nabla^2\phi(X_k)]}Z_{k+1} \right) 
\end{align*}
\EndFor
\end{algorithmic}
\end{algorithm}

\section{Experiments}\label{sec:experiments}
\begin{figure}[!t]
\centering
    \begin{subfigure}{0.31 \textwidth}
         \centering
         \includegraphics[width=0.98\linewidth]{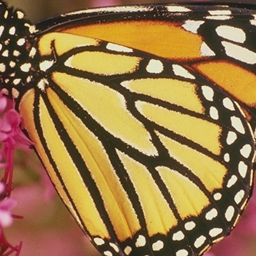}
         \caption{\texttt{Butterfly}}
         \label{fig:butterfly_truth}
    \end{subfigure}
     \begin{subfigure}{0.31 \textwidth}
         \centering
         \includegraphics[width=0.98\linewidth]{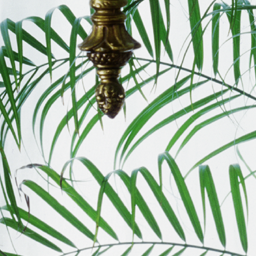}
         \caption{\texttt{Leaves}}
         \label{fig:leaves_truth}
    \end{subfigure}
    \begin{subfigure}{0.31 \textwidth}
         \centering
         \includegraphics[width=0.98\linewidth]{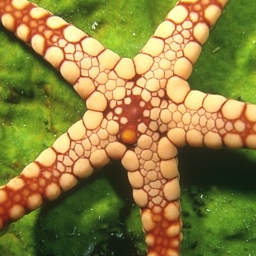}
         \caption{\texttt{Starfish}}
         \label{fig:starfish_truth}
    \end{subfigure}
 \caption{Ground truth images from \texttt{set3c}}
 \label{fig:gt-network-ablation}
 \end{figure}

\begin{figure}[t]
    \centering
    \begin{subfigure}{0.18\linewidth}
    \centering
        \includegraphics[width=\linewidth]{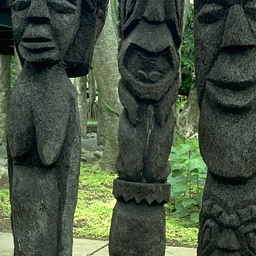}
    \end{subfigure}
    \begin{subfigure}{0.18\linewidth}
    \centering
        \includegraphics[width=\linewidth]{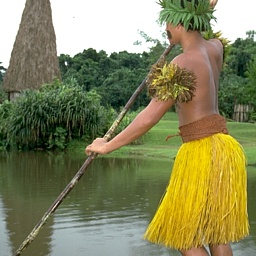}
    \end{subfigure}   
    \begin{subfigure}{0.18\linewidth}
    \centering
        \includegraphics[width=\linewidth]{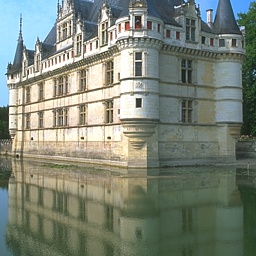}
    \end{subfigure} 
    \begin{subfigure}{0.18\linewidth}
    \centering
        \includegraphics[width=\linewidth]{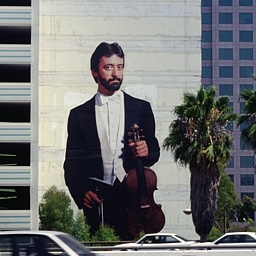}
    \end{subfigure} 
    \begin{subfigure}{0.18\linewidth}
    \centering
        \includegraphics[width=\linewidth]{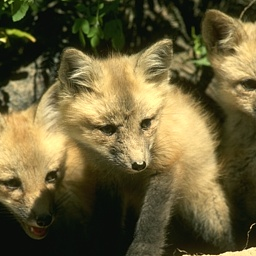}
    \end{subfigure}\vspace{0.1cm}
    
     \begin{subfigure}{0.18\linewidth}
     \centering
        \includegraphics[width=\linewidth]{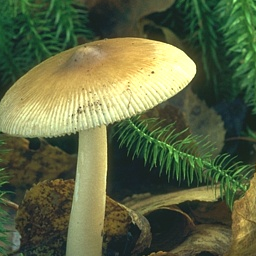}
    \end{subfigure}
    \begin{subfigure}{0.18\linewidth}
    \centering
        \includegraphics[width=\linewidth]{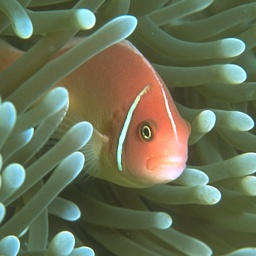}
    \end{subfigure}   
    \begin{subfigure}{0.18\linewidth}
    \centering
        \includegraphics[width=\linewidth]{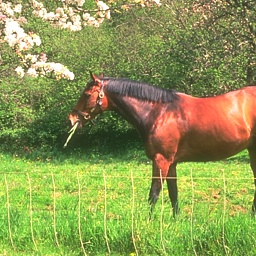}
    \end{subfigure} 
    \begin{subfigure}{0.18\linewidth}
    \centering
        \includegraphics[width=\linewidth]{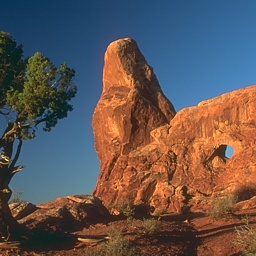}
    \end{subfigure} 
    \begin{subfigure}{0.18\linewidth}
    \centering
        \includegraphics[width=\linewidth]{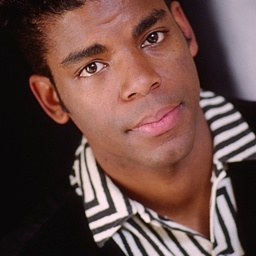}
    \end{subfigure}     
    \caption{\texttt{CBSD10} data set}
    \label{fig:CBSD10}
\end{figure}

We analyze the proposed Bayesian imaging methodologies through a series of numerical experiments related to non-blind Poisson image deconvolution, focusing on the choice of denoiser architecture and Langevin sampling algorithm. We have chosen this deconvolution problem because it provides a flexible framework for comparing different denoiser and computation strategies under various degrees of problem ill-conditioning and levels of noise. More precisely, we consider \eqref{eq:PoissonLikelihoodNoBeta} in the specific case where $A$ is a nearly singular blur operator leading to an ill-conditioned problem with highly noise-sensitive solutions; in the following experiments, we implement $A$ using a range of blur kernels taken from \cite{hurault2023, Levin09}, depicted in Figure \ref{levin_kernels} below. We carry out tests using two open datasets of color images: the \texttt{set3c} dataset \cite{hurault2022gradient}, and a subset of $10$ images from the \texttt{CBSD68} validation set (images of size $256\times 256$ pixels obtained by a centre crop) \cite{MartinFTM01}, which we henceforth refer to as the \texttt{CBSD10} set. The images are displayed in Figure \ref{fig:gt-network-ablation} and Figure \ref{fig:CBSD10}. 

The remainder of this section is organized as follows: 
\begin{itemize}
    \item In Section \ref{subsec:data-driven-priors}, we discuss the different denoisers that we will use in our numerical experiments and their corresponding neural network architectures.
    \item In Section \ref{sec:ablation}, we compare these different image denoiser architectures as PnP image priors. We assess their performance in terms of reconstruction accuracy, uncertainty visualization quality, and the convergence speed of the resulting PnP algorithm.
    \item In Section \ref{sec:convergence-speed}, {we} compare sampling algorithms when used with the most competitive image prior as identified in Section \ref{sec:ablation}; we assess performance in terms of algorithm stability, convergence speed, and reconstruction quality.
    \item In Section \ref{sec:sota}, we compare the proposed algorithms to alternative strategies from the state-of-the-art (SOTA). To capture different aspects of the estimation error, we assess reconstruction accuracy by computing peak signal-to-noise ratio (PSNR)\footnote{$\text{PSNR}(\hat{x}, x)=20\log_{10}\left(\frac{\text{MAX}_{{{x}}}}{\sqrt{\text{MSE}(\hat{x},x)}}\right)$, $\text{MSE}(\hat{x}, x)= \frac{1}{n}\sum_{i=0}^{n-1}\left( x(i)-\hat{x}(i)\right)^2$}, the structural similarity index measure (SSIM) \cite{wang2004image}, and the learned perceptual image patch similarity (LPIPS) \cite{zhang2018perceptual}.
\end{itemize}  
Experiments {were} conducted on a workstation with an Intel i9-9940X CPU (14 cores, 28 threads, 3.30 GHz base clock), 126 GB RAM, NVIDIA GeForce RTX 2080 Ti Rev. A (11 GB VRAM, CUDA 12.6) GPU, Rocky Linux 8.10 operating system and using Python 3.9.18 software with PyTorch 2.1.0 and Numpy 1.24.3 libraries.

\begin{figure}[t]
\centering
    \begin{subfigure}{0.08\textwidth}
        \includegraphics[width=\textwidth]{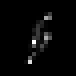}
        \caption{}\label{fig:levin-0}
    \end{subfigure}
    \begin{subfigure}{0.08\textwidth}
        \includegraphics[width=\textwidth]{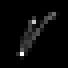}
        \caption{}\label{fig:levin-1}
    \end{subfigure}   
    \begin{subfigure}{0.08\textwidth}
        \includegraphics[width=\textwidth]{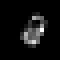}
        \caption{}\label{fig:levin-2}
    \end{subfigure}   
    \begin{subfigure}{0.08\textwidth}
        \includegraphics[width=\textwidth]{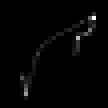}
        \caption{}\label{fig:levin-3}
    \end{subfigure}   
    \begin{subfigure}{0.08\textwidth}
        \includegraphics[width=\textwidth]{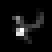}
        \caption{}\label{fig:levin-4}
    \end{subfigure}   
    \begin{subfigure}{0.08\textwidth}
        \includegraphics[width=\textwidth]{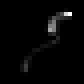}
        \caption{}\label{fig:levin-5}
    \end{subfigure}   
    \begin{subfigure}{0.08\textwidth}
        \includegraphics[width=\textwidth]{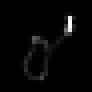}
        \caption{}\label{fig:levin-6}
    \end{subfigure}   
    \begin{subfigure}{0.08\textwidth}
        \includegraphics[width=\textwidth]{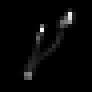}
        \caption{}\label{fig:levin-7}
    \end{subfigure}   
    \begin{subfigure}{0.08\textwidth}
        \includegraphics[width=\textwidth]{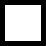}
        \caption{}\label{fig:levin-8}
    \end{subfigure}   
    \begin{subfigure}{0.08\textwidth}
        \includegraphics[width=\textwidth]{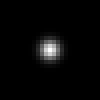}
        \caption{}\label{fig:levin-9}
    \end{subfigure}   
    \caption{Examples of blur kernels (support $25 \times 25$ pixels): motion blur kernels \cite{Levin09} (\ref{fig:levin-0})-(\ref{fig:levin-7}), a box blur (\ref{fig:levin-8}) and an isotropic Gaussian blur (\ref{fig:levin-9}) of bandwidth $1.6$ pixels.}
    \label{levin_kernels}
\end{figure}

\subsection{Choice of  PnP prior for Poisson imaging problems} \label{subsec:data-driven-priors}

As mentioned previously, PnP imaging methods rely predominantly on denoisers that are specifically trained to approximate the posterior mean or MMSE denoiser, with some mild regularization to control the Lipschitz regularity of $D_\epsilon$. These are predominantly denoisers encoded by convolutional neural networks (CNNs), which are trained in a supervised manner using clean and noisy image pairs corrupted by Gaussian noise with level $\epsilon$. Noteworthy examples include the DnCNN architecture (Denoising CNNs) \cite{ZhangZCM016} or the DRUNet (Dilated-Residual U-Net) \cite{zhang22drunet}. Patch-based denoising algorithms such as BM3D \cite{bm3d-2012} are also widely within PnP optimization schemes, but they are too computationally expensive for Langevin PnP sampling. Special attention is paid in the PnP literature to denoisers that are non-expansive (have a Lipschitz constant smaller than $1$), as well as denoisers that are equivalent to gradient steps or proximal steps for a specific potential function, which  allow embedding the denoisers within optimization machinery with convergence guarantees \cite{hurault2022gradient,hurault22a}.

By contrast, the framework \cite{laumont22pnp} for PnP-ULA can be applied with any Lipschitz continuous denoiser $D_\epsilon$ for which \eqref{eq:tweedie-score1} holds approximately, and without the need for $D_\epsilon$ to be contractive or define a gradient or proximal mapping. However, it is still useful to have some control of the Lipschitz constant of $D_\epsilon$, as this leads to Bayesian models with better regularization properties and to PnP-ULA samplers that converge faster. We explore the following choices for $D_\epsilon$:
\begin{enumerate} 
\item The non-expansive denoiser (LMMO) from \cite{Pesquet2020LearningMM}\footnote{\url{https://github.com/matthieutrs/LMMO_lightning}}, based on a modified DnCNN architecture that is trained to provide a maximally monotone operator.
\item The GS-DRUNet denoiser \cite{hurault2022gradient}\footnote{\url{https://github.com/samuro95/GSPnP}}, based on the DRUNet architecture and realized as a trained gradient descent step.
\item The Prox-DRUNet denoiser \cite{hurault22a}\footnote{\url{https://github.com/samuro95/Prox-PnP}}, also based on the DRUNet architecture and realized as a proximal descent step.
\item The B-DRUNet denoiser \cite{hurault2023}\footnote{\url{https://github.com/samuro95/BregmanPnP}}, a Bregman proximal step DRUNet architecture, with a proximal step defined via a (non-Euclidean) Bregman divergence. 
\end{enumerate}
The total number of parameters for the first denoiser is 668k, whereas the three DRUNet denoisers have 17M parameters. The DRUNet gradient-step denoisers have the form $D_\epsilon (x) = x - \nabla g_\epsilon (x) = N_\epsilon (x) + J_{N_\epsilon}^T(x) (x- N_\epsilon (x))$ where $N_\epsilon (x)$ is a DRUNet and $J_{N_\epsilon}(x)$ denotes its Jacobian. As a result, they have a higher computational complexity compared to the DnCNN-type architecture (due to the computation of the Jacobian $J_{N_\epsilon}$ and two applications of $N_\epsilon (x)$). Substituting this formulation into Tweedie's formula Eq. \eqref{eq:tweedie-score1}, we obtain $\nabla \log p_\epsilon(x)= \frac{1}{\epsilon}(x-\nabla g_\epsilon(x)-x)=-\frac{1}{\epsilon} \nabla g_\epsilon(x)$. 

{There exists a main difference between the GS-DRUNet and Prox-DRUNet variants \cite{hurault2022gradient, hurault22a}. While both share the same DRUNet architecture and are initially trained by minimizing a standard $L^2$ loss, 
\begin{equation*}
    \mathcal{L}(\epsilon) = \mathbb{E}_{x \sim p, \nu_\epsilon \sim \mathcal{N}(0, \epsilon^2)} \left[ \left\| D_\epsilon(x + \nu_\epsilon) - x \right\|^2 \right],
\end{equation*}
with $p$ denoting the distribution of clean training images, the Prox-DRUNet proposed in \cite{hurault22a} undergoes an additional fine-tuning stage. This fine-tuning uses a modified objective that includes a regularizer on the spectral norm $||.||_S$ of the Jacobian $\nabla^2g_\epsilon$,
\begin{equation*}
    \mathcal{L}_S(\epsilon) = \mathbb{E}_{x \sim p, \nu_\epsilon \sim \mathcal{N}(0, \epsilon^2)} \left[ \|D_\epsilon(x + \nu_\epsilon) - x\|^2 
     \\ + \mu \max(\|\nabla^2g_\epsilon(x + \nu_\epsilon)\|_S, 1 - \varepsilon) \right],
\end{equation*}
parametrized by scalars $\mu$ and $\varepsilon=0.1$. 
The purpose of this additional term is to explicitly enforce that the operator $\nabla g_\epsilon$ is contractive (i.e., has a Lipschitz constant less than 1). This requires replacing the original ReLU activations with smooth alternatives like softplus to ensure the necessary differentiability. This fine-tuning with spectral norm regularization is the primary difference between Prox-DRUNet and GS-DRUNet, and we hypothesize it contributes to the often more stable and superior performance of Prox-DRUNet.}

We also consider B-DRUNet \cite{hurault2023}, a Bregman generalization of the Euclidean proximal gradient step denoiser \cite{hurault22a}. This denoiser is based a generalized form of Tweedie's identity, suitable for denoising problems with non-Gaussian noise distributions. More precisely, given a noisy observation $z$ of an image $x$, then \cite{hurault2023}
\begin{equation}\label{eq:genTweedie}
    \hat{x}_{\text{MMSE}}(z) = \mathbb{E}[x|z]= z - \frac{1}{\gamma}(\nabla^2 \phi(z))^{-1}\cdot \nabla(-\log p_\gamma)(z)\,,
\end{equation}
where $\phi$ is a $\mathcal{C}^2$ convex potential of Legendre type, related to the distribution of the noise, and $\gamma$ a scalar representing the noise level. Note that when $\phi$ is quadratic, \eqref{eq:genTweedie} reduces to the conventional Tweedie identity \eqref{eq:tweedie-score1} for Gaussian denoising. It is useful to set $\phi$ to match the algebraic form of the likelihood function. In our case, this leads to the Burg entropy \eqref{eq:burg} introduced previously, which is related to inverse gamma noise (see \cite{hurault2023} for details). Given a trained Bregman denoiser $B_\gamma(z)\approx \hat{x}_{\text{MMSE}}(z)$, we obtain the required approximation to the score prior score function as $ - \nabla\log p_\gamma (z) \approx \gamma \nabla^2\phi(z)(z-{B}_\gamma(z))$. Note that in this case, we have $B_\gamma(z)=z-(\nabla\phi(z))^{-1}\cdot \nabla g_\gamma(z)$, which simplifies to $\nabla g_\gamma \approx -\frac{1}{\gamma}\nabla \log p_\gamma$. Thus, we can use our algorithms without any modification by setting $\epsilon = \frac{1}{\gamma}$ and $\nabla \log p_\epsilon(x)= - \frac{1}{\epsilon} \nabla g_{1/\epsilon}(x)$.

Moreover, a common practical issue with PnP optimization and sampling algorithms is that they can suffer from reconstruction artefacts which become more pronounced as iterations progress. Usual strategies to mitigate these issues in PnP optimization schemes are early stopping and iteration-dependent parameter fine-tuning \cite{romano17, ryu19a, Pesquet2020LearningMM}. However, these strategies are ineffective for PnP-ULAs, which aim to explore the posterior distribution to compute statistical estimates of interest, as this requires an algorithm that is ergodic and therefore stable. An interesting alternative, which we adopt herein, is to improve the stability of PnP-ULA by enforcing denoiser equivariance as recommended in \cite{terris2024equivariant}. Based on the intuition that image priors should be invariant to certain groups of transformations, such as rotations or reflections, we define transformations associated with a group $\mathcal{G}$ as $\lbrace T_g \rbrace_{g\in \mathcal{G}} $ with $T_g \in \mathbb{R}^{n \times n}$ denoting unitary matrices to describe a transformation. The resulting averaged denoiser is $D_\mathcal{G}= \frac{1}{|\mathcal{G}|}\sum_{g \in \mathcal{G}}T_g^{-1} D_\epsilon(T_gx)$ (analogous for $\mathcal{B}_\gamma$). For computational efficiency, rather than summing over all transformation, we approximate the sum by a simple one-sample Monte Carlo estimate by drawing $g \sim \mathcal{G}$ and setting $\Tilde{D}_\mathcal{G}(x) = T_{g}^{-1}D_\epsilon(T_{g} x)$. This strategy effectively mitigates artefacts from training the denoiser imperfectly, and leads to PnP schemes with significantly better stability \cite{terris2024equivariant}.

\subsection{Comparison of PnP denoisers} \label{sec:ablation}
\subsubsection{Experimental set up} We compare the different denoisers from Section \ref{subsec:data-driven-priors} focusing on their performance for Bayesian PnP Poisson image deconvolution. 
We evaluate them in terms of the accuracy (PSNR, SSIM, LPIPS) of the reconstruction $\hat{x}_{MMSE}$, the quality of the uncertainty estimates by visually contrasting the posterior standard deviations with the residual reconstruction errors for $\hat{x}_{MMSE}$, and by measuring the number of iterations required to compute the $\hat{x}_{MMSE}$ (we track the PSNR and stop when this statistic is stable).

We use pre-trained checkpoints for GS-DRUNet, Prox-DRUNet and B-DRUNet, and retrain the LMMO denoiser based on the latest released training software\footnote{\url{https://github.com/matthieutrs/LMMO_lightning}\label{note1}}. To ensure fair comparisons, the same training strategy was used for all networks in a shared range of noise levels $\sigma_{net} \in [0,50]$. 
We perform our comparisons  with two different algorithms, PPnP-ULA as an example of a projected  { Euclidean algorithm}\footnote{We have also conducted comparisons with other Euclidean algorithms such as RPnP-ULA (Algorithm \ref{alg:pnpula-reflect}) and R-SKROCK (Algorithm \ref{alg:euclidean-skrock}) -not reported here- and observed very similar results.}, and the  non-Euclidean PnP mirror Langevin algorithm (PnP-MLA). In the case of PnP-MLA we use the mirror map associated with Burg's entropy to ensure that the generated samples remain strictly positive. In addition,
for this experiment, we seek to evaluate the LMMO, GS-DRUNet and Prox-DRUNet denoisers in terms of image reconstruction performance and UQ capabilities, without enforcing equivariance by randomization, as we seek to compare the denoiser architectures without this effect, which can be assessed separately. However, we do apply randomization to B-DRUNet, as the algorithm is unstable otherwise.

{An important aspect of this empirical study is the selection of step size $\delta$ for each algorithm, as theoretical findings are restrictive or unavailable for this problem class. For the Euclidean algorithms (PPnP-ULA, RPnP-SKROCK), existing theoretical bounds \cite{laumont22pnp} are highly conservative for Poisson problems. As our goal is to investigate practical performance limits, we deliberately explore step sizes that exceed these bounds to find a more effective accuracy-speed trade-off. To the best of our knowledge, there is no theory for setting the step size of PnP-MLA (Algorithm \ref{alg:b-pnp}) under a PnP prior or a non-convex setting. This remains an open research problem. Consequently, we adopted an empirical approach, selecting $\delta$ for PnP-MLA via grid search to optimize PSNR. Therefore, the step sizes used herein were chosen to enable a robust and fair empirical comparison of the algorithms' stability and performance at their practical limits.}

Specifically for for PPnP-ULA, \cite{laumont22pnp} suggests setting $\delta\in (0,\delta_{L})$ with {$\delta_{L} = 1/(L_{p^\beta}\newline + L_\epsilon/\epsilon)$}, where $L_{p^\beta}=\alpha^2 \cdot(\max(y)/\beta^2)\cdot||AA^T||$ is the Lipschitz constant of ${\log p^\beta}$ and $L_\epsilon$ is the Lipschitz constant of the denoiser; LMMO and Prox-DRUNet have been trained so that $L_\epsilon = 1$,\footnote{The training loss of LMMO and Prox-DRUNet includes a penalization term on the spectral norm of the Jacobian, for more details see \cite{hurault22a, Pesquet2020LearningMM}.} while for GS-DRUNet $ 1< L_\epsilon < 5$ \cite{hurault22a}. 
We find this choice of step size {is} overly conservative for Poisson problems and set { $\delta = 0.2 \times 10^2 \cdot \delta_{L}$}, which in our experience provides a good accuracy-speed trade-off. We set the step size in PnP-MLA as  $\delta = 10^{-5} \simeq (2\cdot 10^2)\cdot \delta_{L}$ by using grid search and optimizing for PSNR. We run the algorithms for $10^6$ iterations to stress-test the stability of the algorithms. Using grid search, the parameter $\epsilon$ of the LMMO, GS-DRUNet, Prox-DRUNet denoisers $D_\epsilon$ is set to $\epsilon=(20/255)^2$ and $\rho=1$. {Unlike optimization-based PnP methods that may use a decaying schedule, a fixed $\epsilon$ is essential here to ensure the MCMC algorithm targets a stationary distribution.} For the B-DRUNet denoiser $B_\gamma$, we set $\gamma=30$ and $\rho=2$. We set $C=\lbrace x: 0 \leq x_i \leq 1 \rbrace$, as this corresponds to the training domain of the denoisers.

\begin{figure}[!t]
\centering
    \begin{subfigure}{0.31 \textwidth}
         \centering
         \includegraphics[width=\linewidth]{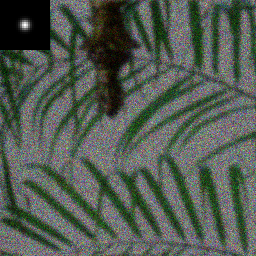}
         \caption{\tiny{Observation}}
         \label{fig:leaves_observation}
    \end{subfigure}
         \begin{subfigure}{0.31 \textwidth}
         \centering
         \includegraphics[width=\linewidth]{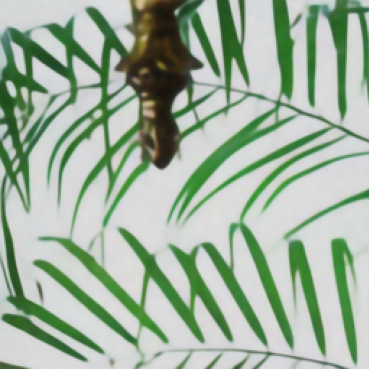}
         \caption{\tiny{PPnP-ULA: Prox-DRUNet}}\label{fig:leaves-1}
    \end{subfigure}
     \begin{subfigure}{0.31 \textwidth}
         \centering
         \includegraphics[width=\linewidth]{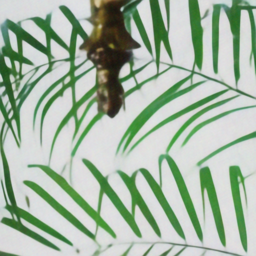}
         \caption{\tiny{PnP-MLA: Prox-DRUNet}}
    \end{subfigure}
    
    \begin{subfigure}{0.31 \textwidth}
         \centering
         \includegraphics[width=\linewidth]{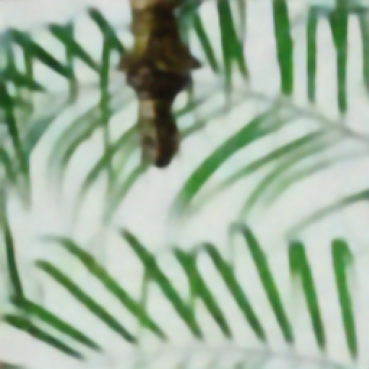}
         \caption{\tiny{PPnP-ULA: LMMO}}
    \end{subfigure}
    \begin{subfigure}{0.31\textwidth}
         \centering
         \includegraphics[width=\linewidth]{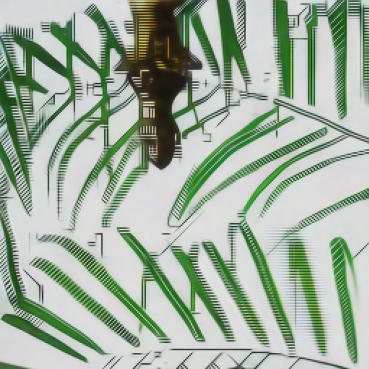}
         \caption{\tiny{PPnP-ULA: GS-DRUNet}}
     \end{subfigure}
    \begin{subfigure}{0.31 \textwidth}
         \centering
         \includegraphics[width=\linewidth]{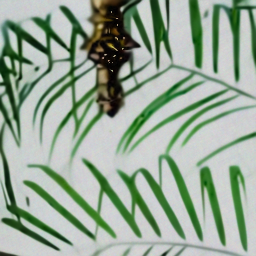}
         \caption{\tiny{PnP-MLA: B-DRUNet}}\label{fig:leaves-end}
     \end{subfigure}
 \caption{Poisson deconvolution problem for photon level $\alpha=20$ and Gaussian blur of size $25 \times 25$ with standard deviation 1.6. using the \texttt{leaves} image. Noisy and blurry observation, and the MMSE reconstructions $\hat{x}_{MMSE}$ for Prox-DRUNet, LMMO, GS-DRUNet and B-DRUNet using PPnP-ULA and PnP-MLA.}
 \label{fig:results_denoisers_leaves}
 \end{figure}

\begin{figure}[p]
\centering
\begin{minipage}[c]{0.89\textwidth}
\centering
    \rotatebox{0}{Standard deviation \phantom{whitespace here} Residual \phantom{more}}
  \vspace{-0.2cm}
    \begin{subfigure}{0.40 \textwidth}
    \centering
        \begin{minipage}{0.10\textwidth} 
        \rotcaption{\tiny{(Prox-DRUNet, PPnP-ULA)}}\label{fig:std-res-proxdrunet-ula}
        \end{minipage}
        \vspace{0.5cm}
        \begin{minipage}{0.85\textwidth} 
        \includegraphics[width=\linewidth]{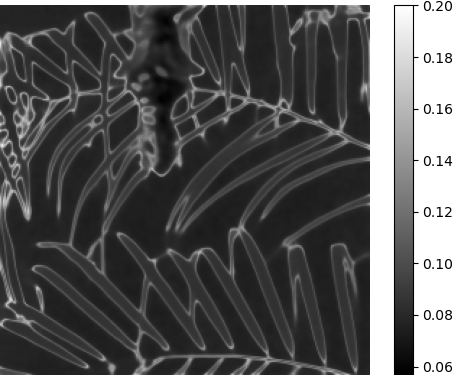}
        \end{minipage}
    \end{subfigure}
    \begin{subfigure}{0.40 \textwidth}
    \centering
         \begin{minipage}{0.10\textwidth} 
         \makebox[0pt]{\rotatebox{90}{\phantom{\tiny{(Prox-DRUNet, PPnP-ULA)}}}}
        \end{minipage}
        \vspace{0.5cm}
        \begin{minipage}{0.85\textwidth} 
         \includegraphics[width=\linewidth]{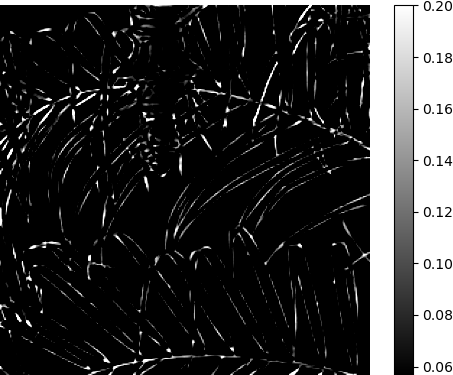}
         \end{minipage}
    \end{subfigure}
    \vspace{-0.2cm}
    \begin{subfigure}{0.4 \textwidth}
    \centering
        \begin{minipage}{0.10\textwidth} 
        \rotcaption{\tiny{(Prox-DRUNet, PnP-MLA)}}\label{fig:std-res-proxdrunet-mla}
        \end{minipage}
        \vspace{0.5cm}
        \begin{minipage}{0.85\textwidth} 
         \includegraphics[width=\linewidth]{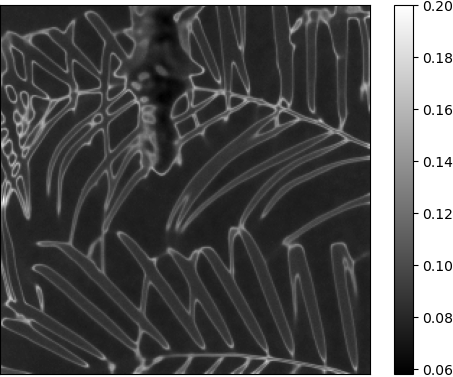}
         \end{minipage}
     \end{subfigure}     
    \begin{subfigure}{0.40 \textwidth}
    \centering
            \begin{minipage}{0.10\textwidth} 
        \makebox[0pt]{\rotatebox{90}{\phantom{\tiny{(Prox-DRUNet, PPnP-ULA)}}}}
        \end{minipage}
        \vspace{0.5cm}
        \begin{minipage}{0.85\textwidth} 
         \includegraphics[width=\linewidth]{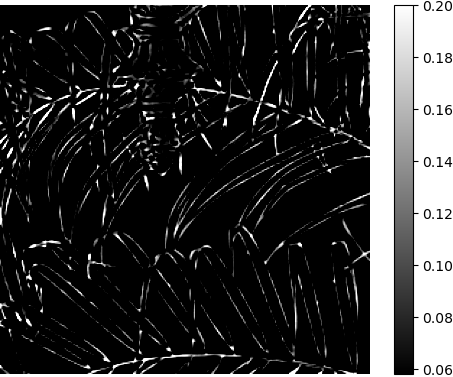}
         \end{minipage}
     \end{subfigure} 
     \vspace{-0.2cm}
    \begin{subfigure}{0.40 \textwidth}
    \centering
         \begin{minipage}{0.10\textwidth} 
        \rotcaption{\tiny{(LMMO, PPnP-ULA)}}\label{fig:std-res-lmmo-ula}
        \end{minipage}
        \vspace{0.5cm}
        \begin{minipage}{0.85\textwidth} 
         \includegraphics[width=\linewidth]{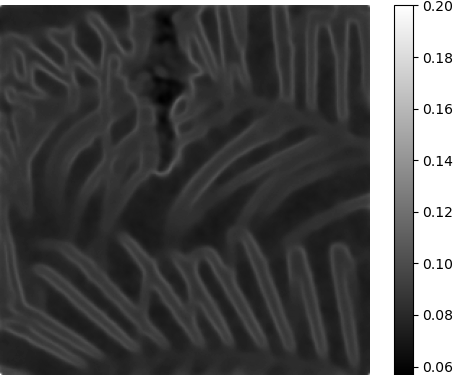}
         \end{minipage}
    \end{subfigure}
    \begin{subfigure}{0.4 \textwidth}
    \centering
         \begin{minipage}{0.10\textwidth} 
       \makebox[0pt]{\rotatebox{90}{\phantom{\tiny{(Prox-DRUNet, PPnP-ULA)}}}}
        \end{minipage}
        \vspace{0.5cm}
        \begin{minipage}{0.85\textwidth} 
         \includegraphics[width=\linewidth]{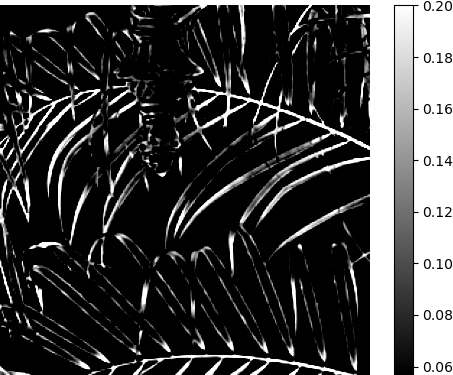}
         \end{minipage}
    \end{subfigure}
    \vspace{-0.2cm}
    \begin{subfigure}{0.4\textwidth}
    \centering
          \begin{minipage}{0.10\textwidth} 
        \rotcaption{\tiny{(GS-DRUNet, PPnP-ULA)}}\label{fig:std-res-gsdrunet-ula}
        \end{minipage}
        \vspace{0.5cm}
        \begin{minipage}{0.85\textwidth} 
         \includegraphics[width=\linewidth]{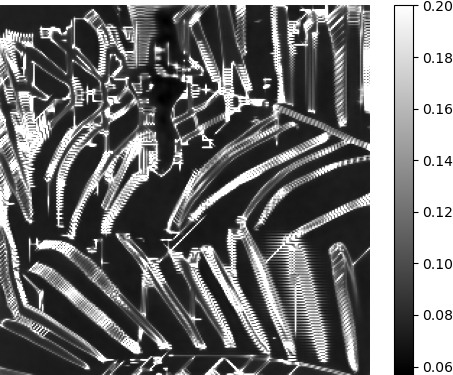}
         \end{minipage}
     \end{subfigure}
     \begin{subfigure}{0.4\textwidth}
     \centering
         \begin{minipage}{0.10\textwidth} 
       \makebox[0pt]{\rotatebox{90}{\phantom{\tiny{(Prox-DRUNet, PPnP-ULA)}}}}
        \end{minipage}
        \vspace{0.5cm}
        \begin{minipage}{0.85\textwidth} 
         \includegraphics[width=\linewidth]{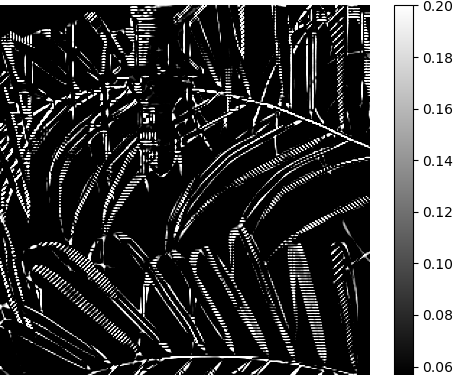}
         \end{minipage}
     \end{subfigure}
     \vspace{-0.2cm}
    \begin{subfigure}{0.4\textwidth}
    \centering
          \begin{minipage}{0.10\textwidth} 
        \rotcaption{\tiny{(B-DRUNet, PnP-MLA)}}\label{fig:std-res-bdrunet-mla}
        \end{minipage}
        \vspace{0.5cm}
        \begin{minipage}{0.85\textwidth} 
         \includegraphics[width=\linewidth]{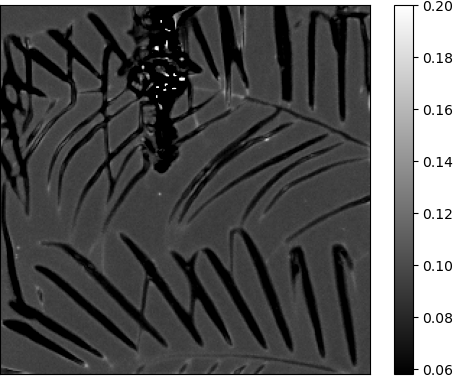}
         \end{minipage}
    \end{subfigure}
    \begin{subfigure}{0.4 \textwidth}
    \centering
          \begin{minipage}{0.10\textwidth} 
        \makebox[0pt]{\rotatebox{90}{\phantom{\tiny{(Prox-DRUNet, PPnP-ULA)}}}}
        \end{minipage}
        \vspace{0.5cm}
        \begin{minipage}{0.85\textwidth} 
         \includegraphics[width=\linewidth]{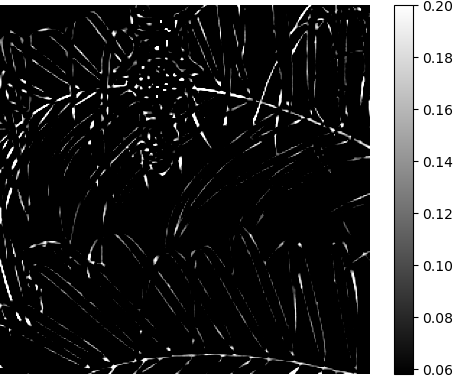}
         \end{minipage}
    \end{subfigure}  
 \caption{Poisson deconvolution problem as in Fig. \ref{fig:results_denoisers_leaves}. Pairwise: Pixelwise standard deviations (left column) and residuals (right column) for all tested denoisers using either PPnP-ULA or PnP-MLA. }
 \label{fig:leaves_st_dev}
 \end{minipage}
 \end{figure}

\begin{table}[t]
\smaller
    \centering
    \renewcommand{\arraystretch}{1.2} 
    \begin{tabular}{llccccc}
        \hline
        \textbf{Method} & \textbf{Denoiser} & \textbf{PSNR} & \textbf{SSIM} & \textbf{LPIPS} & \multicolumn{1}{c}{\textbf{\begin{tabular}[c]{@{}c@{}}Iter. until \\ 98$\%$ peak PSNR\end{tabular}}} & \textbf{sec./Iter.} \\
        \hline
        \multirow{3}{*}{PPnP-ULA} 
        & LMMO       & 17.20 & 0.63 & 0.38 & 4000  & 0.012 \\
        & GS-DRUNet  & 16.06 & 0.65 & 0.35 & 4000  & 0.025 \\
        & Prox-DRUNet & 20.93 & 0.82 & 0.12 & 18000 & 0.025 \\
        \hline
        \multirow{2}{*}{PnP-MLA} 
        & Prox-DRUNet & 21.29 & 0.82 & 0.13 & 12000 & 0.025 \\
        & B-DRUNet    & 17.83 & 0.75 & 0.23 & 136500 & 0.025 \\
        \hline
    \end{tabular}
    \caption{Quantitative results for $\hat{x}_{MMSE}$ calculated by PPnP-ULA and PnP-MLA for the \texttt{leaves} image. Last two columns: The number of iterations required to reach 98\% PSNR and the time per iteration in seconds.}
    \label{tab:ablation-networks-im1}
\end{table}

\subsubsection{Experimental results} 
{We performed a thorough evaluation of the denoisers on the \texttt{set3c} dataset. For clarity and detailed analysis, we present a representative set of results using the \texttt{leaves} image in the main text (Figure \ref{fig:results_denoisers_leaves}). To demonstrate the consistency of our findings, comprehensive results for the \texttt{starfish} image are provided in Appendix \ref{sec:app-ablation}. The results for the \texttt{butterfly} image were qualitatively similar and offered no new insights, and were thus omitted for brevity.}
Figure \ref{fig:leaves_observation} depicts a realization $y$ generated by the forward model Eq. \eqref{eq:PoissonLikelihoodNoBeta} with the operator $A$ modelling a Gaussian blur operator of size $25\times 25$ with standard deviation $1.6$ pixels and a photon level set to $\alpha=20$. Figures \ref{fig:leaves-1} to \ref{fig:leaves-end} show the posterior mean $\hat{x}_{MMSE}$, as calculated for the different denoisers and different algorithms. We observe that Prox-DRUNet outperforms the other denoisers in terms of reconstruction quality {for both Euclidean and non-Euclidean algorithms and see that both algorithms deliver highly similar estimates, which capture
fine detail without noticeable artefacts.} In comparison, the LMMO denoiser produces an estimate with excessive smoothing, while the GS-DRUNet exhibits stripe-like artefacts. These artefacts are also present when respecting the step size bounds recommended in \cite{laumont22pnp} (see Section \ref{sec:RPnP-PPnP}, not shown here), and increase gradually with greater step size. The B-DRUNet also exhibits artefacts, amplifying some structures in the top central part of the image.

Table \ref{tab:ablation-networks-im1} summarizes the performance of the denoisers for the considered image. The LMMO prior leads to faster convergence but achieves low reconstruction quality, while GS-DRUNet is not stable without randomization, so the reconstruction quality decreases as the iterations progress and the artefacts become more pronounced. Similarly, B-DRUNet also leads to instability and reaches its top performance in around $1.3 \cdot 10^5$ iterations before artefacts start to amplify. Conversely, Prox-DRUNet outperforms the other networks in all the image quality metrics considered. To check for artefacts, it can be a helpful diagnostic tool to compare image metrics using the cumulative mean over all samples with the mean computed from a subset of samples (e.g., 100 samples seem to be enough to be representative). Scores will be comparable unless the denoiser produces artefacts, then the scores over the subset of samples will be significantly worse. 

For completeness, Table \ref{tab:ablation-networks-im1} also reports the number of iterations that PPnP-ULA and PnP-MLA require to reach $98\%$ of peak PSNR performance, as an indicator of convergence speed for the posterior mean. We observe that the convergence speed of PPnP-ULA and PnP-MLA are similar for the Prox-DRUNet denoiser. We also report the time in seconds per iteration, which is largely determined by the complexity of one denoiser application.
Note that the Prox-DRUNet yields the best accuracy-speed trade-off, reaching almost top performance under $2 \cdot 10^4$ iterations for both PPnP-ULA and PnP-MLA. We point out that for PPnP-ULA, similar quantitative behavior was observed for the different denoisers even when randomization was used, as well as when using Algorithms \ref{alg:pnpula-reflect} and \ref{alg:euclidean-skrock}. 

Lastly, Figure \ref{fig:leaves_st_dev} shows the pixel-wise posterior standard deviation, as calculated with the LMMO, Prox-DRUNet, GS-DRUNet and B-DRUNet denoisers. For reference, next to each posterior standard deviation plot, we report the residual obtained by comparing $\hat{x}_{MMSE}$ to the true image (these standard deviations represent the models' marginal predictions for these residuals, at the pixel level). We observe that LMMO, Prox-DRUNet, GS-DRUNet produce uncertainty plots that are broadly in agreement with their residuals. For Prox-DRUNet and LMMO, uncertainty concentrates around edges and contours, as expected for a deconvolution problem, whereas the uncertainty estimates of GS-DRUNet are aligned with its reconstruction artefacts. Conversely, B-DRUNet produces uncertainty plots that highlight homogenenous regions, and which do not align well with its residual.  We conclude that Prox-DRUNet is the most appropriate denoising architecture for Bayesian PnP inference in Poisson image deblurring problems and use this as the prior  for the rest of our experiments.

\subsection{Langevin sampling algorithms for PnP Poisson deconvolution} \label{sec:convergence-speed}
\subsubsection{Experimental set up} We now compare different PnP Langevin strategies with Prox-DRUNet as prior. {The primary purpose of this experiment is to provide an empirical and qualitative exploration of the stability boundaries and performance trade-offs of each algorithm. Note that the considered algorithm often demonstrate stability well beyond the range of step sizes predicted by the theory available (an effect we attribute in part to pessimistic stability estimates, to the algorithm's constraints, and the use of a mirror map), hence we aim to characterize their behavior as the step size is further increased. Specifically, we study how different choices of step size affects convergence speed, reconstruction quality, and numerical stability. The results are intended to serve as an illustrative guide demonstrating the visual cues of both stable (well-chosen large step sizes) and unstable (excessively large step sizes) regimes, such as the emergence of artifacts or over-smoothing. We compare MMSE reconstructions computed with different sampling algorithms for a range of step sizes, both quantitatively and qualitatively, to derive practical insights for each algorithm.}
Additionally, we compare the convergence speed between different sampling algorithms for the recommended step size over a set of images by looking at mean cumulative quality metrics. How image quality and convergence speed interrelate for different algorithms can guide the choice of sampling algorithm according to given priorities with regards to metric and computational budget. {Finally, we illustrate the mixing properties of these algorithms by looking at the autocorrelation of the corresponding Markov chains.} 

\makeatletter
\renewcommand{\p@subfigure}{\thefigure.}
\makeatother

\begin{figure}[t!]
    \centering
    \begin{subfigure}{\textwidth}
    \begin{subfigure}[b]{0.19\textwidth}
        \centering
        \includegraphics[width=\textwidth]{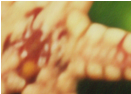}
        \caption{\smaller $c=0.02\cdot10^2$ \\ (20.44, 0.26, 0.67)}
    \end{subfigure}
    \begin{subfigure}[b]{0.19\textwidth}
        \centering
        \includegraphics[width=\textwidth]{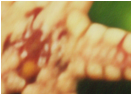}
        \caption{\smaller $c=1.5\cdot10^2$ \\ (\textbf{21.10},\textbf{0.24},\textbf{0.70})}
    \end{subfigure}
    \begin{subfigure}[b]{0.19\textwidth}
        \centering
        \includegraphics[width=\textwidth]{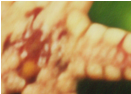}
        \caption{\smaller $c=3\cdot10^2$ \\ (21.05, \textbf{0.24}, \textbf{0.70})}
    \end{subfigure}
    \begin{subfigure}[b]{0.19\textwidth}
        \centering
        \includegraphics[width=\textwidth]{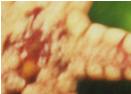}
        \caption{\smaller $c=5\cdot10^2$ \\ (20.84, \textbf{0.24}, \textbf{0.70})}
    \end{subfigure}
    \begin{subfigure}[b]{0.19\textwidth}
        \centering
        \includegraphics[width=\textwidth]{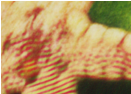}
        \caption{\smaller $c=10^3$ \\ (19.71, 0.28, 0.63)}
    \end{subfigure}
    {\renewcommand*\thesubfigure{\arabic{subfigure}} 
    \setcounter{subfigure}{0}
    \caption{RPnP-ULA. }
    \label{fig:qualitative-detail-rpnp}
    }
\end{subfigure}
\vspace{0.5em} 
\begin{subfigure}{\textwidth}
    \centering
    \setcounter{subfigure}{0}
    \begin{subfigure}[b]{0.19\textwidth}
        \centering
        \includegraphics[width=\textwidth]{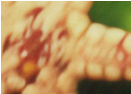}
        \caption{\smaller $c=0.02\cdot10^2$ \\(20.50, 0.26, 0.68)}
    \end{subfigure}
    \begin{subfigure}[b]{0.19\textwidth}
        \centering
        \includegraphics[width=\textwidth]{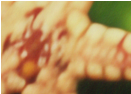}
        \caption{\smaller $c=0.2\cdot10^2$ \\ (21.15, 0.24, 0.70)}
    \end{subfigure}
    \begin{subfigure}[b]{0.19\textwidth}
        \centering
        \includegraphics[width=\textwidth]{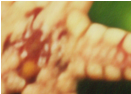}
        \caption{\smaller $c=1.5\cdot10^2$ \\(\textbf{21.42},\textbf{0.23},\textbf{0.71})}
    \end{subfigure}
    \begin{subfigure}[b]{0.19\textwidth}
        \centering
        \includegraphics[width=\textwidth]{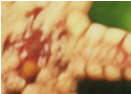}
        \caption{\smaller $c=3\cdot10^2$ \\ (21.37, \textbf{0.23}, \textbf{0.71})}
    \end{subfigure}
    \begin{subfigure}[b]{0.19\textwidth}
        \centering
        \includegraphics[width=\textwidth]{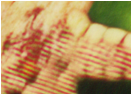}
        \caption{\smaller $c=5\cdot10^2$ \\ (20.62, 0.26, 0.68)}
    \end{subfigure}
    {\renewcommand*\thesubfigure{\arabic{subfigure}}
    \setcounter{subfigure}{1}
    \caption{PPnP-ULA.}
    \label{fig:qualitative-detail-ppnp}
    }
\end{subfigure}
\vspace{0.5em} 
\begin{subfigure}{\textwidth}
    \centering
    \setcounter{subfigure}{0}
    \begin{subfigure}[b]{0.19\textwidth}
        \centering
        \includegraphics[width=\textwidth]{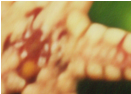}
        \caption{\smaller $c=0.002\cdot10^2$ \\ (21.26, 0.24, 0.71)}
    \end{subfigure}
    \begin{subfigure}[b]{0.19\textwidth}
        \centering
        \includegraphics[width=\textwidth]{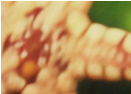}
        \caption{\smaller $c=0.02\cdot10^2$ \\ (21.53, \textbf{0.24}, 0.72)}
    \end{subfigure}
    \begin{subfigure}[b]{0.19\textwidth}
        \centering
        \includegraphics[width=\textwidth]{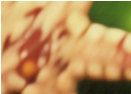}
        \caption{\smaller $c=0.15\cdot10^2$ \\ (21.85, 0.25, 0.72)}
    \end{subfigure}
    \begin{subfigure}[b]{0.19\textwidth}
        \centering
        \includegraphics[width=\textwidth]{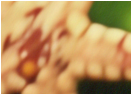}
        \caption{\smaller $c=0.3\cdot10^2$ \\ (\textbf{22.14}, 0.25, \textbf{0.73})}
    \end{subfigure}
    \begin{subfigure}[b]{0.19\textwidth}
        \centering
        \includegraphics[width=\textwidth]{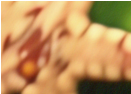}
        \caption{\smaller $c=10^2$ \\ (22.07, 0.27, 0.73)}
    \end{subfigure}    
    {\renewcommand*\thesubfigure{\arabic{subfigure}}
    \setcounter{subfigure}{2}
    \caption{RPnP-SKROCK.}
    \label{fig:qualitative-detail-skrock}
    }
\end{subfigure}
\vspace{0.5em} 
\begin{subfigure}{\textwidth}
    \centering
    \setcounter{subfigure}{0}
    \begin{subfigure}[b]{0.19\textwidth}
        \centering
        \includegraphics[width=\textwidth]{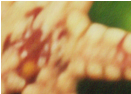}
        \caption{\smaller $c=1.5\cdot10^2$ \\ (20.46, 0.26, 0.66)}
    \end{subfigure}
    \begin{subfigure}[b]{0.19\textwidth}
        \centering
        \includegraphics[width=\textwidth]{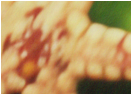}
        \caption{\smaller $c=2\cdot10^2$ \\ (20.50, 0.26, 0.67)}
    \end{subfigure}
    \begin{subfigure}[b]{0.19\textwidth}
        \centering
        \includegraphics[width=\textwidth]{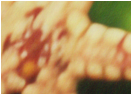}
        \caption{\smaller $c=5\cdot10^2$ \\ (20.63, \textbf{0.25}, {0.67})}
    \end{subfigure}
    \begin{subfigure}[b]{0.19\textwidth}
        \centering
        \includegraphics[width=\textwidth]{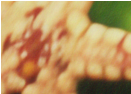}
        \caption{\smaller $c=10^3$ \\(\textbf{20.71},\textbf{0.25},\textbf{0.68})}
    \end{subfigure}
    \begin{subfigure}[b]{0.19\textwidth}
        \centering
        \includegraphics[width=\textwidth]{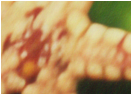}
        \caption{\smaller $c=1.25\cdot10^3$ \\(20.68,\textbf{0.25},0.67)}
    \end{subfigure}
    {\renewcommand*\thesubfigure{\arabic{subfigure}}
    \setcounter{subfigure}{3}
    \caption{PnP-MLA. } 
    \label{fig:qualitative-detail-mirror-noise25}}
\end{subfigure}
\caption{Comparison of MMSE results obtained with different sampling algorithms exemplified on the \texttt{starfish} image. The chosen step size affects the results. (PSNR, LPIPS, SSIM) values averaged over the whole \texttt{set3c} indicated below each image. Best of each algorithm marked in \textbf{bold}. }
\label{fig:TBD}
\end{figure}

For this experiment, we use images from the \texttt{set3c} data set, depicted in Figure \ref{fig:gt-network-ablation}, the blur kernel displayed Figure \ref{fig:levin-1} and photon level $\alpha=20$.  We choose the \texttt{starfish} image to illustrate qualitative results.  With regards to the algorithms and their step sizes, for the RPnP-ULA (Algorithm \ref{alg:pnpula-reflect}) and PPnP-ULA (Algorithm \ref{alg:ppnpula}), we test a range of step sizes $\delta=c\cdot \delta_L$, with $c \in [1,10^3]$. For RPnP-SKROCK (Algorithm \ref{alg:euclidean-skrock}), we test step sizes of the form $\delta=c\cdot \ell_s \delta_{L}$, with $c \in [0.1,10^2]$, where $\ell_s = (s-0.5)^2(2-4/3\eta)-1.5$ with $s=10$ and $\eta=0.05$; for more details see \cite{PVZ20}. For the PnP-MLA (Algorithm \ref{alg:b-pnp}), we test step sizes of the form $\delta=c\cdot \delta_L$, where $c\in[10,10^3]$.  The Prox-DRUNet denoiser parameter $\epsilon$ is set to $\epsilon=(20/255)^2$ for RPnP-ULA, PPnP-ULA and RPnP-SKROCK, whereas we use $\epsilon=(25/255)^2$ for PnP-MLA (see Appendix \ref{sec:app-pnpmla-eps20} for complementary results for PnP-MLA with $\epsilon=(20/255)^2$). {For the Euclidean algorithms we set $\beta$ to $1\%$ of the mean intensity value of the observed data.} In all cases, to improve numerical stability and reduce PnP artefacts, we enforce equivariance to horizontal and vertical flips as well as to rotations of multiples of 90 degrees by randomization. 

\begin{figure}[t!]
\centering
\begin{subfigure}{0.85\textwidth}
    \centering
    \includegraphics[width=\linewidth]{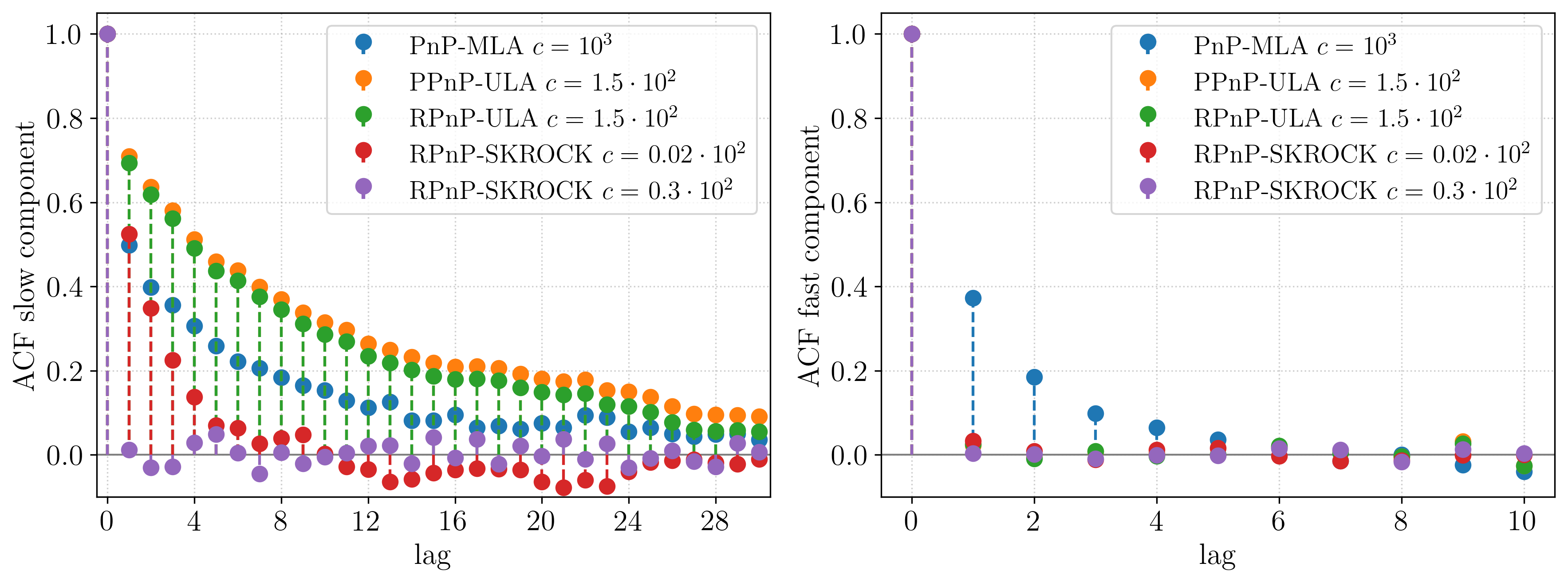}
    \subcaption{\texttt{butterfly}}
    \label{fig:acf_im0}
\end{subfigure}

\begin{subfigure}{0.85\textwidth}
    \centering
    \includegraphics[width=\linewidth]{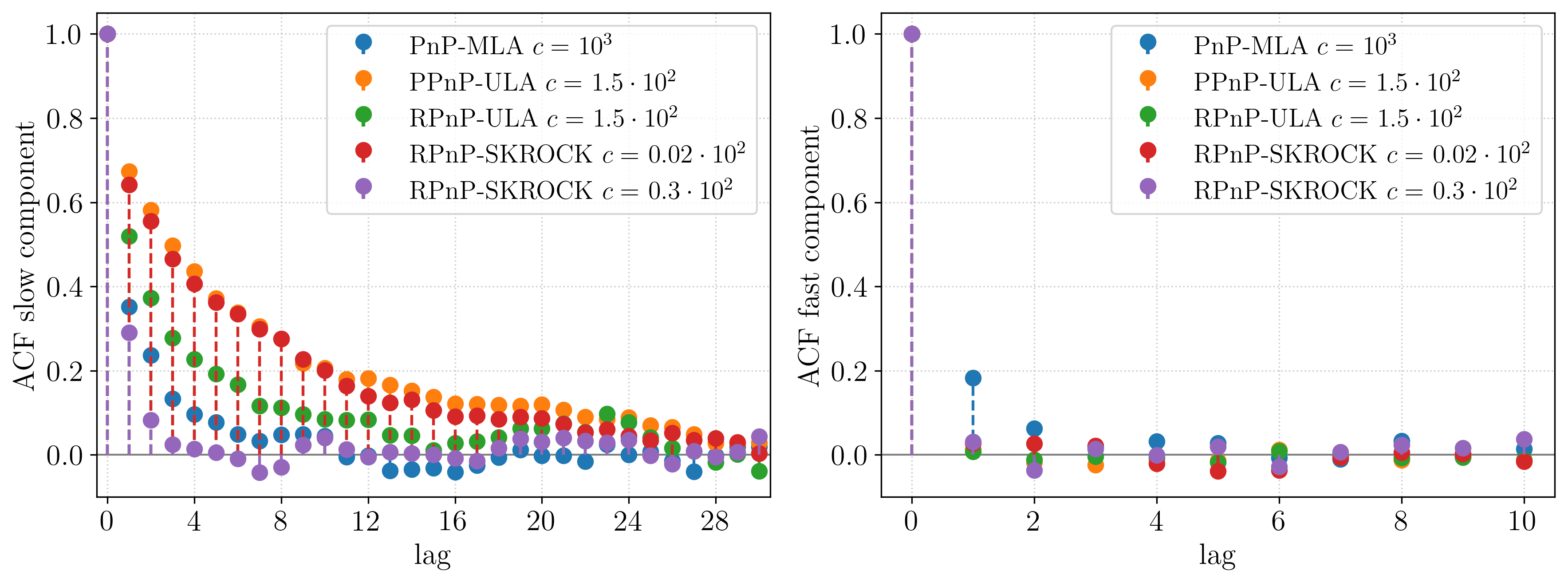}
    \subcaption{\texttt{leaves}}
    \label{fig:acf_im1}
\end{subfigure}

\begin{subfigure}{0.85\textwidth}
    \centering
    \includegraphics[width=\linewidth]{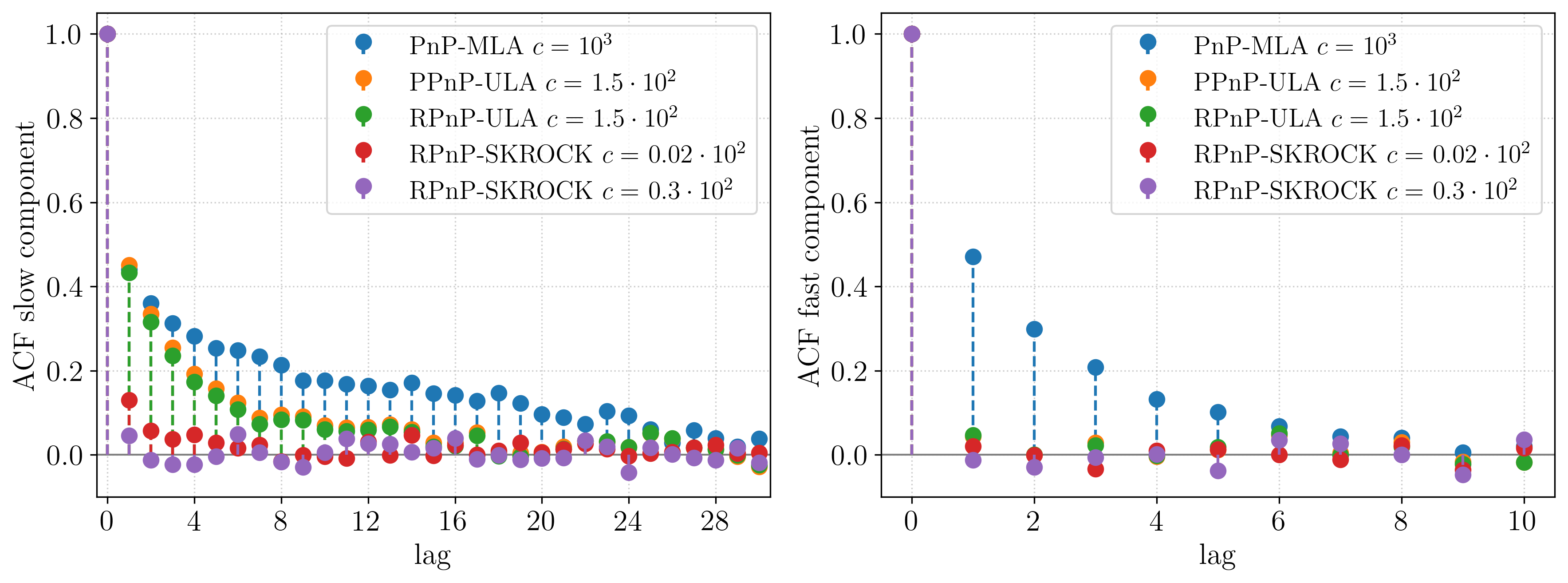}
    \subcaption{\texttt{starfish}}
    \label{fig:acf_im2}
\end{subfigure}
    \caption{Comparing the auto-correlation function (ACF) for PnP-MLA, RPnP-ULA, PPnP-ULA and RPnP-SKROCK for the \texttt{set3c} data set.}
    \label{fig:conv-acf-ess}

\end{figure}

\subsubsection{Experimental results}
\begin{figure}[t!]
    \centering
    \begin{tabular}{cccc}
        \begin{subfigure}[b]{0.21\textwidth}
            \centering
            \includegraphics[width=\textwidth]{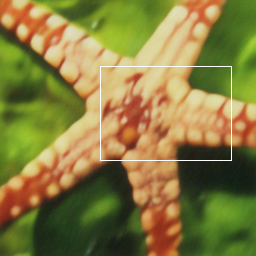}
            \caption{\footnotesize{RPnP-ULA} \\ \hspace*{0.8cm}$c=1.5\cdot10^2$}
        \end{subfigure} &
        \begin{subfigure}[b]{0.21\textwidth}
            \centering
            \includegraphics[width=\textwidth]{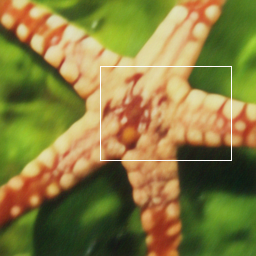}
            \caption{\footnotesize{PPnP-ULA\\\hspace*{0.8cm}$c=1.5\cdot10^2$}}
        \end{subfigure} &
        \begin{subfigure}[b]{0.21\textwidth}
            \centering
            \includegraphics[width=\textwidth]{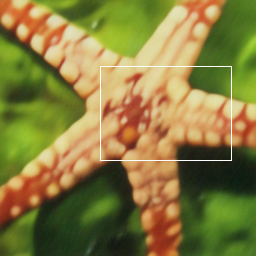}
            \caption{\footnotesize{RPnP-SKROCK}\\ \hspace*{1cm}$c=0.02 \cdot 10^2$}
        \end{subfigure} &
        \begin{subfigure}[b]{0.21\textwidth}
            \centering
            \includegraphics[width=\textwidth]{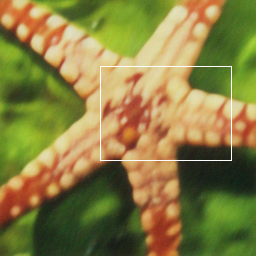}
            \caption{\footnotesize{PnP-MLA} \\\hspace*{0.8cm}$c=10^3$}
        \end{subfigure} \\

    \end{tabular}
    \caption{MMSE estimates under the most competitive step sizes for each algorithm.}
    \label{fig:qualitative-best}
    \vspace{-0.2cm}
\end{figure}

We show qualitative results in Figure \ref{fig:TBD}, comparing MMSE reconstructions $\hat{x}_{MMSE}$ for a range of step sizes for algorithms RPnP-ULA, PPnP-ULA, RPnP-SKROCK, and PnP-MLA respectively; with quantitative results averaged over \texttt{set3c} in the captions. Note that we use a detail of the \texttt{starfish} image in this figure to illustrate the effects of the chosen step size; the full image for the best step size of each algorithm is shown in Figure \ref{fig:qualitative-best}. More detailed quantitative results are deferred to the Appendix \ref{sec:app-conv-speed-extra} in Tables \ref{tab:rpnp-step-size}-\ref{tab:mirror-step-size-eps-25}.

Looking at Figures \ref{fig:qualitative-detail-rpnp} and \ref{fig:qualitative-detail-ppnp}, we find that RPnP-ULA and PPnP-ULA both achieve the best quality at $c=1.5\cdot 10^2$ in all reported metrics (PSNR, LPIPS, SSIM). RPnP-ULA is stable up to $c=5\cdot10^2$,  PPnP-ULA is stable up to $c=3\cdot10^2$. For larger $c$ artefacts in the form of stripes gravely appear, see e.g. Figure \ref{fig:qualitative-detail-rpnp}(e) and \ref{fig:qualitative-detail-ppnp}(e). For both algorithms the performance starts to decrease for $c>1.5\cdot10^2$.

In Figure \ref{fig:qualitative-detail-skrock}, we observe that RPnP-SKROCK is stable as the step size increases and does not show artefacts in its reconstructions. For large step sizes, however, we see significant oversmoothing which is a sign of excessive bias, see e.g.  Figure \ref{fig:qualitative-detail-skrock}(d) and \ref{fig:qualitative-detail-skrock}(e). In this case, PSNR and SSIM continue to improve, while LPIPS worsens, quantitatively confirming a loss of detail. The overall performance deteriorates for a step size for which $c > 0.3\cdot10^2$. Therefore, {for this specific example, we would recommend} to choose a step size for which $c = 0.02\cdot10^2$ or  $c = 0.3\cdot10^2$ depending on whether objective (PSNR, SSIM) or perceptive quality (LPIPS) is more important. {More generally, we advise to check both for qualitative cues such as artifacts or excessive smoothing, and quantitative metrics such as convergence speed, to determine an appropriate step size for the problem at hand.} {Moreover, in Figure \ref{fig:qualitative-detail-mirror-noise25}, we observe that PnP-MLA performs relatively similarly for all the considered step sizes, consistently recovering fine detail without any significant oversmoothing effect.}

{In order to illustrate the mixing properties of the presented algorithms, Fig. \ref{fig:conv-acf-ess} shows the autocorrelation function (ACF) of the slowest and fastest mixing components\footnote{The slow and fast mixing components were identified by selecting the pixel trajectories with the maximum and minimum empirical variance across the chain, respectively. The ACF was computed using the Fast Fourier Transform (FFT).} of the Markov chains for images of the \texttt{set3c} dataset. Efficient sampling algorithms generate samples that are not highly correlated, which can be observed by a rapid decay of the ACF curve towards zero as the lag increases.  The analysis of the slowest component highlights the most significant performance differences. The RPnP-SKROCK (lilac) is often most efficient, showing a very rapid decay in correlation. In contrast, the decay of PnP-MLA, PPnP-ULA and RPnP-ULA varies per image, with a decay within 15-30 lags. For the fast-mixing component, all algorithms perform well, with correlations dropping to near-zero almost immediately (within 5 lags). 
Overall, the RPnP-SKROCK algorithms demonstrate superior performance by effectively handling the more challenging slow-mixing components; the results for the remaining algorithms are mixed but indicate good overall sampling efficiency.}

\begin{figure}[t!]
    {\centering
    \begin{subfigure}{0.5\textwidth}
        \includegraphics[width=\textwidth]{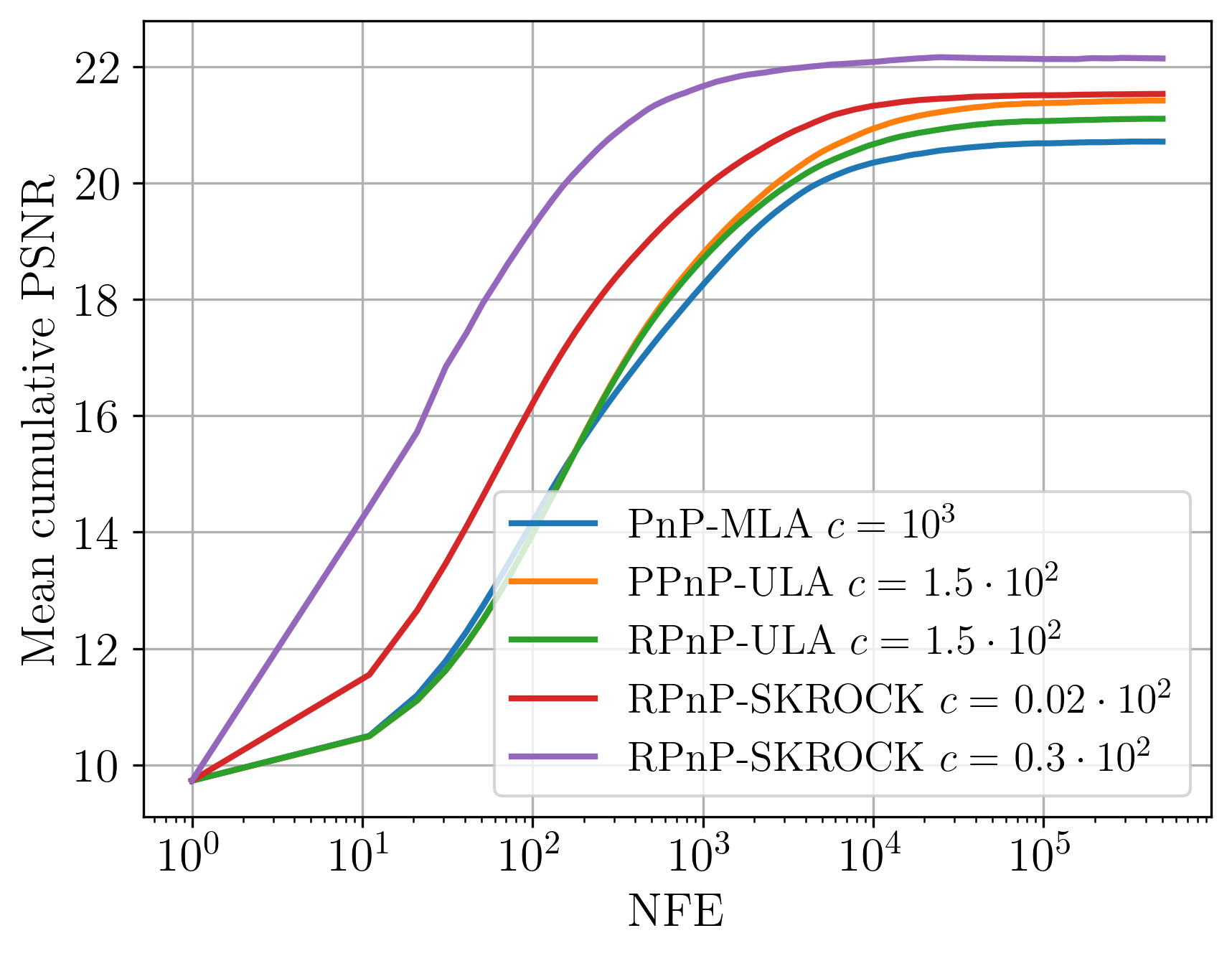}
        \subcaption{PSNR $\uparrow$}\label{fig:conv-psnr-25}
    \end{subfigure}%
    \begin{subfigure}{0.5\textwidth}
        \includegraphics[width=\textwidth]{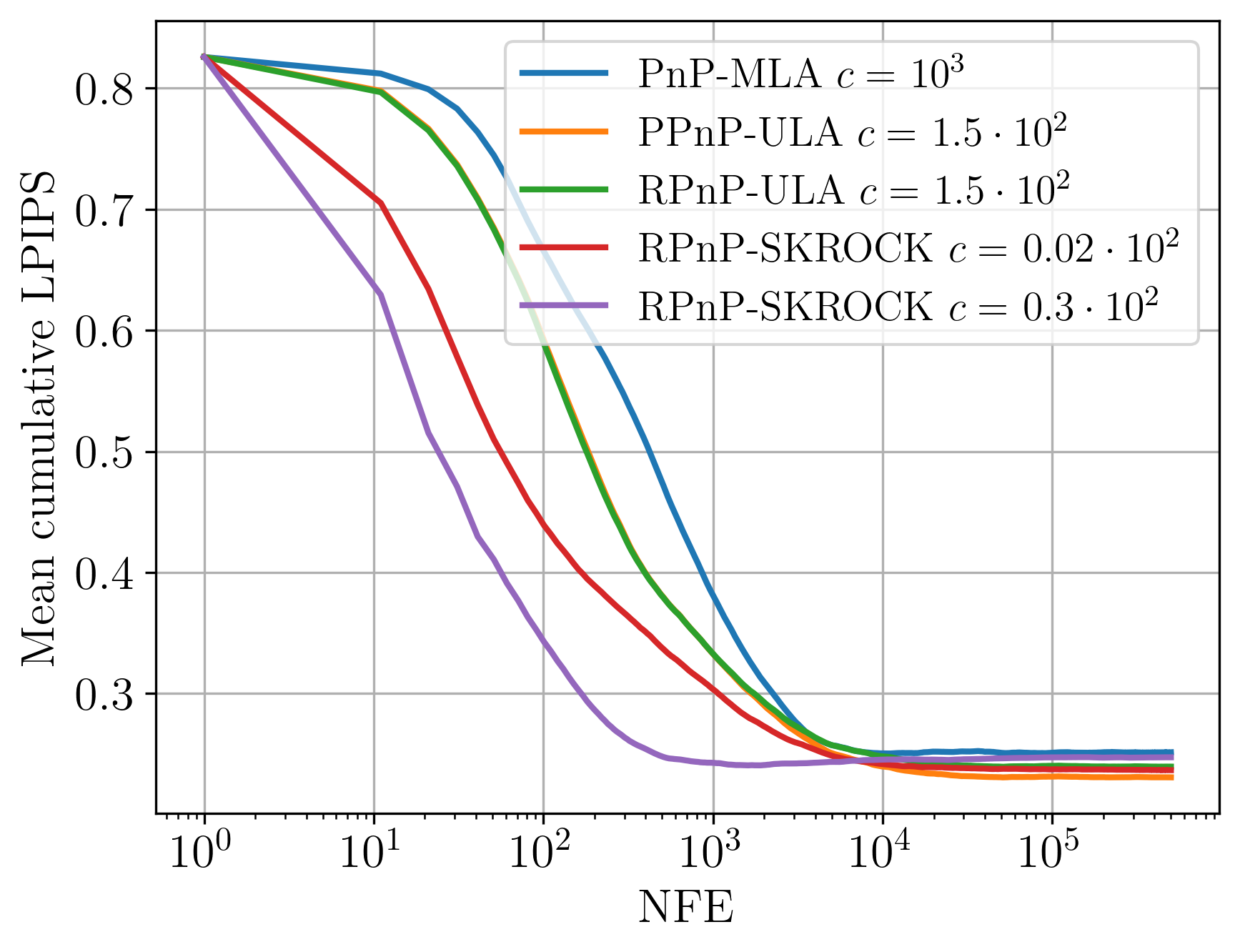}
        \subcaption{LPIPS $\downarrow$}\label{fig:conv-lpips-25}
    \end{subfigure}
    \caption{Mean cumulative metrics for the \texttt{set3c} data set {for photon level $\alpha=20$} using selected time steps, with x-axis in log scale to show the tail behaviour more clearly.} 
    \label{fig:conv-speed-best-timesteps-25}
    }
\end{figure}

Finally, in Figure \ref{fig:conv-speed-best-timesteps-25} we compare quality metrics (PSNR and LPIPS) against neural function evaluations (NFE) averaged over \texttt{set3c} using recommended step sizes to assess the convergence speed. We observe that RPnP-SKROCK outperforms all other algorithms in terms of PSNR in performance and convergence speed. In addition, RPnP-ULA and PPnP-ULA show similar performance and convergence speed while they outperform RPnP-SKROCK and PnP-MLA in terms of LPIPS. 
{These differences relate to the fact that the two algorithms target two slightly different invariant distributions, in particular PnP-MLA targeting \eqref{eq:PoissonLikelihoodNoBeta} and RPnP-SKROCK targeting \eqref{eq:PoissonLikelihood}. In addition, some further bias is introduced by the fact that both algorithms are operating with a finite step-size $\delta$, see for example the discussion in \cite{abdulle2014}.}
PnP-MLA shows slower convergence in terms of PSNR, but it performs similarly with the other algorithms in terms of LPIPS with sharp convergence in $\sim10^{4}$ NFEs.

\begin{figure}[ht!]
\centering
\begin{minipage}[c]{0.9\textwidth}

\begin{center}
\begin{tabular*}{\textwidth}{@{\extracolsep{\fill}} c@{\hspace{1ex}}cccc @{}}
&\makebox[0.18\textwidth]{{\scriptsize\textbf{Blur 1, $\alpha=10$}}}
&\makebox[0.18\textwidth]{{\scriptsize\textbf{Blur 2, $\alpha=10$}}}
&\makebox[0.18\textwidth]{{\scriptsize\textbf{Blur 1, $\alpha=5$}}}
&\makebox[0.18\textwidth]{{\scriptsize\textbf{Blur 2, $\alpha=5$}}}
\\
\addlinespace
\verticalsa{\hspace{2cm}{\scriptsize\textbf{Ground truth}}}
&\begin{subfigure}{0.165\textwidth}
  \includegraphics[width=\linewidth]{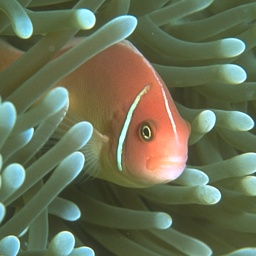}
\end{subfigure}
&\begin{subfigure}{0.165\textwidth}
  \includegraphics[width=\linewidth]{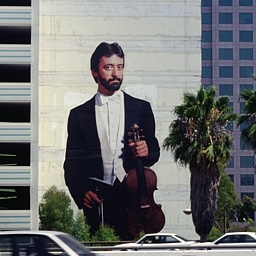}
\end{subfigure}
&\begin{subfigure}{0.165\textwidth}
  \includegraphics[width=\linewidth]{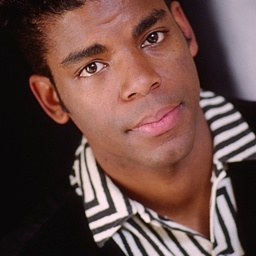}
\end{subfigure}
&\begin{subfigure}{0.165\textwidth}
  \includegraphics[width=\linewidth]{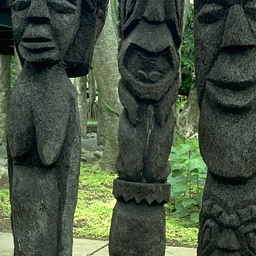}
\end{subfigure}
\vspace{-2cm}
\\
\addlinespace
\verticalsa{\hspace{1.5cm}{\scriptsize\textbf{Observation}}}
&\begin{subfigure}{0.165\textwidth}
  \includegraphics[width=\linewidth]{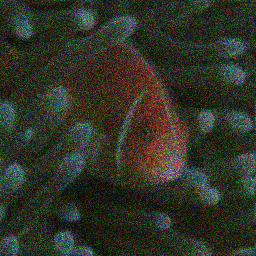}
\end{subfigure}
&\begin{subfigure}{0.165\textwidth}
  \includegraphics[width=\linewidth]{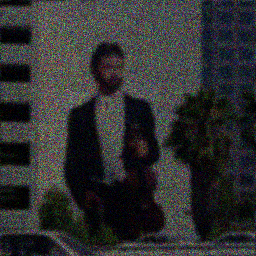}
\end{subfigure}
&\begin{subfigure}{0.165\textwidth}
  \includegraphics[width=\linewidth]{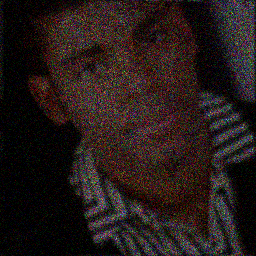}
\end{subfigure}
&\begin{subfigure}{0.165\textwidth}
  \includegraphics[width=\linewidth]{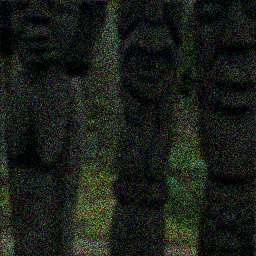}
\end{subfigure}
\vspace{-1.5cm}
\\
\addlinespace
\verticalsa{\hspace{2.8cm}{\scriptsize\textbf{RPnP-SKROCK}}}
&\begin{subfigure}{0.165\textwidth}
  \includegraphics[width=\linewidth]{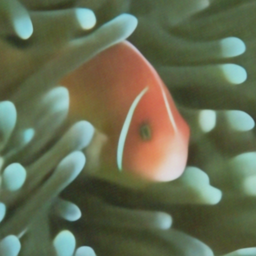}
   \vspace*{-0.65cm}
  \caption*{\footnotesize{(\textbf{26.37}, \textbf{0.21})}}
\end{subfigure}
&\begin{subfigure}{0.165\textwidth}
  \includegraphics[width=\linewidth]{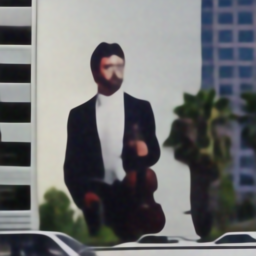}
   \vspace*{-0.65cm}
  \caption*{\footnotesize{(\textbf{22.40}, \textbf{0.36})}}
\end{subfigure}
&\begin{subfigure}{0.165\textwidth}
  \includegraphics[width=\linewidth]{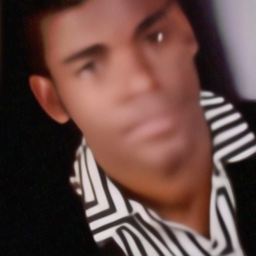}
   \vspace*{-0.65cm}
  \caption*{\footnotesize{(\textbf{25.88}, 0.23)}}
\end{subfigure}
&\begin{subfigure}{0.165\textwidth}
  \includegraphics[width=\linewidth]{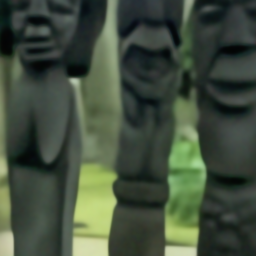}
   \vspace*{-0.65cm}
  \caption*{\footnotesize{(\textbf{22.48}, 0.66)}}
\end{subfigure}
\vspace{-2.6cm}
\\
\addlinespace
\verticalsa{\hspace{2.5cm}{\scriptsize\textbf{PnP-MLA}}}
&\begin{subfigure}{0.165\textwidth}
  \includegraphics[width=\linewidth]{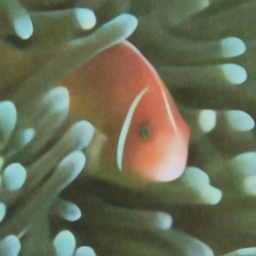}
  \vspace*{-0.65cm}
  \caption*{\footnotesize{(26.00, 0.23)}}
\end{subfigure}
&\begin{subfigure}{0.165\textwidth}
  \includegraphics[width=\linewidth]{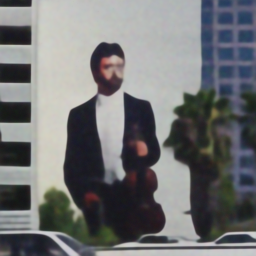}
   \vspace*{-0.65cm}
  \caption*{\footnotesize{(21.77, 0.37)}}
\end{subfigure}
&\begin{subfigure}{0.165\textwidth}
  \includegraphics[width=\linewidth]{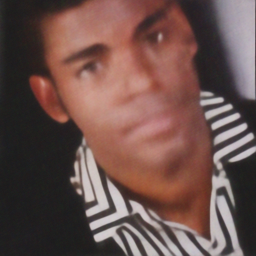}
   \vspace*{-0.65cm}
  \caption*{\footnotesize{(22.71, \textbf{0.21})}}
\end{subfigure}
&\begin{subfigure}{0.165\textwidth}
  \includegraphics[width=\linewidth]{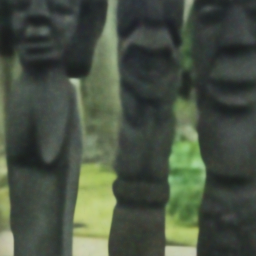}
   \vspace*{-0.65cm}
  \caption*{\footnotesize{(21.81, \textbf{0.60})}}
\end{subfigure}
\vspace{-2cm}
\\
\addlinespace
\verticalsa{\hspace{2.7cm}{\scriptsize\textbf{PIP}}}
&\begin{subfigure}{0.165\textwidth}
  \includegraphics[width=\linewidth]{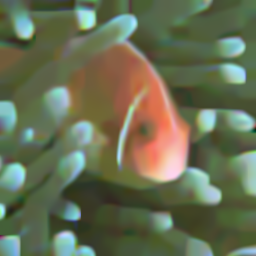}
  \vspace*{-0.65cm}
  \caption*{\footnotesize{(25.00, 0.39)}}
\end{subfigure}
&\begin{subfigure}{0.165\textwidth}
  \includegraphics[width=\linewidth]{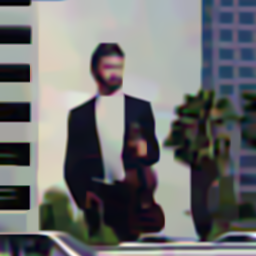}
  \vspace*{-0.65cm}
  \caption*{\footnotesize{(21.28, 0.50)}}
\end{subfigure}
&\begin{subfigure}{0.165\textwidth}
  \includegraphics[width=\linewidth]{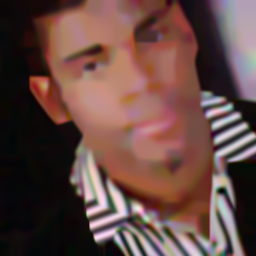}
  \vspace*{-0.65cm}
  \caption*{\footnotesize{(23.29, 0.36)}}
\end{subfigure}
&\begin{subfigure}{0.165\textwidth}
  \includegraphics[width=\linewidth]{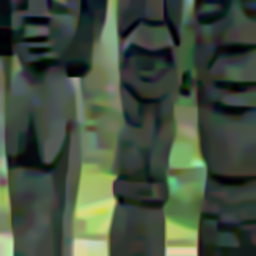}
  \vspace*{-0.65cm}
  \caption*{\footnotesize{(22.07, 0.73)}}
\end{subfigure}
\vspace{-1.7cm}
\\
\addlinespace
\verticalsa{\hspace{3cm}{\scriptsize\textbf{PnP-BPG}}}
&\begin{subfigure}{0.165\textwidth}
  \includegraphics[width=\linewidth]{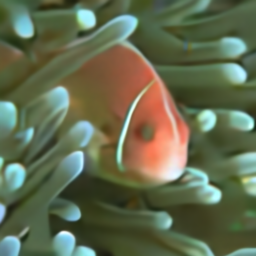}
  \vspace*{-0.65cm}
  \caption*{\footnotesize{(25.92, 0.23)}}
\end{subfigure}
&\begin{subfigure}{0.165\textwidth}
  \includegraphics[width=\linewidth]{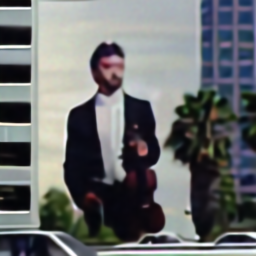}
  \vspace*{-0.65cm}
  \caption*{\footnotesize{(21.97, 0.40)}}
\end{subfigure}
&\begin{subfigure}{0.165\textwidth}
  \includegraphics[width=\linewidth]{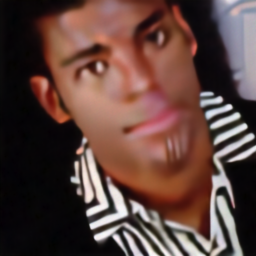}
  \vspace*{-0.65cm}
  \caption*{\footnotesize{(23.41, 0.26)}}
\end{subfigure}
&\begin{subfigure}{0.165\textwidth}
  \includegraphics[width=\linewidth]{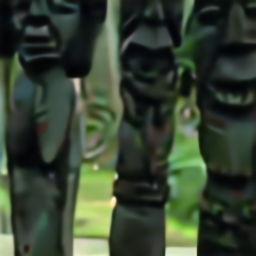}
  \vspace*{-0.65cm}
  \caption*{\footnotesize{(20.51, 0.66)}}
\end{subfigure}
\vspace{-2.3cm}
\\
\addlinespace
\verticalsa{\hspace{3cm}{\scriptsize\textbf{PhD-Net}}}
&\begin{subfigure}{0.165\textwidth}
  \includegraphics[width=\linewidth]{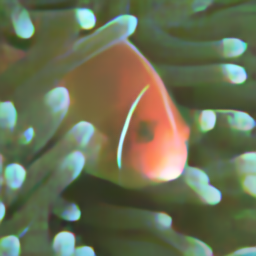}
  \vspace*{-0.65cm}
  \caption*{\footnotesize{(25.51, 0.33)}}
\end{subfigure}
&\begin{subfigure}{0.165\textwidth}
  \includegraphics[width=\linewidth]{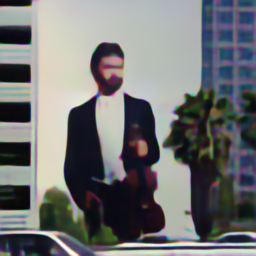}
  \vspace*{-0.65cm}
  \caption*{\footnotesize{(22.18, 0.38)}}
\end{subfigure}
&\begin{subfigure}{0.165\textwidth}
  \includegraphics[width=\linewidth]{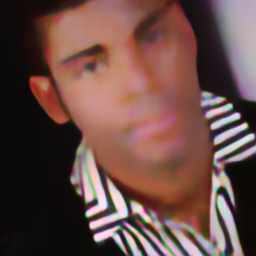}
  \vspace*{-0.65cm}
  \caption*{\footnotesize{(24.21, 0.29)}}
\end{subfigure}
&\begin{subfigure}{0.165\textwidth}
  \includegraphics[width=\linewidth]{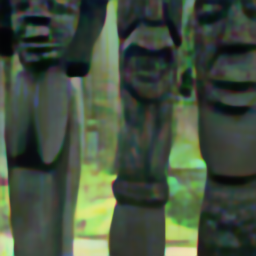}
  \vspace*{-0.65cm}
  \caption*{\footnotesize{(22.15, 0.69)}}
\end{subfigure}

\end{tabular*}
\end{center}
\vspace{-2.5cm}
\caption{Visual results on \texttt{CBSD10} dataset for different methods and simulation settings. Associated (PSNR, LPIPS) values are indicated below each image; best marked in \textbf{bold}.}
\label{fig:sota_results}
    
\end{minipage}
\end{figure}

\subsection{Comparisons with the state-of-the-art}\label{sec:sota}
\subsubsection{Experimental set up}
We are now ready to present comparisons with approaches for Poisson image restoration from the state-of-the-art. These alternative methodologies deliver a single point estimate, rather than a posterior distribution from which point estimates and other forms of inference can be computed. Based on the conclusions of Sections \ref{sec:ablation} and \ref{sec:convergence-speed}, we report Bayesian PnP inference with the RPnP-SKROCK (Algorithm \ref{alg:r_SKROCK}) and PnP-MLA (Algorithm \ref{alg:b-pnp}), and we use ProxDRUNet as PnP prior in both cases with $\epsilon = (20/255)^2$ and $\epsilon = (25/255)^2$, respectively. We set $\delta =  (0.02\cdot10^2)\cdot\ell_s\delta_{L}$ and $\delta = (5\cdot10^2)\cdot\delta_{L}$ for RPnP-SKROCK and PnP-MLA respectively. 

For these comparisons, we consider different shot noise levels and use real-world camera shake kernels extracted from \cite{Levin09}, see Figure \ref{levin_kernels}. We present a selection of these experiments with the kernels from Figure \ref{fig:levin-4} and \ref{fig:levin-9}, which we henceforth denote as \texttt{Blur 1} and \texttt{Blur 2}, respectively, and three different levels of shot noise ($\alpha=5, 10, 20$). We report comparisons with the PnP-ADMM scheme PIP \cite{rond2016-pip} which uses a patch-based BM3D denoiser, the unrolled network PhD-Net \cite{sanghvi2021photonlimited}, and the Bregman proximal gradient method PnP-BPG \cite{hurault2023}. We run our experiments on the \texttt{CBSD10} set.

\begin{figure}[tb!]

\begin{center}
\begin{subfigure}[c]{.2\textwidth}
    \centering
    \includegraphics[width=1\linewidth]{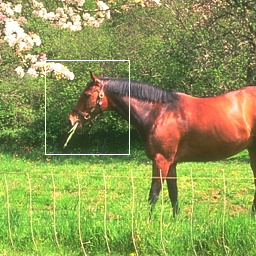}
    \caption{\smaller GT  \\\phantom{-} } 
\end{subfigure}%
\hfill
\begin{subfigure}[c]{.2\textwidth}
    \centering
    \includegraphics[width=1\linewidth]{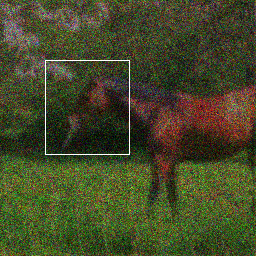}\vspace{0.01cm}
    \caption{\smaller Observation \\\phantom{-}} 
\end{subfigure}%
\hfill
\begin{subfigure}[c]{.2\textwidth}
    \centering
    \includegraphics[width=0.92\linewidth]{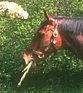}
    \caption{\smaller GT  \\(detail) } 
\end{subfigure}%
\hfill
\begin{subfigure}[c]{.2\textwidth}
    \centering
    \includegraphics[width=0.92\linewidth]{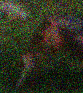}
    \caption{\smaller Observation \\(detail)} 
\end{subfigure}%

\begin{subfigure}{.2\textwidth}
    \centering
    \includegraphics[width=0.95\linewidth]{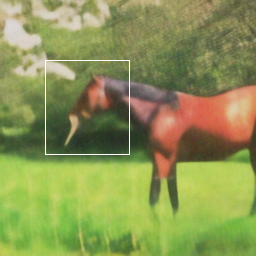}
\end{subfigure}%
\hfill
\begin{subfigure}{.2\textwidth}
    \centering
    \includegraphics[width=0.95\linewidth]{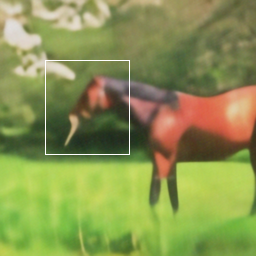}
\end{subfigure}%
\hfill
\begin{subfigure}{.2\textwidth}
    \centering
    \includegraphics[width=0.95\linewidth]{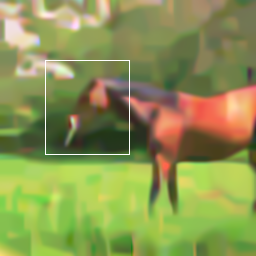}
\end{subfigure}%
\hfill
\begin{subfigure}{.2\textwidth}
    \centering
    \includegraphics[width=0.95\linewidth]{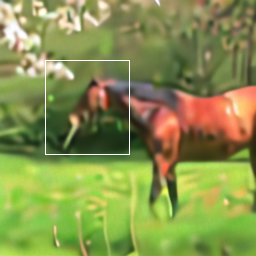}
\end{subfigure}%
\hfill
\begin{subfigure}{.2\textwidth}
    \centering
    \includegraphics[width=0.95\linewidth]{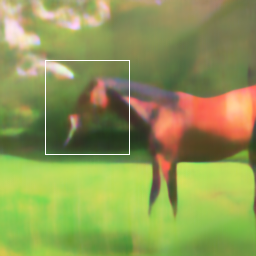}
\end{subfigure}%
\vspace{0.06cm}
\begin{subfigure}{.2\textwidth}
    \centering
    \includegraphics[width=0.95\linewidth]{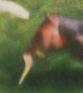}
    \caption{\smaller{PnP-MLA}  \\(\textbf{19.68}, \textbf{0.661}) } 
\end{subfigure}
\hfill
\begin{subfigure}{.2\textwidth}
    \centering
    \includegraphics[width=0.95\linewidth]{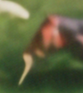}
     \caption{\smaller{\scriptsize{RPnP-SKROCK}}\\ (19.61, 0.782)} 
\end{subfigure}%
\hfill
\begin{subfigure}{.2\textwidth}
    \centering
    \includegraphics[width=0.95\linewidth]{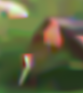}
    \caption{\smaller{PIP  \\(19.37, 0.857)}} 
\end{subfigure}%
\hfill
\begin{subfigure}{.2\textwidth}
    \centering
    \includegraphics[width=0.95\linewidth]{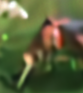}
     \caption{\smaller{PnP-BPG  \\(19.63, 0.787)} }
\end{subfigure}%
\hfill
\begin{subfigure}{.2\textwidth}
    \centering
    \includegraphics[width=0.95\linewidth]{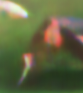}
     \caption{\smaller{PhD-Net\\(19.40, 0.826)}}
\end{subfigure}%
\hfill
\end{center}
    \vspace{-0.2cm}
    \caption{Qualitative results for different algorithms for $\alpha=10$ and Blur 1. In the second row, we zoomed in a specific area, to give more attention to details. Reported metrics: (PSNR, LPIPS). Best result marked in \textbf{bold}.}
    \label{fig:sota-a-10-k-4-im-8}
\end{figure}

\begin{table}[!t]
\setlength\tabcolsep{3pt}
\renewcommand{\arraystretch}{1.2}
\scriptsize
\centering
\begin{tabular}{c|ccccccc}
\hline
\textbf{Level}&\textbf{Kernel}                   &  \textbf{Metrics}   &\textbf{RPnP-SKROCK}  & \textbf{PnP-MLA}    & \textbf{PIP}&\textbf{PnP-BPG}&\textbf{PhD-Net}\\
\hline
\multirow{6}{*}{\makecell{$\alpha=5$}} & \multirow{3}{*}{Blur 1} & PSNR&\textbf{22.77}&21.75 & 21.94&21.24&\underline{22.37}\\
                  &                   & SSIM &\textbf{0.57}&\underline{0.54}&\underline{0.54}&\textbf{0.57}&\underline{0.54}\\
                  &                   & LPIPS&\underline{0.52}&\textbf{0.45}&0.61&\underline{0.52}&0.55\\
\cline{2-8} 
                  & \multirow{3}{*}{Blur 2} &PSNR& \textbf{22.82}&21.81&22.08 &21.23&22.60\\
                  &                   & SSIM & \textbf{0.58} &0.55& 0.55&\underline{0.57}& \underline{0.57} \\
                  &                   & LPIPS & \underline{0.53}&\textbf{0.48}& 0.61 &0.54&\underline{0.53} \\
\hline
\multirow{6}{*}{\makecell{$\alpha=10$}} & \multirow{3}{*}{Blur 1} & PSNR&\textbf{23.42} &22.87&22.86&23.38&23.12\\
                  &                   & SSIM&\underline{0.61} &0.58&0.58&\textbf{0.63}&0.58\\
                  &                   & LPIPS &\underline{0.43}&\textbf{0.40}&0.57&0.47&0.52\\
\cline{2-8}
                  & \multirow{3}{*}{Blur 2} &PSNR & \underline{23.35}&22.88&22.96&23.27&\textbf{23.41}\\
                  &                   & SSIM & \underline{0.61}&0.59&0.59& \textbf{0.63}&\underline{0.61}\\
                  &                   & LPIPS& \underline{0.47}  &\textbf{0.44}& 0.57&0.50&0.50\\
\hline
\end{tabular}
\caption{Quantitative results (averaged over the \texttt{CBSD10} set). For each quality metric, the best result is shown in \textbf{bold} and the second best is \underline{underlined}.}
\label{table:overall_results}
\end{table}

\subsubsection{Experimental results}
Table \ref{table:overall_results} summarizes the results for this experiment. We observe that the proposed RPnP-SKROCK and PnP-MLA Bayesian PnP approaches are very competitive in terms of reconstruction PSNR and SSIM, outperforming or performing similarly to the state-of-the-art in all cases. PnP-MLA consistently outperforms all other methods in perceptual quality, as measured by LPIPS, with RPnP-SKROCK following close in performance. With regards to the alternative approaches, as expected, PIP is the least competitive method, as it is based on a patch-based BM3D denoiser. PnP-BPG performs strongly in SSIM, while PhD-Net is competitive in PSNR. However, they are less accurate when errors are measured via LPIPS. Detailed quantitative results for each image of \texttt{CBSD10} are depicted in scatter plots in Fig. \ref{fig:scatter} in Appendix \ref{sec:app-sota}.

Figure \ref{fig:sota_results} displays a selection of results. We observe that RPnP-SKROCK and PnP-MLA achieve reconstructions with better detail and no colorization or color blocking artefacts; see the second and fourth columns of Figure \ref{fig:sota_results} as an example. 
Finally, in Figure \ref{fig:sota-a-10-k-4-im-8} we present another example image from the \texttt{CBSD10} dataset. Observe that the Bayesian PnP methods are able to better recover sharp details, and PnP-MLA even recovers some of the texture in the background, whereas the alternative methods struggle to recover perceptual details and textures.

{
\subsection{Low-dose computer tomography experiment}\label{sec:low-dose-ct}
\subsubsection{Experimental set up}
We now present an experiment on low-dose Computed Tomography (CT) reconstruction, a canonical inverse problem commonly encountered in medical imaging and non-destructive testing. In the low-dose setting, CT measurements exhibit strong Poisson noise, making it an ideal test case for the proposed algorithms. We consider a standard parallel-beam CT configuration, where the forward operator $A$ is a discretization of the Radon transform in a full-view setup. While this full-view setup avoids sparse-view and limited-angle artifacts, the problem remains severely ill-conditioned and hence requires significant regularization to produce solutions that are well posed. To simulate low-dose acquisitions, we apply this operator to ground truth lung images from the the LIDC-IDRI dataset \cite{armato2011lung} and then corrupt the resulting sinograms with aggressive Poisson noise, effectively limiting the photons per view. For the PnP prior, we trained a new GS-DRUNet denoiser $D_\epsilon$ specifically for this task. 
The network was trained on 2D slices from the LIDC-IDRI training set, following the data preparation and training strategy of \cite{tachella2023equivariant} and \cite{hurault22a}.
We set the photon level $\alpha=10$, the regularization parameter $\rho=4$, the denoiser level $\epsilon=(20/255)^2$, and determined appropriate step sizes for PnP-MLA ($\delta_{MLA}=10^{-4}$) and RPnP-SKROCK ($\delta_{SKR}=5 \cdot 10^{-5}$) via a grid search. For RPnP-SKROCK, which requires a Lipschitz-continuous likelihood, we used the regularized Poisson likelihood \eqref{eq:PoissonLikelihood} as in Sections \ref{sec:convergence-speed}, with $\beta$ is set to $1\%$ of the mean intensity value of the observed data. In contrast, PnP-MLA uses the same likelihood, however with a much smaller $\beta = 10^{-8}$ which is required to ensure numerical stability.

\subsubsection{Experimental results}

Figure \ref{fig:low-dose-ct} presents an example of a reconstruction result for an image from the LIDC-IDRI test set, alongside the standard Filtered Back-Projection (FBP) for reference. Both RPnP-SKROCK and PnP-MLA successfully recover high-quality images from the low-dose CT measurements, demonstrating a significant improvement over FBP. We also show the joint marginal standard deviations for pixel scale and groups of pixels of size $2\times 2$, $4\times 4$ and $8\times 8$ pixels, which display the uncertainty about structures at different sizes and in different regions of the reconstruction. These maps provide valuable insights for interpretation, highlighting uncertainty in areas rich in fine anatomical structures. 

\begin{figure}[t]
    \centering

\begin{tikzpicture}[every node/.style={inner sep=0,outer sep=0}]
   
    \node (a) at (0,3.5) {\includegraphics[width=3cm]{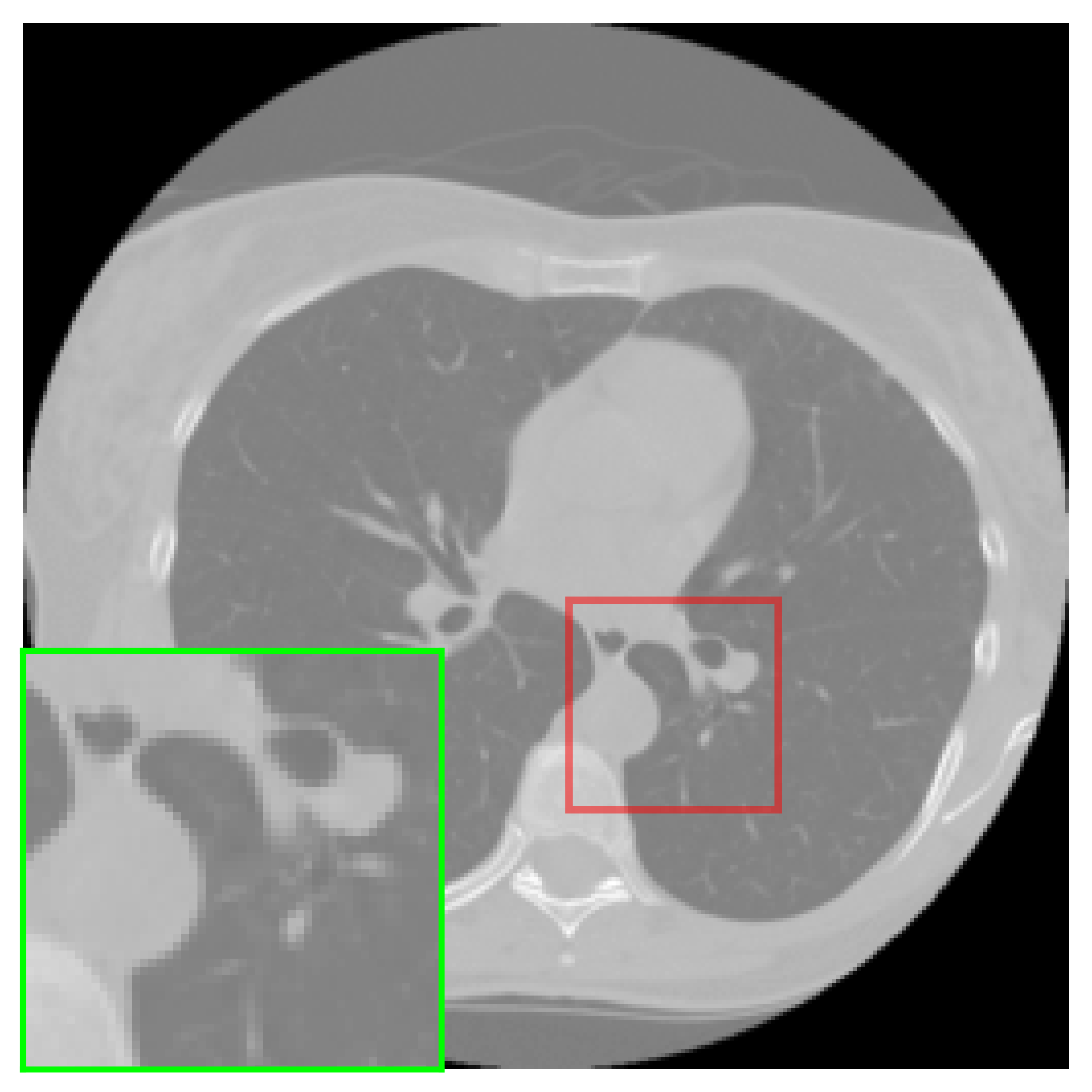}};
    \node (b) at (3.2,3.5) {\includegraphics[width=2.9cm]{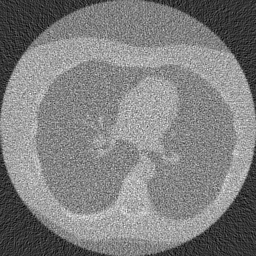}};
    \node (c) at (6.4,3.5) {\includegraphics[width=3cm]{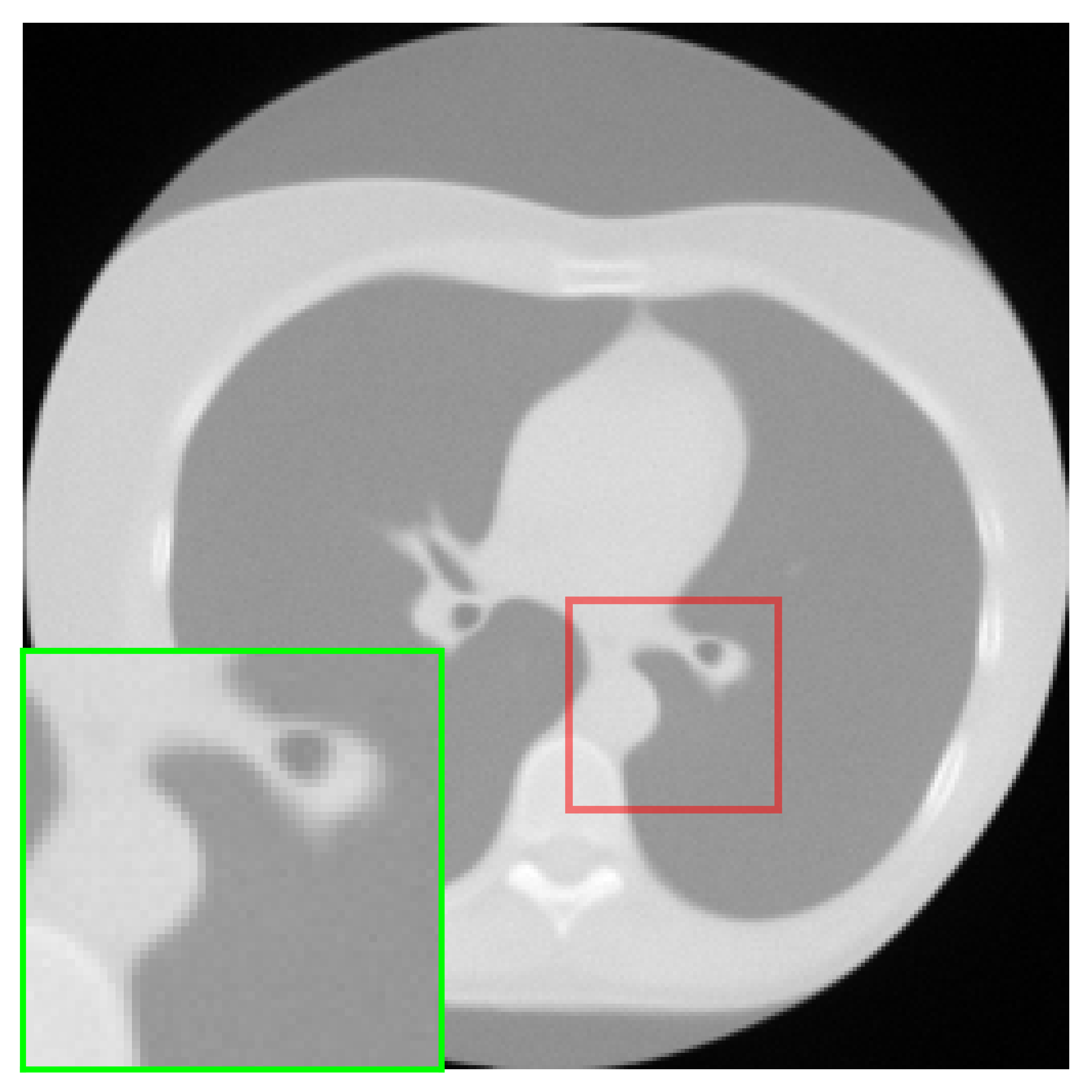}};
    \node (d) at (9.6,3.5) {\includegraphics[width=3cm]{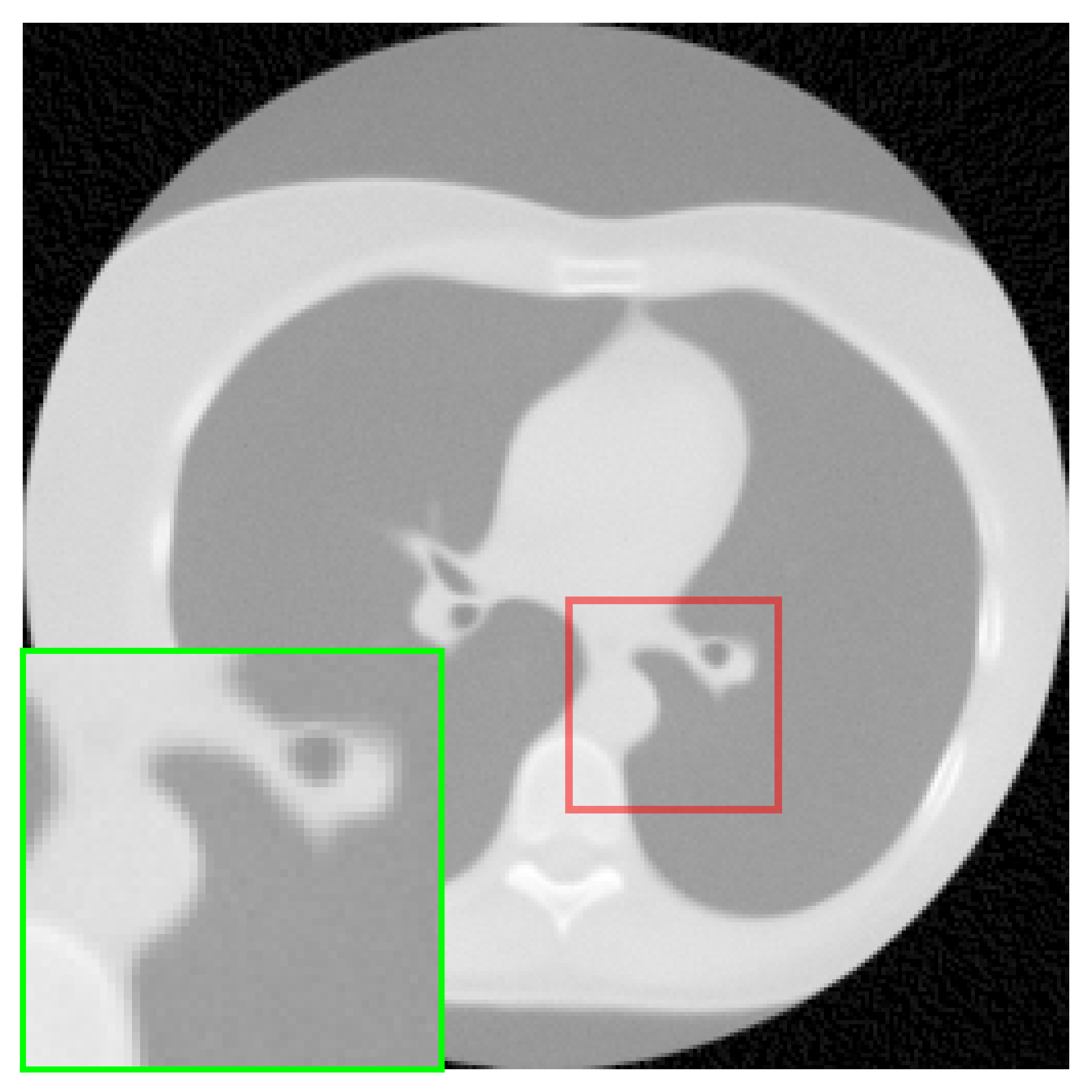}};

    \node[above=0.1cm of a] {\smaller GT};
    \node[above=0.1cm of b] {\smaller FBP};


    \node (e) at (0,-0.2) {\includegraphics[width=3cm]{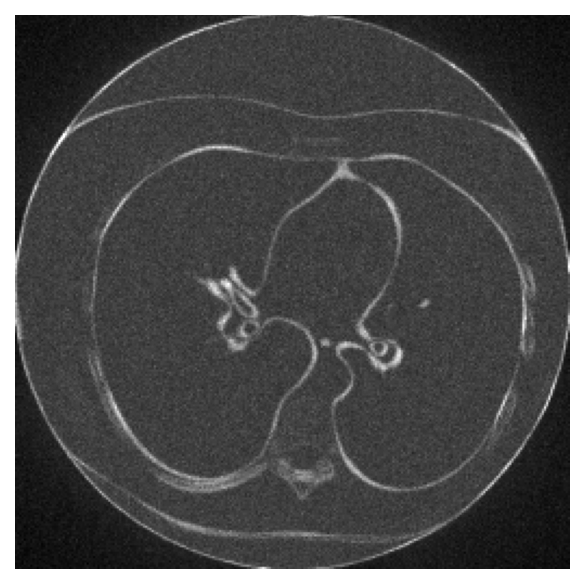}};
    \node (f) at (3.2,-0.2) {\includegraphics[width=3cm]{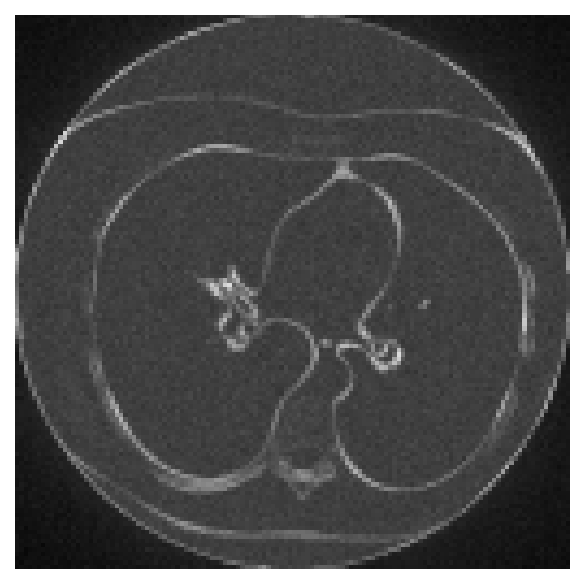}};
    \node (g) at (6.4,-0.2) {\includegraphics[width=3cm]{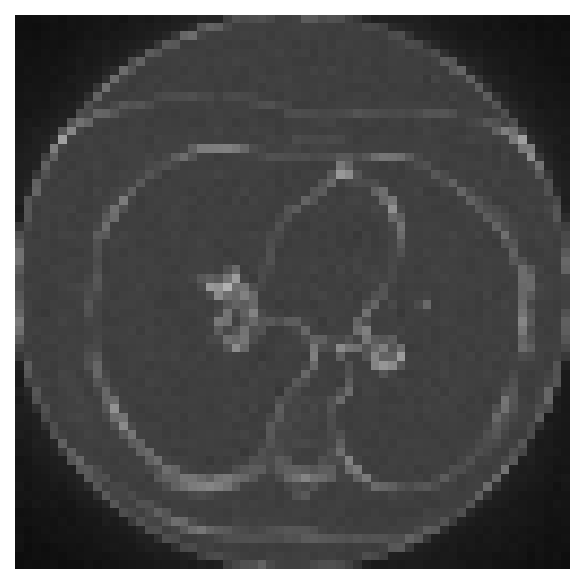}};
    \node (h) at (9.6,-0.2) {\includegraphics[width=3cm]{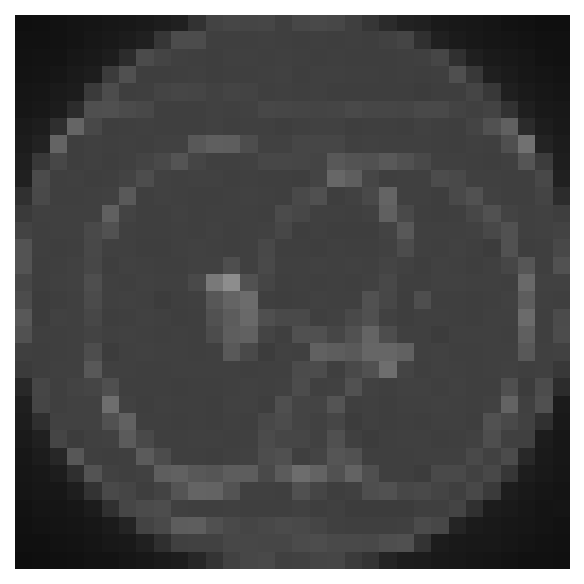}};
    
    \node[above=0.1cm of c] { \smaller \textbf{RPnP-SKROCK}};
    \node[below=0.1cm of c] {\scriptsize (29.2,0.10)};
    \node[above=0.1cm of d] { \smaller \textbf{PnP-MLA}};
    \node[below=0.1cm of d] {\scriptsize (29.8,0.15)};
    \node[above=0.1cm of e] { \smaller \textbf{RPnP-SKROCK}};

    \node[draw, fit=(c)(d), inner ysep=11pt,
    inner xsep=2pt, 
    thick, rounded corners,label={[label distance=1mm]above:$\hat{x}_{MMSE}$}] {};

    \node[below=0.1cm of e] {\scriptsize Std. dev. 1$\times 1$};
    \node[below=0.1cm of f] {\scriptsize Std. dev. 2$\times 2$};
    \node[below=0.1cm of g] {\scriptsize Std. dev. 4$\times 4$};
    \node[below=0.1cm of h] {\scriptsize Std. dev. 8$\times 8$};

    

    \node (i) at (0,-3.9) {\includegraphics[width=3cm]{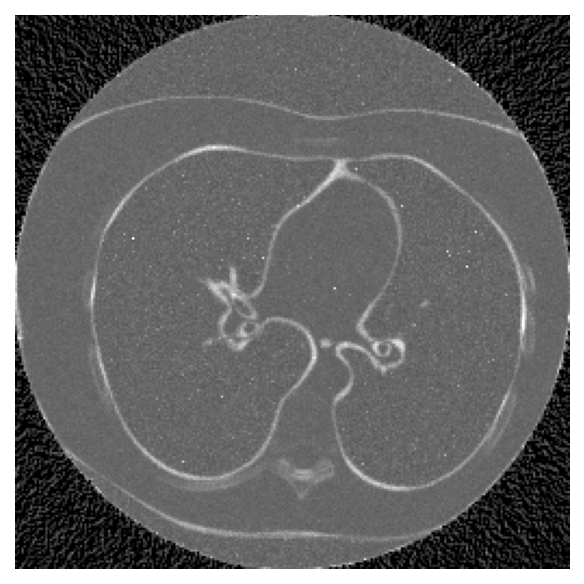}};
    \node (j) at (3.2,-3.9) {\includegraphics[width=3cm]{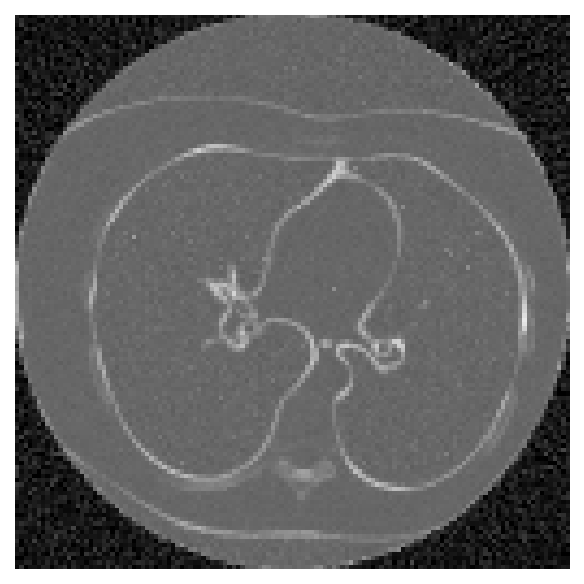}};
    \node (k) at (6.4,-3.9) {\includegraphics[width=3cm]{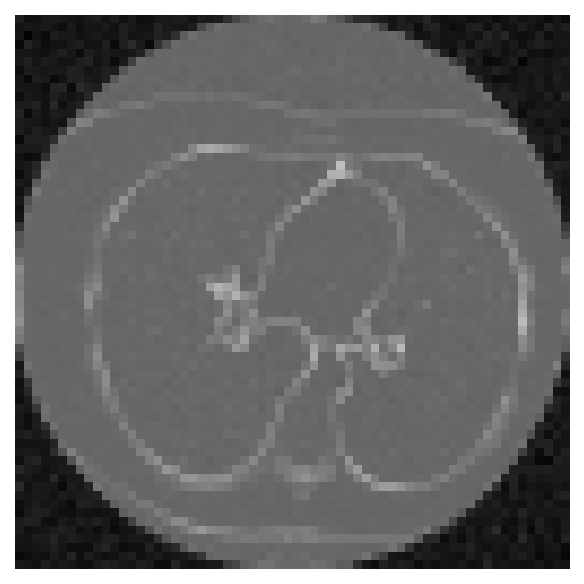}};
    \node (l) at (9.6,-3.9) {\includegraphics[width=3cm]{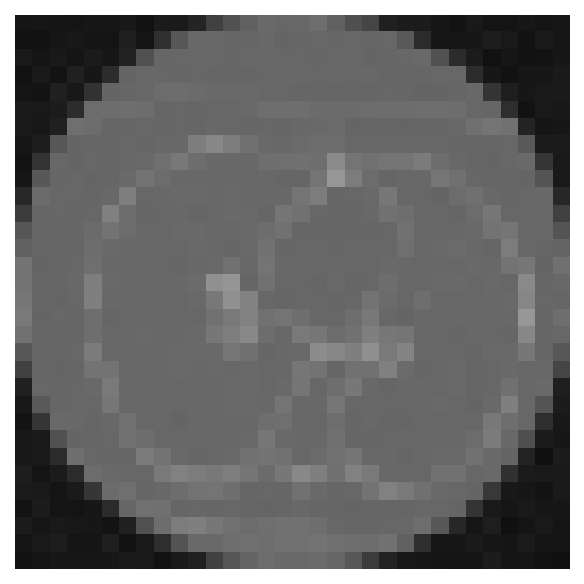}};

    \node[above=0.1cm of i] { \smaller \textbf{PnP-MLA}};

    \node[below=0.1cm of i] {\scriptsize Std. dev. $1\times 1$};
    \node[below=0.1cm of j] {\scriptsize Std. dev. $2\times 2$};
    \node[below=0.1cm of k] {\scriptsize Std. dev. $4\times 4$};
    \node[below=0.1cm of l] {\scriptsize Std. dev. $8\times 8$};

\end{tikzpicture}
    \caption{Results of the low-dose CT experiment for $\alpha=10$. First row: Ground truth, FBP, $\hat{x}_{MMSE}$, with reported metrics (PSNR, LPIPS) for two select methods (RPnP-SKROCK and PnP-MLA). Second and third row: The marginal standard deviations of the pixel-wise scale $1\times 1$, and for groups of pixels of size $2\times 2$, $4\times 4$ and $8\times 8$ pixels, for both methods.}
    \label{fig:low-dose-ct}
\end{figure}

}

{
\section{Discussion}\label{sec:discussion}
A natural question arising from this work is whether the explicit acceleration of SKROCK could be combined with the geometric adaptation of Mirror Langevin (MLA) to create a further improved sampler. While this presents an intriguing prospect, we understand that these two methodologies represent fundamentally different, and potentially conflicting, paradigms for addressing ill-conditioned problems. SKROCK provides explicit acceleration by using a sophisticated integrator to improve convergence rates as a function of the Euclidean condition number $\kappa$. In contrast, MLA provides implicit acceleration by reconditioning the problem via a mirror map, using a geometry where the effective condition number $\tilde{\kappa}$ is significantly smaller, thus allowing a simpler integrator to perform efficiently.
Combining these approaches is non-trivial and opens significant theoretical questions. It is unclear what additional benefit an explicit $\mathcal{O}(\kappa)$-type acceleration scheme would provide if the mirror map has already driven the effective condition number $\tilde{\kappa} \approx 1$. Furthermore, defining and analyzing Nesterov-style momentum within the dual space geometry of a mirror map, particularly for a stochastic process, is highly non trivial. In particular, this extension would require extending to the SDE setting the theories of accelerated mirror descent from optimization \cite{krichene2015accelerated} and continuous-time ODE perspectives on non-Euclidean acceleration \cite{dobson2024accelerated, alimisis2020continuous}. Given these complexities, we consider the development of a provably accelerated mirror Langevin algorithm a research direction in its own right, and an interesting avenue for future work. Our contribution rigorously studies the practical performance of these two distinct state-of-the-art approaches for the challenging problem of PnP Poisson imaging.
}

\section{Conclusion}\label{sec:conclusion}
{This paper addressed an important gap in Bayesian imaging methodology by making advanced Plug-and-Play (PnP) MCMC a viable and powerful tool for Poisson inverse problems. We systematically adapted and extended state-of-the-art Langevin samplers to handle the poor regularity and constraints of the Poisson likelihood. We focused on two algorithms: the explicitly accelerated RPnP-SKROCK, using reflections and a likelihood approximation, and PnP-MLA, the first PnP mirror Langevin algorithm for imaging, which leverages a problem-adapted geometry to simultaneously deal with the constraints and with the Poisson likelihood without approximations.}

{Our main contribution is a comprehensive empirical study comparing these distinct geometric and accelerated MCMC methodologies. Using Poisson image deblurring as a testbed, we evaluated a wide range of denoiser architectures, concluding that Prox-DRUNet provides a robust prior for this problem class. The extensive numerical experiments demonstrated that in challenging low-photon settings (small $\alpha$), the proposed sampling algorithms significantly outperform optimization-based PnP strategies in reconstruction accuracy while also providing uncertainty quantification. {The results yielded a set of practical recommendations: for low-photon imaging problems, we advocate for PnP-MLA in applications that require superior perceptual quality (LPIPS) and hence a more faithful reconstruction of fine detail and small structures with edges and contours. Conversely, we expect RPnP-SKROCK to deliver a {stronger} MMSE performance (PSNR) with a comparatively lower computational cost. We expect both methods to produce similarly detailed uncertainty quantification results}.

{These methodologies also open several avenues for future research. An important direction is to embed Markov kernels like PnP-MLA and RPnP-SKROCK within empirical Bayesian strategies for automatic parameter calibration in semi-blind problems \cite{vidal-2020, kemajou24}. } 
In addition, while in this paper we have focused on relatively conventional PnP denoiser architectures, we see great potential in using a PnP prior derived from a denoising diffusion model (DM) within a Langevin sampling scheme, like in \cite{mbakam2024empiricalbayesianimagerestoration}. Extending the DM-based PnP-ULA of \cite{mbakam2024empiricalbayesianimagerestoration} to Poisson imaging problems by using either PnP-MLA or RPnP-SKROCK is also an important perspective for future work. Lastly, a highly promising but more challenging direction for further research would be to explore accelerated PnP-MLA sampling, combining the strengths of PnP-MLA and RPnP-SKROCK.

{
Finally, we acknowledge the computational cost of the proposed MCMC methods. Unlike optimization techniques that yield a single point estimate, MCMC samplers inherently require a significant number of iterations to explore the posterior distribution for uncertainty quantification, particularly in challenging low-photon regimes. However, there are several promising directions to mitigate this cost. The development of explicitly accelerated mirror Langevin schemes, as mentioned above, could reduce the required number of iterations. Furthermore, the cost per iteration could be lowered by replacing large denoisers like Prox-DRUNet with more lightweight yet powerful architectures, such as those based on model distillation \cite{tang25} or learned (weakly) convex regularizers \cite{goujon2022crrnn, weakly24}. Scalability could also be significantly improved by adapting divide-and-conquer MCMC strategies \cite{vyner23} to imaging inverse problems. These approaches represent important next steps toward making advanced Bayesian inference both powerful and computationally practical for a broader range of applications.
}

\clearpage

\bibliographystyle{siamplain}
\bibliography{references} 

\clearpage

\makeatletter
\renewcommand{\p@subfigure}{\thefigure} \makeatother
\appendix

\section{Additional information for Section \ref{sec:ablation}}\label{sec:app-ablation}

\noindent\textbf{Additional results.} 
 We present results based on the \texttt{starfish} image from  \texttt{set3c}, to complement the results for the \texttt{leaves} image presented in Section \ref{sec:ablation}. Figure \ref{fig:starfish_truth} presents the ground truth and Figure \ref{fig:noisy_starfish} presents a realization $y$ generated by the forward model Eq. \eqref{eq:PoissonLikelihoodNoBeta} with operator $A$ modeling the motion blur kernel in Figure \ref{fig:levin-4}, while the photon level of the Poisson noise is set to $\alpha=20$. 

 \begin{figure}[h!]
\centering
    \begin{subfigure}{0.31 \textwidth}
         \centering
         \includegraphics[width=\linewidth]{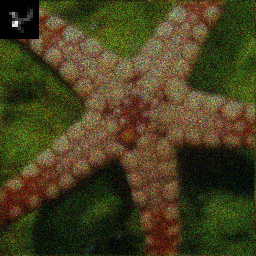}
         \caption{\tiny{Observation}}
         \label{fig:noisy_starfish}
    \end{subfigure}
    \begin{subfigure}{0.31 \textwidth}
         \centering
         \includegraphics[width=\linewidth]{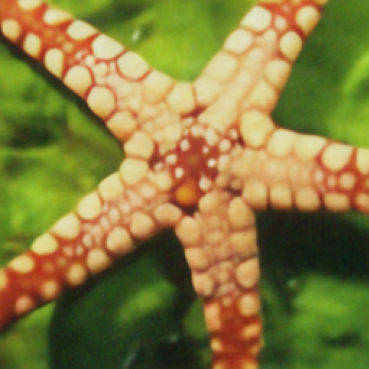}
         \caption{\tiny{PPnP-ULA: Prox-DRUNet}}
         \label{fig:starfish_prox_ppnp6.1}
    \end{subfigure}
    \begin{subfigure}{0.31 \textwidth}
         \centering
         \includegraphics[width=\linewidth]{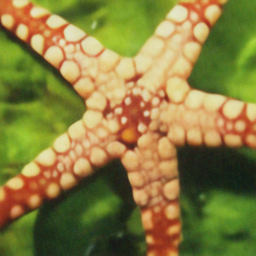}
         \caption{\tiny{PnP-MLA: Prox-DRUNet}}
         \label{fig:starfish_prox_mirror6.1}
    \end{subfigure}

    \begin{subfigure}{0.31 \textwidth}
         \centering
         \includegraphics[width=\linewidth]{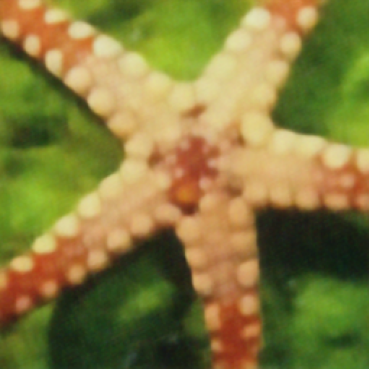}
         \caption{\tiny{PPnP-ULA: LMMO}}
        \label{fig:starfish_lmmo}
    \end{subfigure}
    \begin{subfigure}{0.31 \textwidth}
         \centering
         \includegraphics[width=\linewidth]{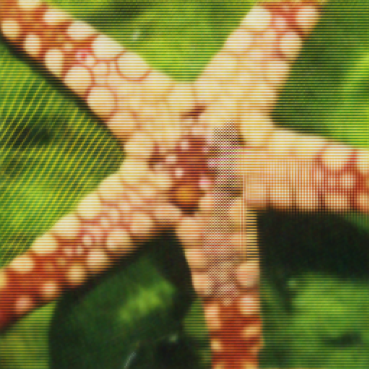}
         \caption{\tiny{PPnP-ULA: GS-DRUNet}}
        \label{fig:starfish_gsdrunet}
    \end{subfigure}
    \begin{subfigure}{0.31 \textwidth}
         \centering
         \includegraphics[width=\linewidth]{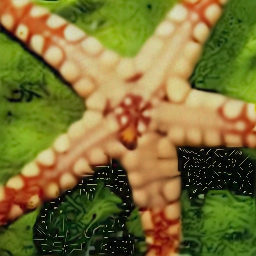}
         \caption{\tiny{PnP-MLA: B-DRUNet}}
        \label{fig:starfish_bdrunet_106}
    \end{subfigure}    
 \caption{Poisson deconvolution problem for $\alpha=20$ and Blur 1 using the \texttt{starfish} image. Noisy and blurry observation, and the MMSE reconstructions $\hat{x}_{MMSE}$ for Prox-DRUNet, LMMO, GS-DRUNet and B-DRUNet using PPnP-ULA and PnP-MLA.}
 \label{fig:results_denoisers_starfish}
 \end{figure}

Figures \ref{fig:starfish_prox_ppnp6.1} to \ref{fig:starfish_bdrunet_106} show the posterior mean $\hat{x}_{MMSE}$, as calculated for the different denoisers and different algorithms using $10^6$ iterations to stress-test the stability of the algorithms. Prox-DRUNet proved to be the most competitive in terms of stability and performance among all the considered denoisers. More specifically, we observe that PPnP-ULA with the LMMO denoiser produces an estimate of the mean with excessive smoothing, while with the GS-DRUNet exhibits checkerboard-like and stripe-like reconstruction artefacts on large parts of the \texttt{starfish} and the background. The B-DRUNet denoiser progressively develops artefacts after approximately $10^5$ iterations which deteriorate gravely over time, see Figure \ref{fig:starfish_bdrunet_106}. 

Table \ref{tab:ablation-networks-im2} summarizes the performance of the denoisers for the considered image. The LMMO prior leads to faster convergence but achieves low reconstruction quality, while GS-DRUNet is not stable, so the reconstruction quality decreases as the iterations progress and the artefacts become more pronounced. Similarly, B-DRUNet also leads to instability and reaches its top performance in around $10^5$ iterations before artefacts start to amplify. Conversely, Prox-DRUNet outperforms the other networks in all the image quality metrics considered. For completeness, Table \ref{tab:ablation-networks-im2} also reports the number of iterations that PPnP-ULA and PnP-MLA require to reach $98\%$ of peak PSNR performance, as an indicator of convergence speed for the posterior mean. Convergence speed is comparable between PPnP-ULA and PnP-MLA using Prox-DRUNet.
Note that the Prox-DRUNet yields the best accuracy-speed trade-off, reaching almost top performance at approximately $3 \cdot 10^4$ iterations for PPnP-ULA and PnP-MLA. We point out that for PPnP-ULA similar quantitative behavior was observed for the different denoisers even when equivariance was used, as well as when using Algorithms \ref{alg:pnpula-reflect} and \ref{alg:euclidean-skrock}. We do not report results for PnP-MLA without randomization as this technique was necessary to stabilize PnP-MLA when using B-DRUNet.

Figure \ref{fig:starfish_st_dev} shows the pixel-wise posterior standard deviation, as calculated with the LMMO, Prox-DRUNet, GS-DRUNet and B-DRUNet denoisers. For reference, next to each posterior standard deviation plot, we also report the residual obtained by comparing $\hat{x}_{MMSE}$ to the true image (note that these standard deviations represent the models' marginal predictions for these residuals, at the pixel level). We observe that LMMO and Prox-DRUNet produce uncertainty plots that are broadly in agreement with their residuals. For Prox-DRUNet and LMMO, uncertainty concentrates around edges and contours, as expected for a deconvolution problem, whereas the uncertainty estimates of GS-DRUNet are aligned with its reconstruction artefacts. Conversely, B-DRUNet produces uncertainty plots that highlight homogenenous regions, and which do not align well with its residual, and has particularly high uncertainty in the areas where artefacts develop.  We conclude that Prox-DRUNet is the most appropriate denoising architecture for Bayesian PnP inference in Poisson image deblurring problems regardless of the sampling algorithm.

 \begin{figure}[p!]
 \centering
\begin{minipage}[c]{0.89\textwidth}
\centering
 \rotatebox{0}{Standard deviation \phantom{whitespace here} Residual \phantom{more}}
   \vspace{-0.2cm}
    \begin{subfigure}{0.4 \textwidth}
         \centering
          \begin{minipage}{0.10\textwidth} 
        \rotcaption{\tiny{(Prox-DRUNet, PPnP-ULA)}}\label{fig:std-res-proxdrunet-ppnp-starfish}
        \end{minipage}
        \vspace{0.5cm}
        \begin{minipage}{0.85\textwidth} 
         \includegraphics[width=\linewidth]{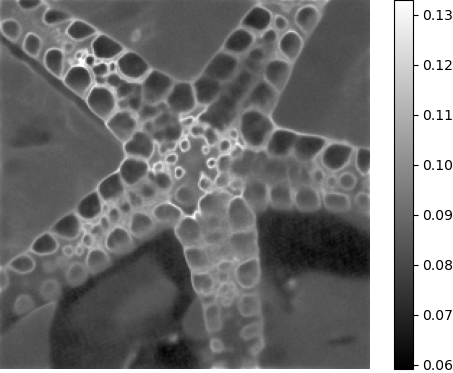}
         \end{minipage}
    \end{subfigure}
    \begin{subfigure}{0.4 \textwidth}
         \centering
          \begin{minipage}{0.10\textwidth} 
           \makebox[0pt]{\rotatebox{90}{\phantom{\tiny{(Prox-DRUNet, PPnP-ULA)}}}}
        \end{minipage}
        \vspace{0.5cm}
        \begin{minipage}{0.85\textwidth} 
         \includegraphics[width=\linewidth]{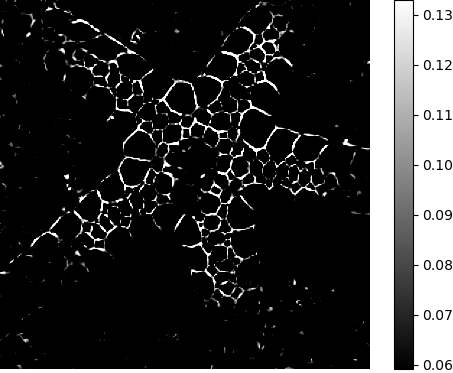}
         \end{minipage}
    \end{subfigure}
  \begin{subfigure}{0.40 \textwidth}
         \centering
        \begin{minipage}{0.10\textwidth} 
        \rotcaption{\tiny{(Prox-DRUNet, PnP-MLA)}}\label{fig:std-res-proxdrunet-mla-starfish}
        \end{minipage}
        \vspace{0.5cm}
        \begin{minipage}{0.85\textwidth} 
         \includegraphics[width=\linewidth]{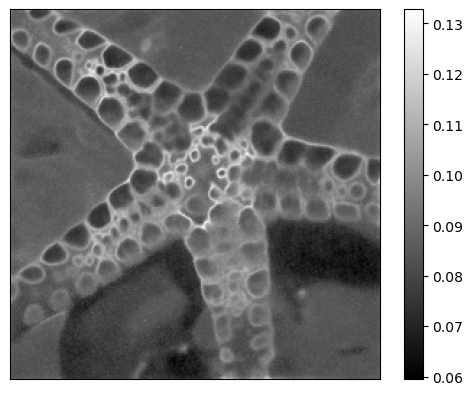}
         \end{minipage}
    \end{subfigure}
    \begin{subfigure}{0.4 \textwidth}
         \centering
          \begin{minipage}{0.10\textwidth} 
           \makebox[0pt]{\rotatebox{90}{\phantom{\tiny{(Prox-DRUNet, PnP-MLA)}}}}
        \end{minipage}
        \vspace{0.5cm}
        \begin{minipage}{0.85\textwidth} 
         \includegraphics[width=\linewidth]{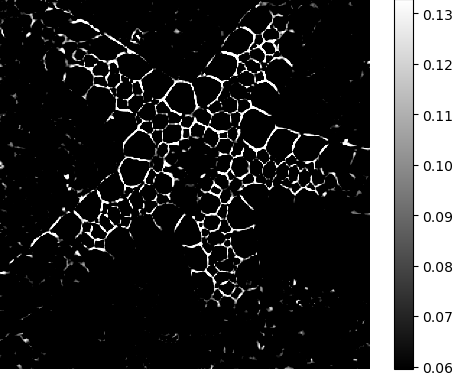}
         \end{minipage}
    \end{subfigure}
    
    \begin{subfigure}{0.4 \textwidth}
         \centering
          \begin{minipage}{0.10\textwidth} 
        \rotcaption{\tiny{(B-DRUNet, PnP-MLA)}}\label{fig:std-res-bdrunet-mla-starfish}
        \end{minipage}
        \vspace{0.5cm}
        \begin{minipage}{0.85\textwidth} 
         \includegraphics[width=\linewidth]{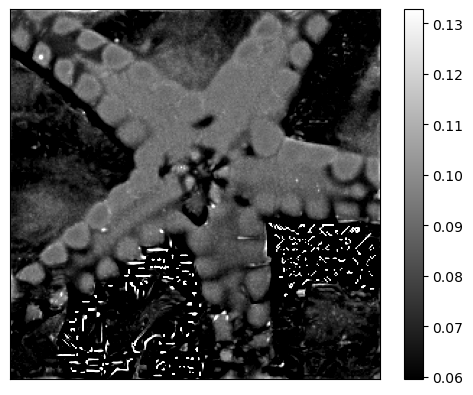}
         \end{minipage}
     \end{subfigure}    
     \begin{subfigure}{0.4 \textwidth}
         \centering
          \begin{minipage}{0.10\textwidth} 
           \makebox[0pt]{\rotatebox{90}{\phantom{\tiny{(B-DRUNet, PnP-MLA)}}}}
        \end{minipage}
        \vspace{0.5cm}
        \begin{minipage}{0.85\textwidth} 
         \includegraphics[width=\linewidth]{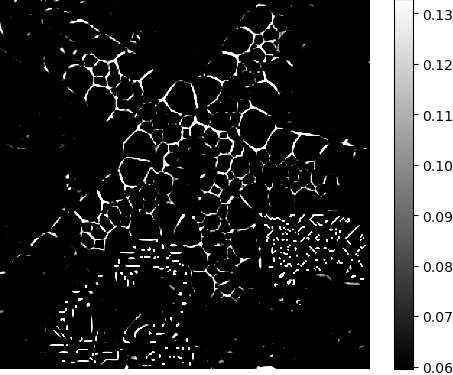}
         \end{minipage}
     \end{subfigure}
    \begin{subfigure}{0.4 \textwidth}
         \centering
          \begin{minipage}{0.10\textwidth} 
        \rotcaption{\tiny{(LMMO, PPnP-ULA)}}\label{fig:std-res-lmmo-ppnp-starfish}
        \end{minipage}
        \vspace{0.5cm}
        \begin{minipage}{0.85\textwidth} 
         \includegraphics[width=\linewidth]{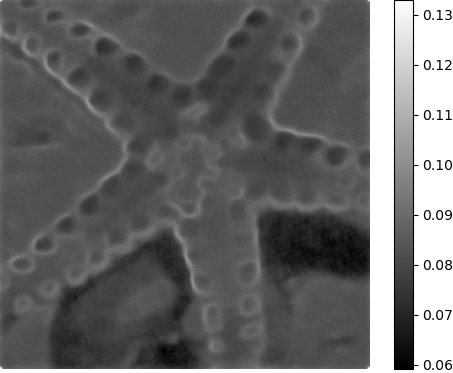}
         \end{minipage}
    \end{subfigure}
    \begin{subfigure}{0.4 \textwidth}
         \centering
          \begin{minipage}{0.10\textwidth} 
           \makebox[0pt]{\rotatebox{90}{\phantom{\tiny{(LMMO, PPnP-ULA)}}}}
        \end{minipage}
        \vspace{0.5cm}
        \begin{minipage}{0.85\textwidth} 
         \includegraphics[width=\linewidth]{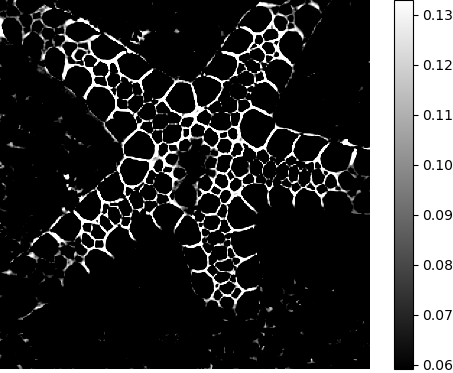}
        \end{minipage}
    \end{subfigure}
    
    \begin{subfigure}{0.4 \textwidth}
         \centering
          \begin{minipage}{0.10\textwidth} 
        \rotcaption{\tiny{(GS-DRUNet, PPnP-ULA)}}\label{fig:std-res-gsdrunet-ppnp-starfish}
        \end{minipage}
        \vspace{0.5cm}
        \begin{minipage}{0.85\textwidth} 
         \includegraphics[width=\linewidth]{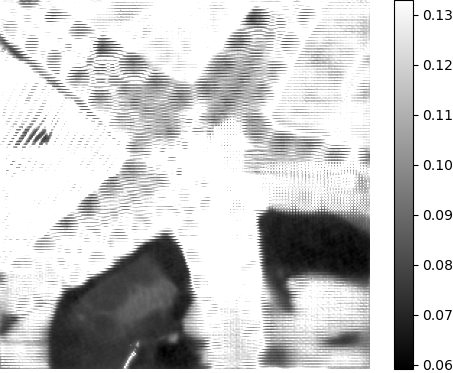}
         \end{minipage}
     \end{subfigure} 
    \begin{subfigure}{0.4 \textwidth}
         \centering
          \begin{minipage}{0.10\textwidth} 
           \makebox[0pt]{\rotatebox{90}{\phantom{\tiny{(GS-DRUNet, PPnP-ULA)}}}}
        \end{minipage}
        \vspace{0.5cm}
        \begin{minipage}{0.85\textwidth} 
         \includegraphics[width=\linewidth]{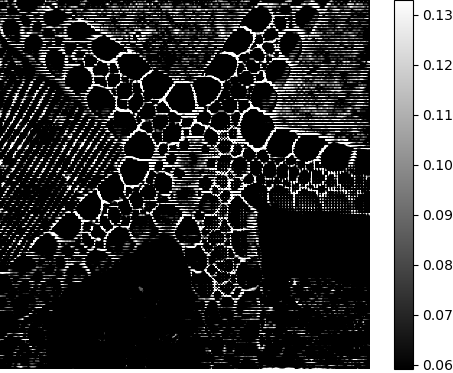}
         \end{minipage}
    \end{subfigure}
 \caption{Poisson deconvolution problem as in Fig. \ref{fig:results_denoisers_starfish}. Pairwise: Pixelwise standard deviations (left column) and residuals (right column) for all tested denoisers using either PPnP-ULA or PnP-MLA.}
 \label{fig:starfish_st_dev}
 \end{minipage}
 \end{figure}

\begin{table}[t]
\smaller
    \centering
    \renewcommand{\arraystretch}{1.2} 
    \begin{tabular}{llccccc}
\hline
\textbf{Method}           & \multicolumn{1}{c}{\textbf{Denoiser}} & \multicolumn{1}{c}{\textbf{PSNR}} & \textbf{SSIM} & \textbf{LPIPS} & \multicolumn{1}{c}{\textbf{\begin{tabular}[c]{@{}c@{}}Iter. until \\ 98$\%$ PSNR\end{tabular}}} \\ \hline
\multirow{3}{*}{PPnP-ULA} & \textbf{LMMO}                          & 21.63                              & 0.60          & 0.41           & 11500                                                                                         \\ 
                          & \textbf{GS-DRUNet}                     & 18.68                              & 0.46          & 0.41           & 12000                                                                                         \\ 
                          & \textbf{Prox-DRUNet}                   & 23.54                              & 0.69          & 0.29           & 25000                                                                                         \\ \hline
\multirow{2}{*}{PnP-MLA}  & \textbf{Prox-DRUNet}                   & 23.68                              & 0.69          & 0.29           & 33000                                                                                          \\ 
                          & \textbf{B-DRUNet}                      & 19.53                              & 0.49          & 0.47           & 111500                                                                                         \\ \hline
\end{tabular}
\caption{Quantitative results for $\hat{x}_{MMSE}$ (calculated by PPnP-ULA and PnP-MLA) for the \texttt{starfish} image. Last column: The number of iterations required such that $98\%$ PSNR is reached. }
\label{tab:ablation-networks-im2}
\end{table}

\clearpage

\section{Additional information for Section \ref{sec:convergence-speed}}\label{sec:app-convergence}

\subsection{Additional details for MMSE reconstructions of \texttt{set3c}}\label{sec:app-conv-speed-extra}
To support the results in Figure \ref{fig:TBD}, we include here the quantitative results averaged over \texttt{set3c} in Tables \ref{tab:rpnp-step-size}-\ref{tab:mirror-step-size-eps-25}.
\begin{table}[h!]
	\smaller
	\centering
	\renewcommand{\arraystretch}{1.2} 
	\begin{tabular}{ccccc}
		\hline
		\textbf{$c$} & \multicolumn{1}{c}{\textbf{PSNR}} & \multicolumn{1}{c}{\textbf{LPIPS}} & \multicolumn{1}{c}{\textbf{SSIM}} & \textbf{Stability} \\ \hline
		{$\mathbf{0.02\cdot 10^2}$}                      & 20.44                              & 0.26                               & 0.67                       &\ding{51}       \\ 
		{$\mathbf{0.1 \cdot 10^2}$}                     & 20.95                              & 0.25                               & 0.70               & \ding{51}              \\ 
		{$\mathbf{0.2\cdot10^2}$}                     & 21.02                              & 0.25                               & 0.70                 & \ding{51}             \\ 
		{$\mathbf{1.5\cdot10^2}$}                    & \textbf{21.10}                              & \textbf{0.24}                               & \textbf{0.70}         & \ding{51}                    \\ 
		{$\mathbf{3\cdot10^2}$}                    & 21.05                              & 0.24                               & 0.70            & \ding{51}                 \\ 
		{$\mathbf{5\cdot10^2}$}                    & 20.84                              & 0.24                               & 0.70                     & \ding{51}         \\ 
		$\mathbf{10^3   }$                & 19.71                              & 0.28                               & 0.63                & \ding{55}              \\ 
	\end{tabular}
	\caption{Quantitative results for RPnP-ULA under different step sizes on \texttt{set3c}.}
	\label{tab:rpnp-step-size}
\end{table}

\begin{table}[h!]
	\smaller
	\centering
	\renewcommand{\arraystretch}{1.2} 
	\begin{tabular}{ccccc}
		\hline
		\textbf{$c$} & \multicolumn{1}{c}{\textbf{PSNR}} & \multicolumn{1}{c}{\textbf{LPIPS}} & \multicolumn{1}{c}{\textbf{SSIM}} & \textbf{Stability} \\ \hline
		$\mathbf{0.02\cdot10^2}$                    & 20.50                              & 0.26                               & 0.68        &  \ding{51}                   \\ 
		$\mathbf{0.1\cdot10^2}$                    & 21.04                              & 0.24                               & 0.70         &     \ding{51}                \\ 
		$\mathbf{0.2\cdot10^2}$                     & 21.15                              & 0.24                               & 0.70       &       \ding{51}                \\ 
		$\mathbf{1.5\cdot10^2}$                   & \textbf{21.42}                              & \textbf{0.23}                               & \textbf{0.71}    &  \ding{51}                        \\ 
		$\mathbf{3\cdot10^2}$                   & 21.37                              & {0.23}                               & 0.71       &     \ding{51}                  \\ 
		{$\mathbf{5\cdot10^2}$}                    & 20.62                              & 0.26                               & 0.68      &    \ding{55}                    \\ 
		{$\mathbf{10^3}$ }                  & 19.36                              & 0.30                               & 0.61        &         \ding{55}             \\ 
	\end{tabular}
	\caption{Quantitative results for PPnP-ULA under different step sizes on \texttt{set3c}.}
	\label{tab:ppnp-step-size}
\end{table}
\begin{table}[h!]
	\smaller
	\centering
	\renewcommand{\arraystretch}{1.2} 
	\begin{tabular}{cccccc}
		\hline
		\textbf{$c$} & \multicolumn{1}{c}{\textbf{PSNR}} & \multicolumn{1}{c}{\textbf{LPIPS}} & \multicolumn{1}{c}{\textbf{SSIM}} & \textbf{Stability} & \textbf{Smoothing} \\ \hline
		$\mathbf{0.002\cdot10^2}$                     & 21.26                              & 0.24                               & 0.71             & \ding{51}    & -              \\ 
		$\mathbf{0.01\cdot10^2}$                   & 21.44                              & 0.24                               & 0.71                 & \ding{51} & -             \\ 
		$\mathbf{0.02\cdot10^2}$                    & 21.53                              & \textbf{0.24}                               & 0.72           & \ding{51} & -                    \\ 
		$\mathbf{0.15\cdot10^2}$                 & 21.85                              & 0.25                               & 0.72                    & \ding{51} & -          \\
		$\mathbf{0.3\cdot10^2}$                  & \textbf{22.14}                              & 0.25                               & \textbf{0.73}        & \ding{51}   & \ding{51}                    \\ 
		$\mathbf{0.5\cdot10^2}$                  & 22.05                              & 0.26                               & 0.72                 & \ding{51}   & \ding{51}           \\ 
		{$\mathbf{10^2}$  }                 & 22.07                              & 0.27                               & 0.73            & \ding{51}     & \ding{51}              \\ 
	\end{tabular}
	\caption{Quantitative results for RPnP-SKROCK under different step sizes on \texttt{set3c}.}
	\label{tab:rskrock-step-size}
\end{table}
\begin{table}[h!]
	\smaller
	\centering
	\renewcommand{\arraystretch}{1.2} 
	\begin{tabular}{ccccc}
		\hline
		\textbf{$c$} & \multicolumn{1}{c}{\textbf{PSNR}} & \multicolumn{1}{c}{\textbf{LPIPS}} & \multicolumn{1}{c}{\textbf{SSIM}} & \textbf{Stability} \\ \hline
		$\mathbf{0.1\cdot10^2}$                   & 19.74                              & 0.26                               & 0.60             &     \ding{51}                 \\ 
		$\mathbf{0.5\cdot10^2}$                     & 20.29                              & {0.26}                               & 0.65              &     \ding{51}               \\ 
		$\mathbf{10^2}$                  & 20.40                              & 0.26                              & 0.66                  &     \ding{51}            \\ 
		$\mathbf{1.5\cdot10^2}$                    & 20.46                              & 0.26                               & 0.66                   &     \ding{51}           \\ 
		$\mathbf{2\cdot10^2}$                & 20.50                                 & 0.26                               & 0.67                   &     \ding{51}           \\ 
		$\mathbf{5\cdot10^2}$                  & 20.63                              & 0.25                               & 0.67         &    \ding{51}                  \\ 
		$\mathbf{7.5\cdot10^2}$                    & 20.68                              & 0.25                               & 0.67                 &     \ding{51}             \\ 
		$\mathbf{10^3}$                  & 20.71                              & 0.25                               & 0.68              &     \ding{51}                \\ 
{        $\mathbf{1.25\cdot10^3}$    }              & {20.68}                              & {0.25}                               & {0.67}             &     {\ding{51}       }         \\
	\end{tabular}
	\caption{Quantitative results for PnP-MLA under different step sizes for $\epsilon=25$ on \texttt{set3c}.}
	\label{tab:mirror-step-size-eps-25}
\end{table}

\subsection{Additional results for PnP-MLA using Prox-DRUNet with $\epsilon=20$} \label{sec:app-pnpmla-eps20}

Here, additional results to Section \ref{sec:convergence-speed} are presented. We use a detail of the \texttt{starfish} image shown in Figure \ref{fig:qualitative-best-appendix} to illustrate the effects of the chosen step size. Results for the PnP-MLA using $\epsilon=20$ are summarized in Table \ref{tab:mirror-step-size}, yielding similar results to PPnP-ULA in Table \ref{tab:ppnp-step-size} in terms of best metric and step size, both quantitatively and qualitatively (compare Figures \ref{fig:qualitative-detail-ppnp} and \ref{fig:qualitative-detail-mirror}, with the best results in Figure \ref{fig:detail-mirror-150}). With increasing step size the algorithm shows instability by introducing artefacts that only appear in the \texttt{starfish} image (not in any other image in \texttt{set3c} - not shown here), see Figure \ref{fig:detail-mirror-500} and \ref{fig:detail-mirror-750}. We find that by choosing $\epsilon=25$ for the neural network denoiser (these results are described in the main text for Section \ref{sec:convergence-speed}), and thus increasing the denoising strength slightly, all artefacts can be mitigated.

\begin{figure}[t]
    \centering
    \begin{tabular}{cccc}
        \begin{subfigure}[b]{0.21\textwidth}
            \centering
            \includegraphics[width=\textwidth]{figures_stepsize_eval_qualitative_im2_SKROCK20.png}
            \caption{\footnotesize{RPnP-SKROCK}\\\hspace*{0.8cm}$c=0.02\cdot10^2$}
        \end{subfigure} &
        \begin{subfigure}[b]{0.21\textwidth}
            \centering
            \includegraphics[width=\textwidth]{figures_stepsize_eval_qualitative_im2_ppnp_False150.png}
            \caption{\footnotesize{RPnP-ULA} \\\hspace*{0.8cm}$c=1.5\cdot10^2$}
        \end{subfigure} &
        \begin{subfigure}[b]{0.21\textwidth}
            \centering
            \includegraphics[width=\textwidth]{figures_stepsize_eval_qualitative_im2_ppnp_True150.png}
            \caption{\footnotesize{PPnP-ULA} \\\hspace*{0.8cm}$c=1.5\cdot10^2$}
        \end{subfigure} &
        \begin{subfigure}[b]{0.21\textwidth}
            \centering
            \includegraphics[width=\textwidth]{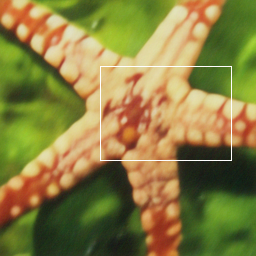}
            \caption{\footnotesize{PnP-MLA} \\\hspace*{0.8cm}$c=2\cdot10^2$}
        \end{subfigure} \\

    \end{tabular}
    \caption{Overview of best qualitative MMSE results for all algorithms with $\epsilon=20$.}
    \label{fig:qualitative-best-appendix}
\end{figure}

\begin{figure}[t!]
    \centering
    \begin{subfigure}[b]{0.19\textwidth}
        \centering
        \includegraphics[width=\textwidth]{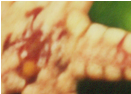}
        \caption{$c=0.5\cdot10^2$}
        \label{fig:detail-mirror-50}
    \end{subfigure}
    \begin{subfigure}[b]{0.19\textwidth}
        \centering
        \includegraphics[width=\textwidth]{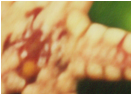}
        \caption{$c=1.5\cdot10^2$}
        \label{fig:detail-mirror-150}
    \end{subfigure}
    \begin{subfigure}[b]{0.19\textwidth}
        \centering
        \includegraphics[width=\textwidth]{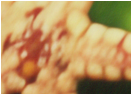}
        \caption{$c=2\cdot10^2$}
        \label{fig:detail-mirror-200}
    \end{subfigure}
    \begin{subfigure}[b]{0.19\textwidth}
        \centering
        \includegraphics[width=\textwidth]{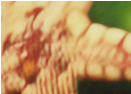}
        \caption{$c=5\cdot10^2$}
        \label{fig:detail-mirror-500}
    \end{subfigure}
    \begin{subfigure}[b]{0.19\textwidth}
        \centering
        \includegraphics[width=\textwidth]{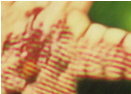}
        \caption{$c=7.5\cdot10^2$}
        \label{fig:detail-mirror-750}
    \end{subfigure}
    \caption{PnP-MLA with $\epsilon=20$.}
    \label{fig:qualitative-detail-mirror}
\end{figure}

\begin{table}[t!]
\smaller
    \centering
    \renewcommand{\arraystretch}{1.2} 
    \begin{tabular}{ccccc}

\hline
\textbf{$c$} & \multicolumn{1}{c}{\textbf{PSNR}} & \multicolumn{1}{c}{\textbf{LPIPS}} & \multicolumn{1}{c}{\textbf{SSIM}} & \textbf{Stability} \\ \hline
$\mathbf{0.1\cdot10^2}$                    & 20.22                              & \textbf{0.23}                               & 0.62             &     \ding{51}                 \\ 
$\mathbf{0.5\cdot10^2}$                  & 21.06                              & {0.23}                               & 0.69               &     \ding{51}               \\ 
$\mathbf{10^2}$                 & 21.20                              & 0.24                               & 0.70                  &     \ding{51}            \\ 
$\mathbf{1.5\cdot10^2}$                 & 21.27                              & 0.24                               & 0.70                   &     \ding{51}           \\ 
$\mathbf{2\cdot10^2}$                & 21.31                              & 0.24                               & 0.70                   &     \ding{51}           \\ 
$\mathbf{5\cdot10^2}$                  & \textbf{21.36}                              & 0.24                               & \textbf{0.71}          &    $\sim$                   \\ 
$\mathbf{7.5\cdot10^2}$                   & 21.18                              & 0.24                               & 0.70                 &     $\sim$             \\ 
\end{tabular}
\caption{PnP-MLA, mean over \texttt{set3c}, $\epsilon=20$.  Instability shows through artefacts (stripes) that appear in one of the images in the data set, which increase in strength as the step size increases.}
\label{tab:mirror-step-size}
\end{table}

We compare the convergence speed for selected step sizes in Figure \ref{fig:conv-speed-best-timesteps}. We compute the cumulative metrics (depicted here are PSNR and LPIPS) against neural function evaluations (NFE) and take the mean over the \texttt{set3c} data set. We observe that in terms of convergence speed  RPnP-SKROCK outperforms RPnP-ULA and PPnP-ULA (which are fairly similar in speed) and the PnP-MLA for both chosen step sizes. Furthermore, the behaviour in terms of PSNR is rather monotonic as NFEs increase, but this is not necessarily the case for LPIPS with the most visible non-monotonic behaviour occurring for PnP-MLA with the best LPIPS obtained overall around $10^{4}$ NFEs.

\begin{figure}[t!]
    \centering
    \begin{subfigure}{0.5\textwidth}
        \includegraphics[width=\textwidth]{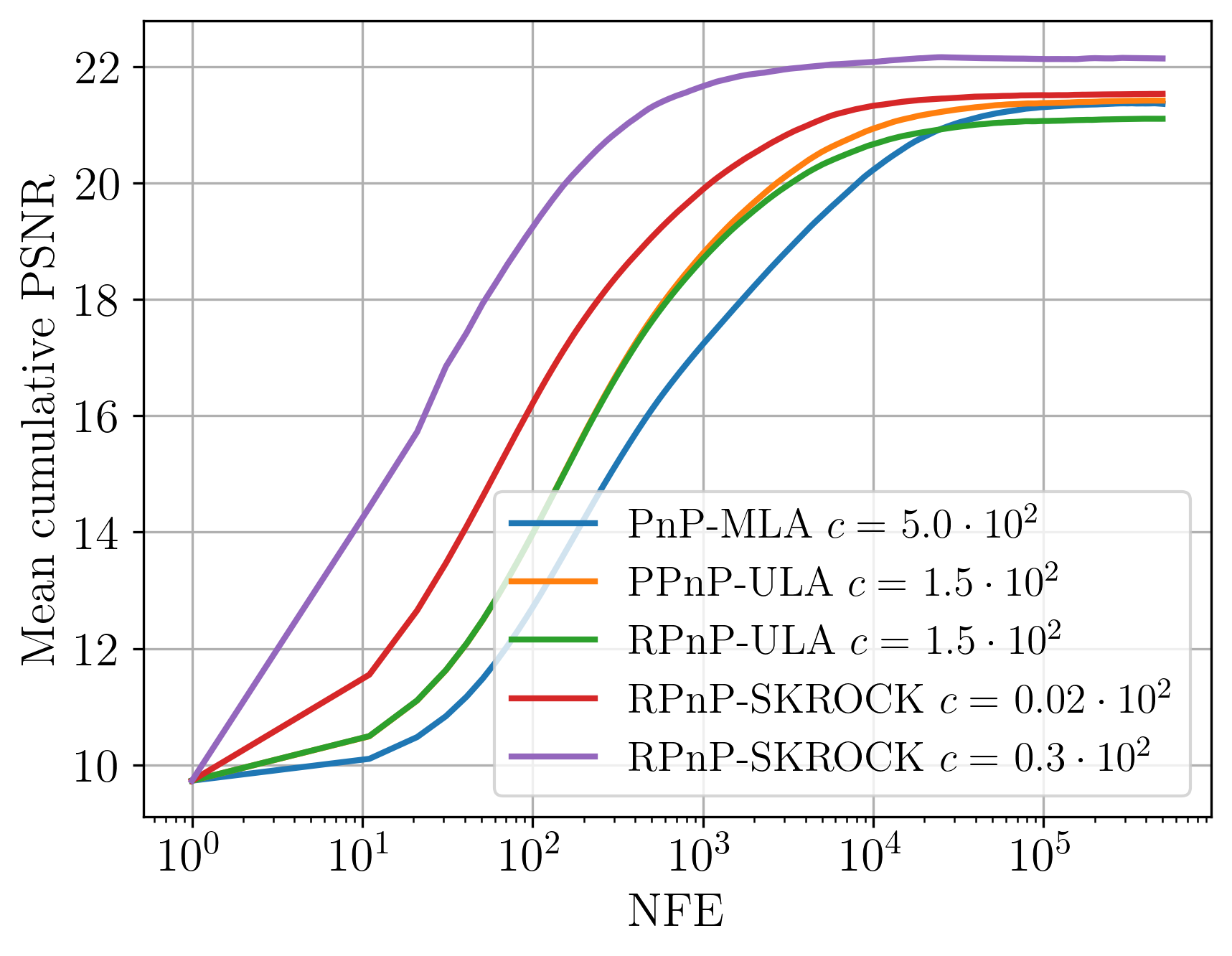}
        \subcaption{PSNR $\uparrow$}
    \end{subfigure}%
    \begin{subfigure}{0.5\textwidth}
        \includegraphics[width=\textwidth]{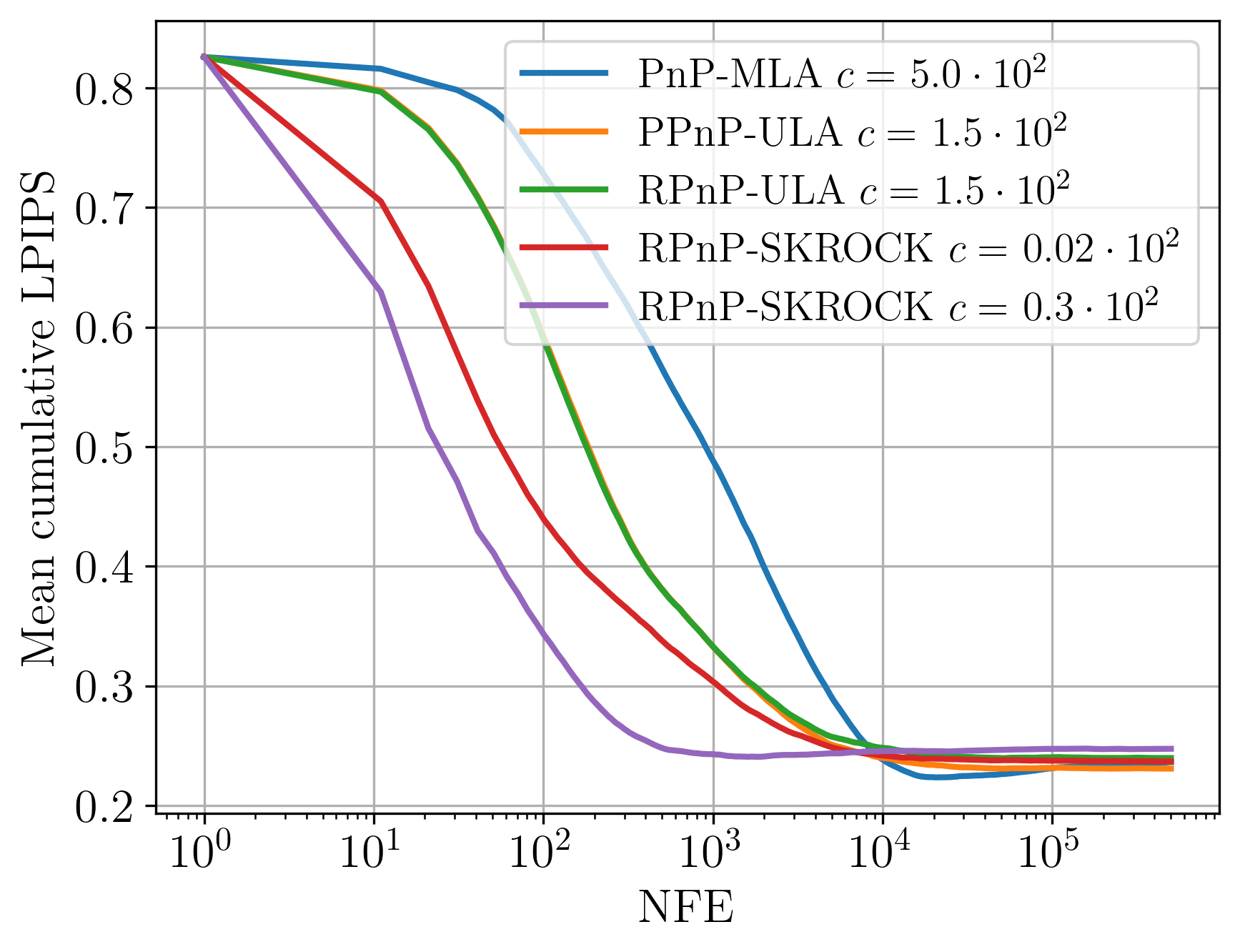}
        \subcaption{LPIPS $\downarrow$}
    \end{subfigure}
    \caption{Mean cumulative metrics for the \texttt{set3c} data set using selected time steps ($\epsilon=20$ for all algorithms). }
    \label{fig:conv-speed-best-timesteps}
\end{figure}

\section{Additional information for Section \ref{sec:sota}}\label{sec:app-sota}
\subsection{Scatter plots} 
The plots in Figure \ref{fig:scatter} depict detailed results of each image in the data set used for the experiments in Section \ref{sec:sota}. Each symbol denotes a different image, and each algorithm is depicted in a distinct color.

\begin{figure}[h!]
    \centering

\begin{minipage}{0.85\linewidth}
    \begin{subfigure}[b]{0.45\textwidth}
        \centering
        \includegraphics[width=\textwidth]{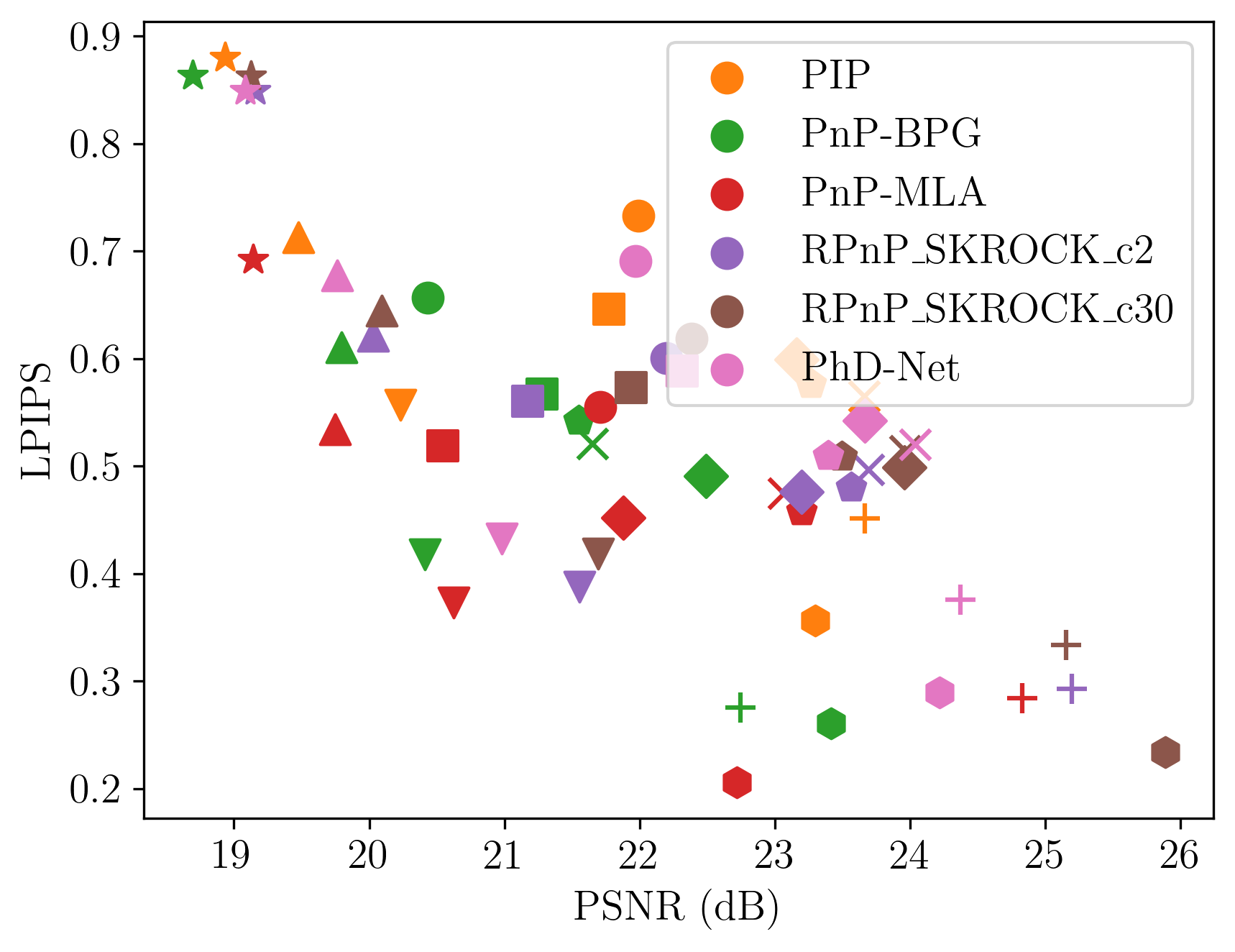}
        \caption{$\alpha=5$, Blur 1}
        \label{fig:scatter-a5k4}
    \end{subfigure}
    \hfill
    \begin{subfigure}[b]{0.45\textwidth}
        \centering
        \includegraphics[width=\textwidth]{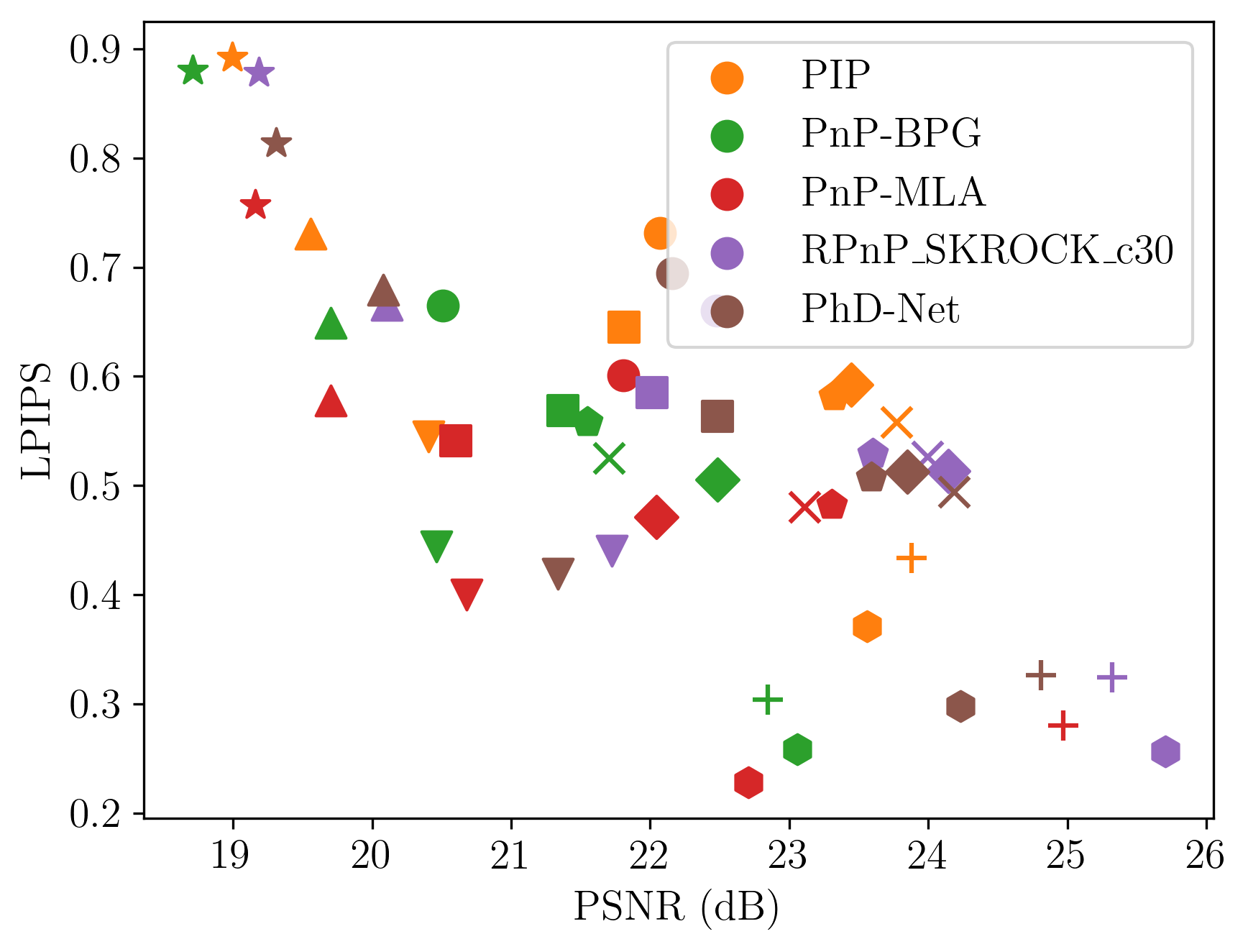}
        \caption{$\alpha=5$, Blur 2}
        \label{fig:scatter-a5k9}
    \end{subfigure}
    
    \vskip\baselineskip
    \begin{subfigure}[b]{0.45\textwidth}
        \centering
        \includegraphics[width=\textwidth]{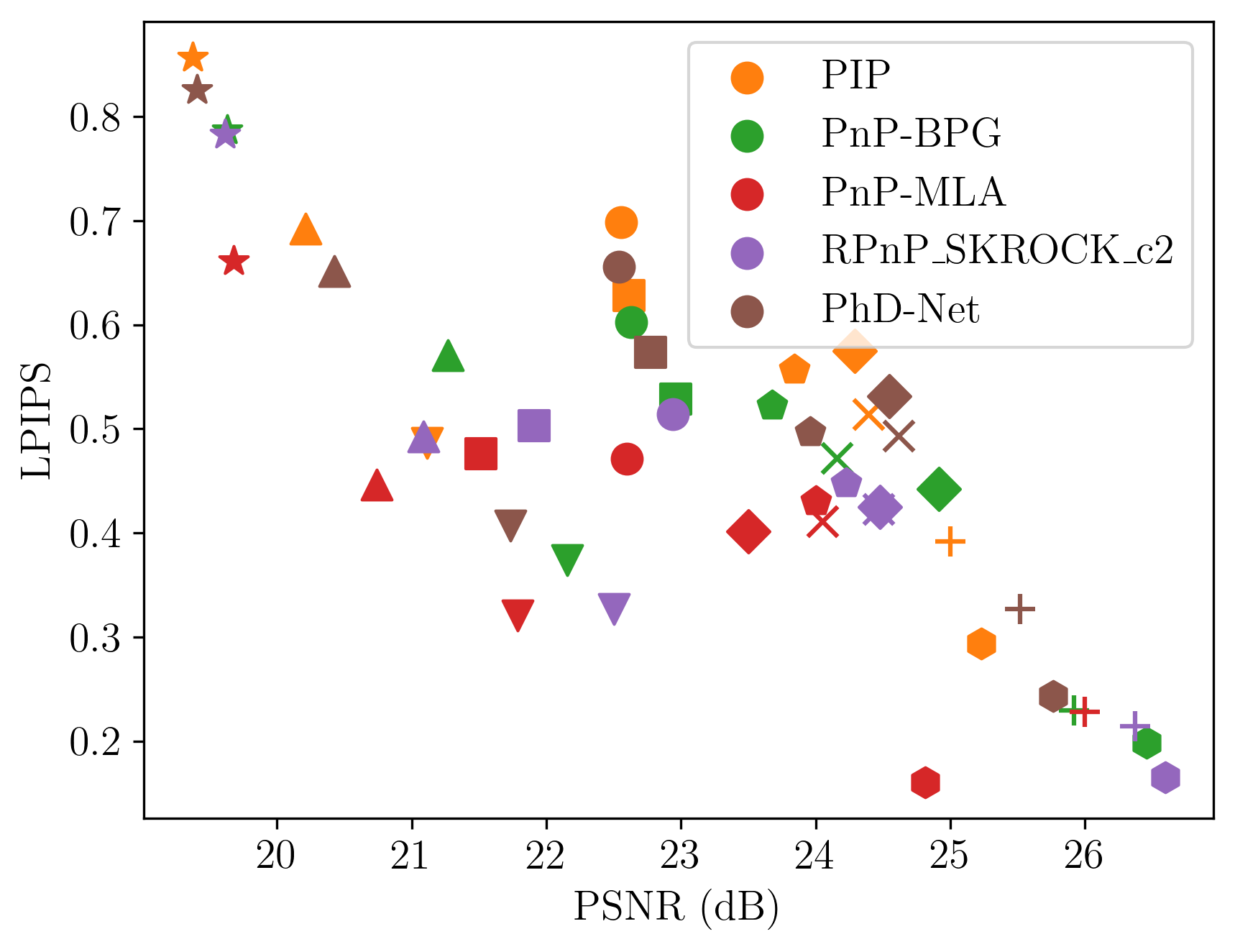}
        \caption{$\alpha=10$, Blur 1}
        \label{fig:scatter-a10k4}
    \end{subfigure}
    \hfill
    \begin{subfigure}[b]{0.45\textwidth}
        \centering
        \includegraphics[width=\textwidth]{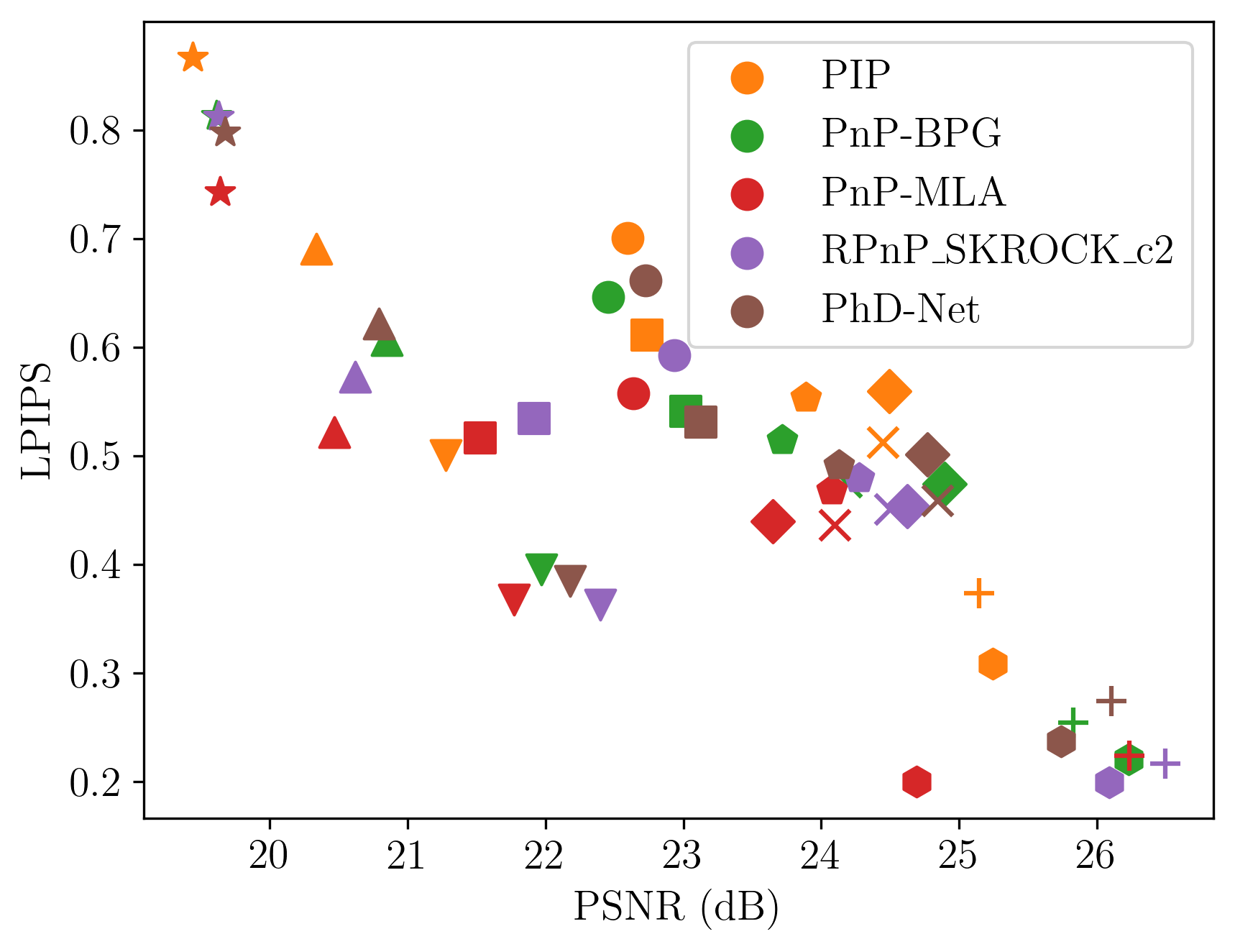}
        \caption{$\alpha=10$, Blur 2}
        \label{fig:scatter-a10k9}
    \end{subfigure}
    
    \vskip\baselineskip
    \begin{subfigure}[b]{0.45\textwidth}
        \centering
        \includegraphics[width=\textwidth]{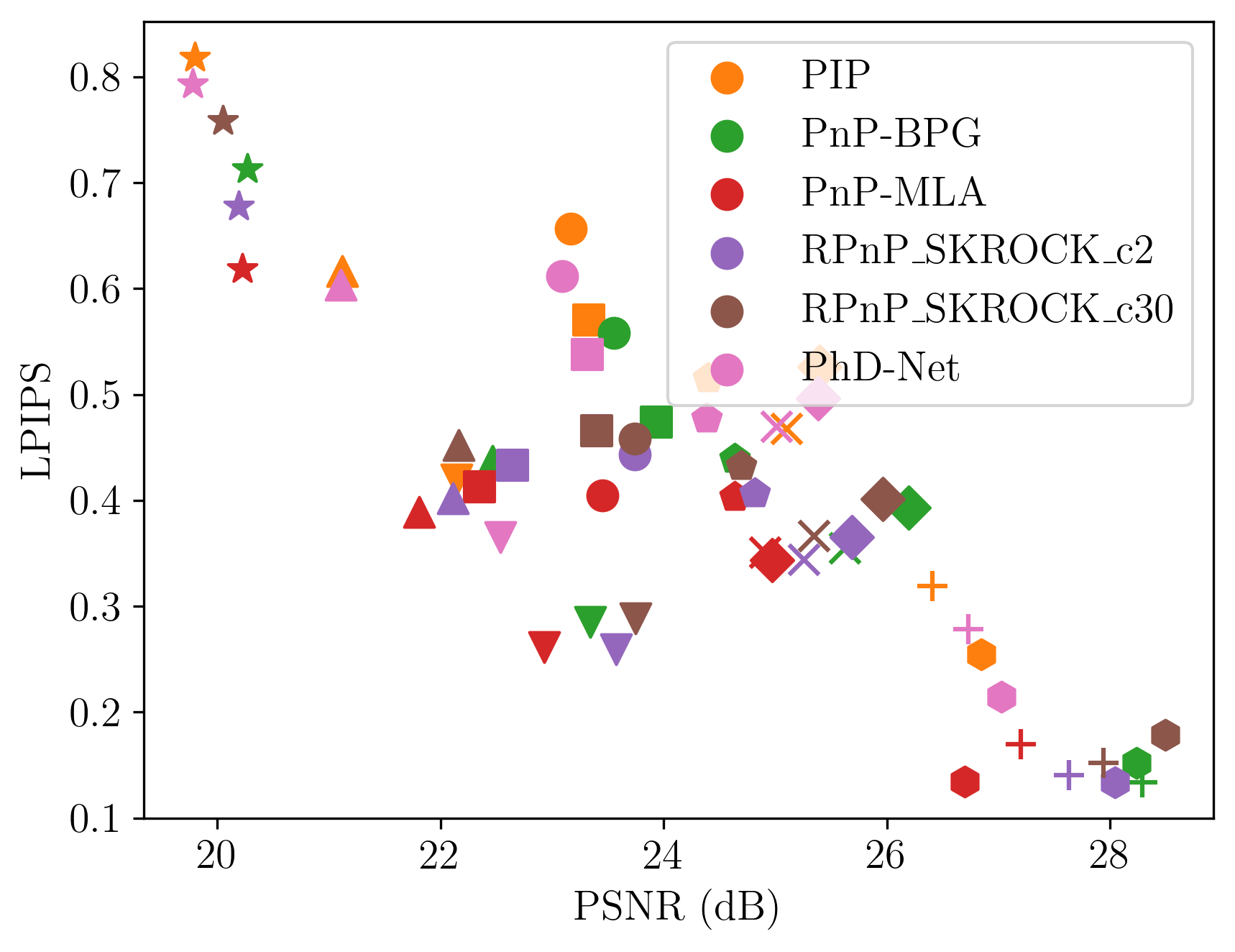}
        \caption{$\alpha=20$, Blur 1}
        \label{fig:scatter-a20k4}
    \end{subfigure}
    \hfill
    \begin{subfigure}[b]{0.45\textwidth}
        \centering
        \includegraphics[width=\textwidth]{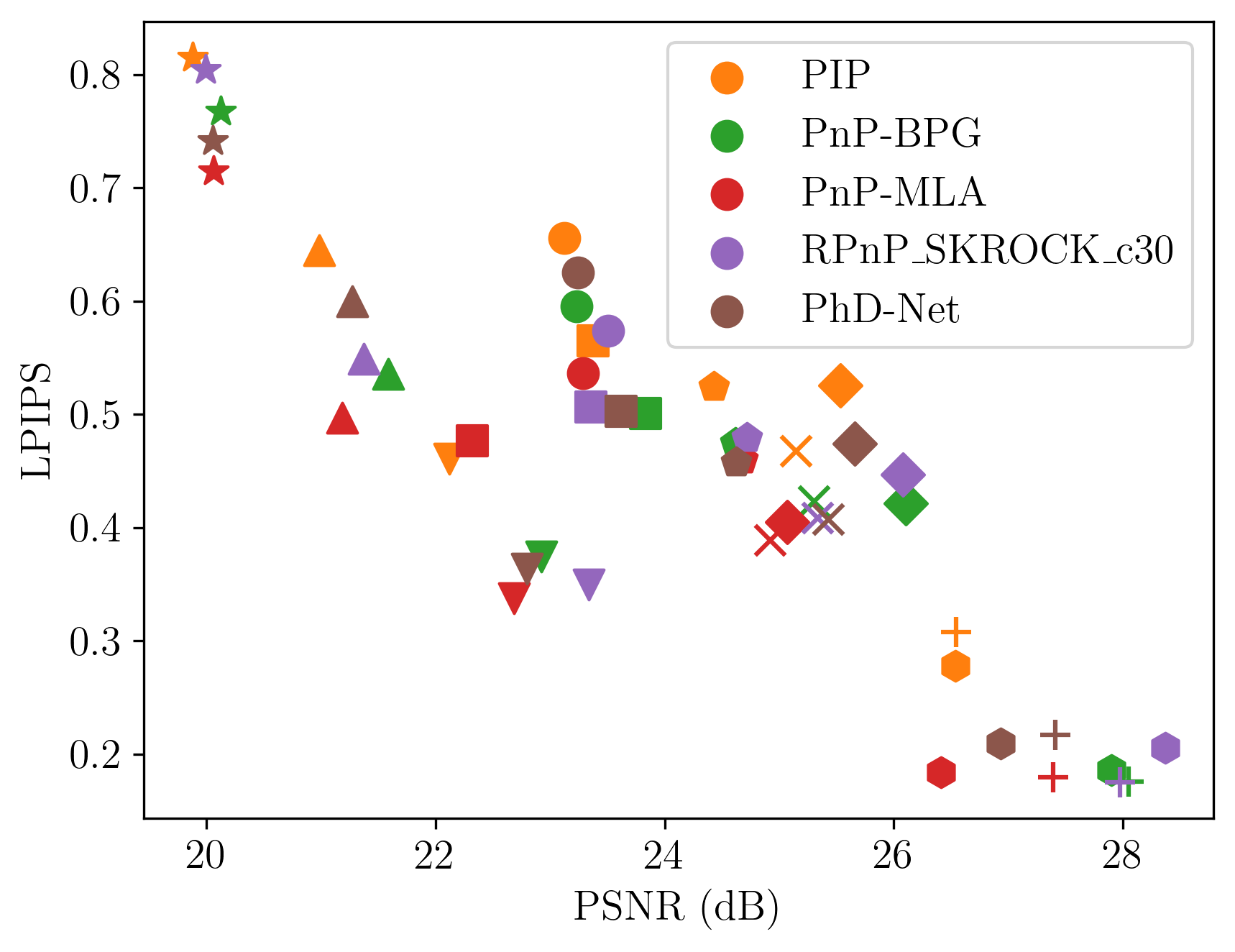}
        \caption{$\alpha=20$, Blur 2}
        \label{fig:scatter-a20k9}
    \end{subfigure}
    
    \caption{Scatter plots to support Section \ref{sec:sota}. Each marker marks a distinct image, the colors correspond to different methods. On the LPIPS axis, lower is better; on the PSNR axis, higher is better.}
    \label{fig:scatter}

\end{minipage}
\end{figure}

\end{document}